%% file: clusters_idr5_v2.tex
\documentclass[a4paper,fleqn,usenatbib]{mnras}
\usepackage[T1]{fontenc}
\usepackage{ae,aecompl}
\usepackage{graphicx}
\usepackage{amsmath}
\usepackage{amssymb}
\usepackage{natbib}
\usepackage{epstopdf}
\include{latex_macros}

\usepackage{float}

\usepackage{times}
\usepackage{amsmath}
\usepackage{amssymb}
\usepackage{amsfonts}
\usepackage{mathrsfs}
\usepackage{graphicx}

\usepackage{rotating}

\title[Membership probabilities in 32 GES open clusters]{The Gaia-ESO Survey: Membership probabilities for stars in 32 open clusters from 3D kinematics}

\author[R. J. Jackson, R. D. Jeffries, N. J. Wright et al.]{R. J.~Jackson$^1$, R. D.~Jeffries$^1$, N. J.~Wright$^1$, S.~Randich$^2$, G.~Sacco$^2$, E. Pancino$^2$,
\newauthor T.~Cantat-Gaudin$^3$, G.~Gilmore$^4$, A.~Vallenari$^5$, T.~Bensby$^6$,  A.~Bayo$^7$, M.~T.~Costado$^8$, \newauthor E.~Franciosini$^2$, A.~Gonneau$^4$, A.~Hourihane$^4$, J.~Lewis$^4$, L. Monaco$^{9}$, L.~Morbidelli$^2$, \newauthor and  C.~Worley$^4$\\
\\
    $^1$ Astrophysics Group, Keele University, Keele, Staffordshire ST5 5BG\\
    $^2$ INAF -- Osservatorio Astrofisco di Arcetri, Largo E. Fermi, 5, 50125 Firenze, Italy\\
    $^3$ Institut de Ci\`encies del Cosmos, Universitat de Barcelona (IEEC-UB), Mart\'i i Franqu\`es 1, E-08028 Barcelona, Spain\\  
    $^4$ Institute of Astronomy, Cambridge University, Madingley Road, Cambridge CB3 0H\\
    $^5$ INAF -- Osservatorio Astronomico di Padova, Vicolo dell’Osservatorio 5, 35122 Padova, Italy\\
    $^6$ Lund Observatory, Department of Astronomy and Theoretical Physics, Box 43, SE-221 00 Lund, Sweden\\
    $^7$ Instituto  de  F\'isica  y Astronom\'ia,  Facultad  de Ciencias,  Universidad de Valpara\'iso, Av. Gran Breta\~na 1111, 5030 Casilla, Valpara\'iso, Chile\\
    $^8$ Departamento de Did\'actica, Universidad de C\'adiz, 11519 Puerto Real, C\'adiz, Spain\\
    $^{9}$ Departamento de Ciencias Fisicas, Universidad Andres Bello, Fernandez Concha 700, Las Condes, Santiago, Chile}

\date{Accepted for MNRAS}

\pagerange{\pageref{firstpage}--\pageref{lastpage}} \pubyear{2020}

\def\LaTeX{L\kern-.36em\raise.3ex\hbox{a}\kern-.15em
    T\kern-.1667em\lower.7ex\hbox{E}\kern-.125emX}

\begin{document}
\label{firstpage}
\maketitle

\begin{abstract}

The {\it Gaia}-ESO Survey (GES) observed many open clusters as part of its programme to spectroscopically characterise the various Milky Way populations. GES spectroscopy and {\it Gaia} astrometry from its second data release are used here to assign membership probabilities to targets towards 32 open clusters with ages from 1--3800 Myr, based on maximum likelihood modelling of the 3D kinematics of the cluster and field populations. From a parent catalogue of 14398 individual targets, 5033 stars with uniformly determined 3D velocities, $T_{\rm eff}$, $\log g$ and chemistry are assigned cluster membership with probability $>0.9$, and with an average probability of 0.991. The robustness of the membership probabilities is demonstrated using independent membership criteria (lithium and parallax) in two of the youngest clusters. The addition of radial velocities improves membership discrimination over proper motion selection alone, especially in more distant clusters. The kinematically-selected nature of the membership lists, independent of photometry and chemistry, makes the catalogue a valuable resource for testing stellar evolutionary models and investigating the time evolution of various parameters.

\end{abstract}


\section{Introduction}
\label{intro}
Open star clusters and associations play a fundamental role in our
understanding of stellar evolution, in the testing of stellar models
and in anchoring the age scale of stars. They offer samples of stars at
a range of masses and evolutionary stages, but with very similar ages
and (initial) compositions. Stars in a single cluster can be used to
test the mass-dependence predicted by models or to use models to
estimate masses; whilst the comparison of clusters across a range of
ages can be used to test the time- or chemical composition-dependence predicted
by models and to explore phenomena empirically that are poorly
understood from a physical point of view. A non-exhaustive list of
examples would include: testing how models predict the positions of
stars in the Hertzsprung-Russell diagram; estimating the stellar
initial mass function and identifying substellar objects; calibrating
white dwarf cooling models; following the spin-down of stars and
calibrating gyrochronology; and investigating the depletion of light
elements in stellar interiors.

The {\it Gaia}-ESO Survey (GES) is a large public survey
programme executed on the 8-m UT2-{\it Kueyen} telescope of the Very Large Telescope facility. The
survey rationale, methodology and calibration strategy are detailed in \cite{Gilmore2012a}, \cite{Randich2013a} and \cite{Pancino2017a}. Over the course of about 6 years, beginning
on 31 December 2011, medium ($R \sim 17\,000$) and high ($R \sim
47\,000$) resolution multi-fibre spectroscopy were obtained, using FLAMES \citep[Fiber Large Array Multi-Element Spectrograph,][]{Pasquini2002a} combined with the
GIRAFFE and UVES \citep[Ultraviolet and Visual Echelle Spectrograph,][]{Dekker2000a} spectrographs, of about $10^5$ and $10^4$ stars in our Galaxy. The survey
had the aim of understanding, through measurements of kinematics and
chemical abundances, the formation and evolution of all the components
of our Galaxy, and included a significant proportion ($\sim$40 per
cent) of time devoted to studying star clusters and associations at a
range of ages. At the time of writing, GES has internally delivered
radial velocities and chemical abundances for 32 clusters as part of
the internal Data Release 5 (hereafter GESiDR5).

  A pre-requisite for most studies
using star clusters is to accurately assess which stars are actually
members, in the presence of contaminating sources. Many
different methods can be used to filter stars -- positions, kinematics,
spectroscopic parameters, abundances, photometry, but it is important
that the filtering criteria are understood and do not bias any
subsequent investigation of cluster properties by using those same
properties to select cluster members. The addition of {\it Gaia}
astrometric data \citep{Gaia2016a}, in the form of its first and second data 
releases \citep[{\it Gaia} DR1 and {\it Gaia} DR2,][]{Gaia2016a, Gaia2018a}, has
dramatically enhanced our capability to separate cluster members from
unrelated field stars using proper motion and parallax \citep[][]{Gaia2018c, CantatGaudin2018a}.

The work presented here follows on from \cite{Randich2018a}, where
{\it Gaia} astrometry from {\it Gaia} DR1 
was used in conjunction with spectroscopic
parameters from GES to define samples of high probability cluster members
for eight open clusters. Here we describe a closely related methodology that
uses temperatures, gravities and radial velocities from GESiDR5, together
with astrometry from {\it Gaia} DR2, to define membership probabilities for
 sources in 32 GES clusters based on their three-dimensional kinematics. The inclusion of the third dimension of radial velocity from GES in stars as faint as $V \sim 19$ improves our ability to separate cluster members from contaminants over studies using proper motion alone, especially in the more distant clusters. The aim is to provide rigorously determined membership lists, with quantitative membership probabilities, that can be used for a host of follow-up investigations.

\section{Potential cluster members}
\label{2}
\subsection{Source data}
\label{2.1}
GES Data for 32 open clusters was taken from the GESiDR5 analysis iteration in the GES archive at the Wide Field Astronomy Unit of Edinburgh University\footnote{http//ges/roe.ac.uk/}.
Table~\ref{table1} shows a list of cluster names together with initial values of age, distance modulus and reddening reported in the literature. Also shown are the number of targets in each cluster that were observed using the GIRAFFE 665\,nm filter (HR15n) and/or the UVES 520\,nm or 580\,nm filters. A summary of the target selection strategy for the GES clusters can be found in \cite{Randich2018a} and summaries of the spectroscopic data and analyses can be found in \cite{Sacco2014a}, \cite{Damiani2014a}, \cite{Jeffries2014a}, \citet{Smiljanic2014a} and \cite{Jackson2015a}.

The GES data are not complete in any sense. Only a (variable) fraction of members will have been observed in each cluster, either because of the inability to cover the full spatial extent of the cluster (particularly those that are nearby and of large angular extent), the inability to assign fibers to all the targets or in a few cases because the data quality were insufficient to provide the necessary parameters for further analysis (see below). Our philosophy for membership selection is therefore not to strive to be as complete as possible, but to aim to provide a secure list of kinematically selected members where any contamination is accurately accounted for by the membership probabilities. 

\begin{table*}
\caption{Cluster data. Columns 2--4 show ages, intrinsic distance moduli and reddening from the literature (superscripts refer to references listed below the Table). Columns 5--7 show the numbers of targets observed, the numbers of targets with a full set of the required data (see Section~\ref{2.2}) and the number fitted in the  membership analysis. Column 8 shows the spectral resolving power measured from arc lamp line widths (see Appendix~\ref{appa1}) and columns 9-10 show the final mean values of distance modulus and reddening for the clusters, determined from high probability cluster members; the first error bar on the distance modulus is a statistical uncertainty, the second is a systematic uncertainty corresponding to $\pm 0.1$ mas in parallax (see Section~\ref{3.4}).} \label{table1}
\begin{tabular}{lllrrrrrrr}
\hline 
Cluster	 &	Age	&	$(M-m)_0$	&	$E(B-V)$	&	Number	&	Number	&	Number	&	Resolving	&	$(M-m)_0^{\rm c}$	&	$E(B-V)^{\rm c}$	\\
&	(Myr)	&	Literature	&	Literature	&	observed	&	complete	&	fitted	&	power	&	members	&	members	\\\hline
Trumpler 14  	&	1--3$^{19}$	&	12.3$^{19}$	&	0.4--0.9$^{19}$	&	1118	&	1063	&	729	&	12951	&	12.15$\pm$0.03$\pm$0.60	&	0.71$\pm$0.14	\\
Chamaeleon I 	&	2$^{25}$	&	6.02$^{44}$	&	$\sim$1$^{25}$	&	720	&	649	&	148	&	13897	&	6.39$\pm$0.01$\pm$0.04	&	0.19$\pm$0.10	\\
NGC 6530     	&	1--7$^{32}$	&	10.48$^{32}$	&	0.35$^{38}$	&	1980	&	1294	&	1075	&	13561	&	10.66$\pm$0.03$\pm$0.30	&	0.48$\pm$0.10	\\
NGC 2264     	&	3$^{40}$	&	9.4$^{37}$	&	0.07$^{41}$	&	1884	&	1738	&	1344	&	14968	&	9.35$\pm$0.01$\pm$0.16	&	0.03$\pm$0.07	\\
Rho Ophiuchus	&	3$^{15}$	&	5.4$^{24}$	&	---	&	313	&	298	&	70	&	15854	&	5.72$\pm$0.01$\pm$0.03	&	0.58$\pm$0.20	\\
Lambda Ori   	&	6$^{14}$	&	7.9$^{13}$	&	0.12$^{11}$	&	618	&	546	&	296	&	17281	&	8.02$\pm$0.01$\pm$0.09	&	0.06$\pm$0.03	\\
Gamma 2 Vel  	&	5--10$^{22}$	&	7.72$^{22}$	&	0.04$^{22}$	&	1283	&	1283	&	496	&	14301	&	7.76$\pm$0.01$\pm$0.08	&	0.02$\pm$0.03	\\
NGC 2232     	&	32$^{26}$	&	7.56$^{16}$	&	0.03$^{43}$	&	1769	&	1764	&	760	&	12402	&	7.56$\pm$0.01$\pm$0.07	&	0.06$\pm$0.04	\\
NGC 2547     	&	35$^{21}$	&	7.97$^{16}$	&	0.06$^{29}$	&	480	&	480	&	267	&	13862	&	7.97$\pm$0.01$\pm$0.09	&	0.08$\pm$0.05	\\
IC 4665      	&	42$^{5}$	&	7.69$^{16}$	&	0.17$^{5}$	&	567	&	567	&	300	&	15332	&	7.71$\pm$0.02$\pm$0.08	&	0.13$\pm$0.04	\\
IC 2602      	&	46$^{12}$	&	5.91$^{16}$	&	0.03$^{17}$	&	1861	&	1794	&	116	&	13542	&	5.91$\pm$0.01$\pm$0.03	&	0.01$\pm$0.02	\\
NGC 2451b    	&	50$^{18}$	&	7.84$^{18}$	&	0.01$^{31}$	&	1657	&	1655	&	418	&	13862	&	7.84$\pm$0.02$\pm$0.08	&	0.07$\pm$0.03	\\
IC 2391      	&	53$^{3}$	&	5.9$^{16}$	&	0.01$^{43}$	&	438	&	420	&	67	&	12963	&	5.92$\pm$0.01$\pm$0.03	&	0.02$\pm$0.02	\\
NGC 2451a    	&	50-80$^{18}$	&	6.44$^{16}$	&	0.01$^{31}$	&	1657	&	1655	&	354	&	13862	&	6.44$\pm$0.01$\pm$0.04	&	0.02$\pm$0.02	\\
NGC 2516     	&	125$^{26}$	&	8.09$^{16}$	&	0.11$^{39}$	&	764	&	764	&	643	&	13440	&	8.08$\pm$0.01$\pm$0.09	&	0.13$\pm$0.03	\\
NGC 6067     	&	120$^{43}$	&	10.76$^{43}$	&	0.38$^{43}$	&	532	&	530	&	489	&	17279	&	11.79$\pm$0.02$\pm$0.50	&	0.32$\pm$0.04	\\
Blanco 1     	&	100--150$^{28}$	&	6.88$^{16}$	&	0.01$^{43}$	&	468	&	468	&	326	&	17282	&	6.89$\pm$0.01$\pm$0.05	&	0.01$\pm$0.04	\\
NGC 6259     	&	210$^{27}$	&	11.61$^{9}$	&	0.66$^{27}$	&	447	&	447	&	373	&	17359	&	11.91$\pm$0.03$\pm$0.53	&	0.69$\pm$0.06	\\
NGC 6705     	&	250$^{35}$	&	11.37$^{23}$	&	0.42$^{35}$	&	1070	&	1070	&	963	&	14393	&	11.90$\pm$0.01$\pm$0.53	&	0.36$\pm$0.08	\\
NGC 4815     	&	500$^{6}$	&	11.99$^{6}$	&	0.7$^{6}$	&	126	&	126	&	105	&	14012	&	12.87$\pm$0.11$\pm$0.85	&	0.67$\pm$0.11	\\
NGC 6633     	&	575$^{42}$	&	7.99$^{16}$	&	0.17$^{42}$	&	1600	&	1598	&	143	&	14532	&	7.97$\pm$0.01$\pm$0.09	&	0.16$\pm$0.02	\\
Trumpler 23  	&	900$^{4}$	&	11.71$^{7}$	&	0.58$^{4}$	&	89	&	89	&	77	&	13600	&	12.29$\pm$0.04$\pm$0.64	&	0.74$\pm$0.04	\\
NGC 6802     	&	950$^{20}$	&	11.28$^{20}$	&	0.84$^{20}$	&	103	&	103	&	94	&	14309	&	12.72$\pm$0.13$\pm$0.79	&	0.82$\pm$0.07	\\
Berkeley 81  	&	1000$^{33}$	&	12.39$^{33}$	&	1.0$^{33}$	&	203	&	203	&	169	&	13849	&	12.89$\pm$0.10$\pm$0.86	&	0.94$\pm$0.05	\\
Ruprecht 134 	&	1000$^{7}$	&	12.66$^{7}$	&	0.5$^{7}$	&	680	&	680	&	415	&	17299	&	12.09$\pm$0.05$\pm$0.58	&	0.47$\pm$0.07	\\
NGC 6005     	&	1200$^{30}$	&	12.16$^{30}$	&	0.45$^{30}$	&	355	&	355	&	275	&	13771	&	12.38$\pm$0.10$\pm$0.67	&	0.27$\pm$0.08	\\
Pismis18     	&	1200$^{30}$	&	11.75$^{30}$	&	0.5$^{30}$	&	101	&	101	&	86	&	13965	&	12.43$\pm$0.04$\pm$0.69	&	0.69$\pm$0.05	\\
Trumpler 20  	&	1400$^{8}$	&	12.39$^{8}$	&	0.35$^{8}$	&	557	&	557	&	447	&	13743	&	13.09$\pm$0.04$\pm$0.96	&	0.38$\pm$0.10	\\
NGC 2420     	&	2200$^{34}$	&	11.97$^{36}$	&	0.05$^{2}$	&	563	&	563	&	514	&	12408	&	12.20$\pm$0.02$\pm$0.61	&	0.01$\pm$0.03	\\
Berkeley 31  	&	2900$^{10}$	&	14.4$^{10}$	&	0.19$^{10}$	&	616	&	616	&	454	&	13048	&	14.39$\pm$0.23$\pm$2.14	&	0.05$\pm$0.12	\\
Berkeley 44  	&	2900$^{20}$	&	12.46$^{20}$	&	0.98$^{20}$	&	93	&	93	&	82	&	13600	&	12.63$\pm$0.08$\pm$0.76	&	0.87$\pm$0.07	\\
NGC 2243     	&	3800$^{1}$	&	12.96$^{1}$	&	0.05$^{1}$	&	705	&	705	&	564	&	14051	&	13.41$\pm$0.02$\pm$1.14	&	0.01$\pm$0.11	\\
\hline
 \end{tabular}
\begin{flushleft}
$^{1}$\cite{Anthony-Twarog2005a}
$^{2}$\cite{Anthony-Twarog2006a}
$^{3}$\cite{Barrado1999a}
$^{4}$\cite{Bonatto2007a}
$^{5}$\cite{Cargile2010a}
$^{6}$\cite{Carraro1994a}
$^{7}$\cite{Carraro2006a}
$^{8}$\cite{Carraro2010a}
$^{9}$\cite{Ciechanowska2006a}
$^{10}$\cite{Cignoni2011a}
$^{11}$\cite{Diplas1994a}
$^{12}$\cite{Dobbie2010a}
$^{13}$\cite{Dolan1999a}
$^{14}$\cite{Dolan2002a}
$^{15}$\cite{Erickson2011a}
$^{16}$\cite{Gaia2018a}
$^{17}$\cite{Hill1969a}
$^{18}$\cite{Hunsch2003a}
$^{19}$\cite{Hur2012a}
$^{20}$\cite{Janes2011a}
$^{21}$\cite{Jeffries2005a}
$^{22}$\cite{Jeffries2009a}
$^{23}$\cite{Jeffries2017a}
$^{24}$\cite{Loinard2008a}
$^{25}$\cite{Luhman2007a}
$^{26}$\cite{Lyra2006a}
$^{27}$\cite{Mermilliod2001a}
$^{28}$\cite{Moraux2007a}
$^{29}$\cite{Naylor2006a}
$^{30}$\cite{Piatti1998a}
$^{31}$\cite{Platais2001a}
$^{32}$\cite{Prisinzano2005a}
$^{33}$\cite{Sagar1998a}
$^{34}$\cite{Salaris2004a}
$^{35}$\cite{Santos2005a}
$^{36}$\cite{Sharma2006a}
$^{37}$\cite{Sung1997a}
$^{38}$\cite{Sung2000a}
$^{39}$\cite{Sung2002a}
$^{40}$\cite{Sung2004a}
$^{41}$\cite{Turner2012a}
$^{42}$\cite{vanLeeuwen2009a}
$^{43}$WEBDA \citep{Dias2002a}
$^{44}$\cite{Whittet1997a}
\end{flushleft}
\label{cluster properties}
\end{table*}

Table~\ref{table2} lists the parameters associated with the summed spectrum for a particular target observed with a particular instrumental setup. Most targets were observed with either GIRAFFE or UVES though a single wavelength filter; the 26000 lines in Table~\ref{table2} represent 25417 unique targets. Wherever possible, values for effective temperature ($T_{\rm eff}$ in K), surface gravity ($\log g$, with $g$ in cm\,s$^{-2}$) and the gravity-sensitive spectroscopic index  $\gamma$ \citep{Damiani2014a} were taken from the {\it Recommended$\_$Astro$\_$Analysis} database. If no value is given in that database then the parameter is taken from the {\it  Astro$\_$Analysis} data base. If the {\it Astro$\_$Analysis} database shows multiple values from different working groups then Table~\ref{table2} shows their median value. 

Table~\ref{table2} shows the GESiDR5 values of radial velocity (RV) and uncertainty for each target/filter combination. For targets observed with the GIRAFFE 665~nm filter, an improved empirical precision, $S_{\rm RV}$, is calculated from the target signal-to-noise ratio ($S/N$) and projected equatorial velocity ($v\sin i$), following the method described in \cite{Jackson2015a}, and using empirical constants determined from the analysis of GESiDR5 cluster data (see Appendix~\ref{appa}). The values of $S_{\rm RV}$ are shown in Table~\ref{table2} and used in the subsequent membership analysis.

\begin{figure*}
	\centering
	\begin{minipage}[ht]{\textwidth}
	\centering
  \includegraphics[width = 178mm]{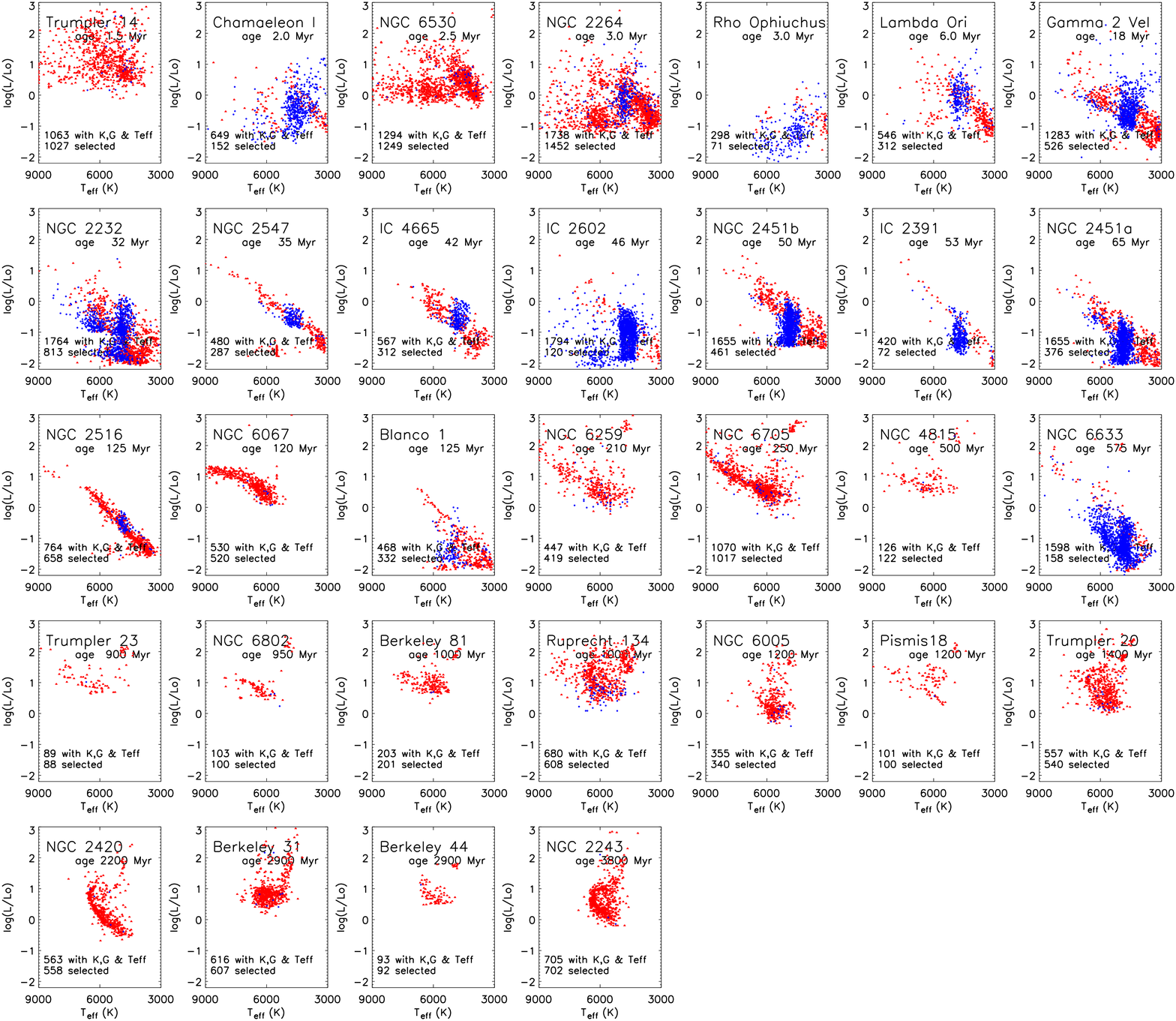}
	\end{minipage}
	\caption{Cluster HR diagrams based on targets that have a 2MASS $K_s$ magnitude, {\it Gaia} $G$ magnitude and GES $T_{\rm eff}$ estimate. Blue points are classified as probable field giants or other distant field stars, red points are potential cluster members selected for further analysis (see Section~\ref{2.2}) Text on the plots shows the number of targets with complete data ($K_s$ and $G$ magnitude and $T_{\rm eff}$ ) and the number selected as potential cluster members having no strong indication of being a background giant or other distant field star.}
	\label{fighrd}
\end{figure*}

GES values of RA and Dec in Table~\ref{table2} were cross-matched with the {\it Gaia} DR2 catalogue to obtain photometry, parallax and proper motion data. Table~\ref{table3} shows the tangential velocities in units of km\,s$^{-1}$ calculated as
\begin{eqnarray}
V_{\rm RA} & = & 4.74 d_{\rm c}\,{\rm pm_{\rm RA}}\, , \nonumber \\
V_{\rm Dec} & = & 4.74 d_{\rm c}\,{\rm pm_{\rm Dec}}\, ,
\end{eqnarray}
where pm$_{\rm RA}$ and pm$_{\rm Dec}$ are the proper motions in units of mas\,yr$^{-1}$, and $d_c$ (in pc) is the cluster distance used for the analysis of cluster membership. The implicit assumption here is that all stars are at a common distance. This will be a good approximation for cluster members, but for unassociated field stars $d_{\rm c}$ is simply a scaling constant that multiplies their proper motions prior to the analysis. The {\it Gaia} DR2 data were not filtered for possible problems with the astrometry \citep{Lindegren2018a}. The issue of astrometric reliability is discussed further in Section~\ref{5.1}.

\subsection{Selecting potential cluster members}
\label{2.2}
Potential cluster members were selected from the list of observed targets in Table~\ref{table2}. Targets were first selected as having reported values of 2MASS $K_s$ magnitude, {\it Gaia} $G$ magnitude and cross-match radius $<2$\,arcsec, $T_{\rm eff}$ and a spectral $S/N\geq 5$. The GES cluster targets were based on 2MASS coordinates and all have 2MASS data, but 75 were excluded here with no {\it Gaia} match. Figure~\ref{fighrd} shows Hertzsprung-Russell (HR) diagrams of targets in each cluster. The luminosity is estimated from the $K_s$ magnitude as 
\begin{equation}
\log L/L_{\odot} = (4.75 - M_{K_s} - BC_{K_s})/2.5\, ,
\end{equation}
where 
\begin{equation}
M_{K_s} = K_s - (M-m)_0^{\rm c} - R_{K_s}\, {E(B-V)}^{\rm c}\, ,
\end{equation}
where $(M-m)_0^{\rm c}$ and ${E(B-V)}^{\rm c}$ are the adopted intrinsic distance modulus and reddening of the cluster.

The bolometric correction $BC_{K_s}$ is estimated from the de-reddened $G-K_s$ using solar-metallicity Pisa model isochrones \citep{Tognelli2011a} at the cluster age , $(M-m)_0^{\rm c}$ and $E(B-V)^{\rm c}$, and assuming  extinctions in the $G$ and $K_s$ bands of  $R \times E(B-V)^{\rm c}$ where $R_G=2.50$ \citep{Chen2019a} and $R_{K_s}=0.35$ \citep{Yuan2013a}.  These coefficients are an approximate average for all spectral types and the possibility of disks and anomalous reddening in the youngest clusters is ignored. Since the only purpose (in this paper) of calculating luminosities is to estimate approximate masses to use in a small correction to the binary RV uncertainty distribution (see Section~\ref{3.1}), further detail is not warranted.

In the first analysis pass, the intrinsic distance modulus and reddening of each cluster were taken from the literature (columns 3-4 in Table~\ref{table1}). A subsequent iteration used revised values derived from an initial list of cluster members (see Section~\ref{3.4}).

The next step was to screen out targets that, based on their surface gravity and distance, were highly likely to be field giants rather than cluster members. For younger clusters ($<1$\,Gyr), field giants were identified as having $\log g\leq 3.4$, $4000< {\rm T}_{\rm eff}<7000$\,K,  {\it and} with a parallax smaller than (by at least $2\sigma$) a value corresponding to the intrinsic distance modulus of the cluster $+2$ mag. For the few targets that had $T_{\rm eff}$ but no available $\log g$,  a modified version of the $\gamma$ index $\gamma' = \gamma+\tau/6$ \citep[see][]{Damiani2014a} was used as a proxy for $\log g$ with a threshold value of $\gamma' \geq 1.335$ corresponding to $\log g \leq 3.4$ \citep{Randich2018a} and where the temperature index $\tau$  was interpolated as a cubic function of $T_{\rm eff}$. 

For all clusters a further screening was then made to cut out distant targets whatever their gravity (including distant giants in the older clusters). Stars were removed if their parallax was smaller than (by at least 4$\sigma$) a value corresponding to the intrinsic distance modulus of the cluster $+2$ mag. Finally, a few targets were rejected that had reported values of $V_{\rm RA}$, $V_{\rm Dec}$ or RV outside a 150\,km\,s$^{-1}$ window located approximately +/-75\,km\,s$^{-1}$ either side of the median velocity of the remaining targets. This effectively rejected targets with bad velocity data whilst retaining almost the entire velocity spectrum of the field population.  Since the UVES observations have larger systematic RV uncertainties associated with their wavelength calibration \citep{Sacco2014a, Jackson2015a}, any UVES observations were discarded if a GIRAFFE observation was present for the same target, ensuring that each target was modelled only once in the membership analysis. Targets rejected for whatever reason are flagged as ${\rm mem} = -1$ in Table~\ref{table3}. The number of targets  with good photometry, temperature, no strong indication of being a background giant or other distant field star {\it and} lying within all three 150\,km\,s$^{-1}$ velocity windows are shown in Table~\ref{table1} (as the "Number fitted"), whereas all targets with complete data are shown in Fig.~\ref{fighrd}. 

The fraction of targets excluded prior to the membership analysis varies from about 1 per cent (e.g. NGC 2243, Berkeley 44) to 93 per cent (IC 2602). The reasons for high exclusion rates in the younger clusters are because the GES target lists were designed to be as inclusive as possible, selecting from very broad regions in colour-magnitude diagrams usually including the lower main sequence and as a result usually including large numbers of distant background giants at similar colours. Conversely, in the older clusters, the target selection was much more focused, using lists of targets that were more likely to be cluster members based on their photometry. In addition, since these older clusters tended to be more distant and usually included cluster giants as targets, there was little scope for the observed targets to be of similar brightness but much further away than the cluster.

\begin{table*}
\caption{{\it Gaia}-ESO Survey data used to estimate the probability of cluster membership for targets observed in the 32 clusters of Table~\ref{table1}. Data are from the iDR5 iteration of the GES analysis, obtained at the GES archive of the Wide Field Astronomy Unit at Edinburgh University  (see Section~\ref{2.1}). A sample of the table is shown here. The full table is available electronically.}\label{table2}
\begin{tabular}{lllllrrrrrrrr}
\hline
Target	&	Filter	&	Cluster	&	RA	&	Dec	&	$S/N$	&	$T_{\rm eff}$	&	$\log g$	&	$\gamma$	&	$K_s$	&	RV	&	$\sigma_{\rm RV}$  	&	$\log L/L_{\odot}$	\\
cname	&	(nm)	&	DB name	&	(deg)	&	(deg)	&		&	(K)	&		&		&	(mag)	&	\multicolumn{2}{c}{(km\,s$^{-1}$)}	&		\\\hline
10532815-7710268	&	665	&	Cha\_I	&	163.36729	&	-77.1741	&	17.140	&	5698	&	3.435	&	1.007	&	11.385	&	26.77	&	0.54	&	-1.20	\\
10563044-7711393	&	665	&	Cha\_I	&	164.12683	&	-77.1942	&	57.060	&	3989	&	4.590	&	0.968	&	8.631	&	16.84	&	0.30	&	-0.06	\\
10573004-7620097	&	665	&	Cha\_I	&	164.37517	&	-76.3360	&	197.930	&	4590	&	2.700	&	1.02	&	8.022	&	41.03	&	0.24	&	0.33	\\
11022491-7733357	&	580	&	Cha\_I	&	165.60379	&	-77.5599	&	56.500	&	4544	&	4.510	&	-999	&	8.199	&	15.67	&	0.33	&	0.30	\\
11100704-7629377	&	580	&	Cha\_I	&	167.52933	&	-76.4938	&	35.220	&	4267	&	4.550	&	-999	&	8.451	&	14.94	&	0.32	&	-0.02	\\
\hline

 \end{tabular}
\end{table*}

\begin{table*}
\caption{{\it Gaia} data for targets observed in the 32 clusters of Table~\ref{table1} (see Section~\ref{2.1}) and other calculated parameters. Parallax and proper motion data are from the {\it Gaia} DR2 catalogue. The RV precision, $S_{\rm RV}$, is calculated from the  GESiDR5 data  as described in Appendix~\ref{appa}. A Gaia flag of zero indicates targets with potentially unreliable astrometric data. The final two columns shows the probability that the target is a member of its given cluster using  the full data set, $p$, or the same probability computed using a data set filtered to remove any targets with a Gaia flag of zero (see Section~\ref{5.1}). Targets with a membership probability of -1 were excluded from the membership analysis. A sample of the table is shown here. The full table is available electronically.}\label{table3}

\begin{tabular}{llrrrrrrrrrrrr}
\hline
Target	&	Filter	&	Cluster	&	PLX	&	$\sigma_{\rm PLX}$	&	$V_{\rm RA}$	&	$\sigma_{\rm VRA}$	&	$V_{\rm Dec}$	&	$\sigma_{\rm VDec}$	&	RV	&	$S_{\rm RV}$	&	Gaia	&	\multicolumn{2}{c}{Membership Probability}	\\
cname	&	(nm)	&	DB name	&	(mas)	&	(mas)	&	\multicolumn{2}{c}{(km\,s$^{-1}$)}	&	\multicolumn{2}{c}{(km\,s$^{-1}$)}	&	\multicolumn{2}{c}{(km\,s$^{-1}$)}	& flag$^{*}$		& $p$  & $p_{\rm filter}$	\\\hline
10532815-7710268	&	665	&	Cha\_I	&	1.89	&	0.08	&	-20.62	&	0.12	&	5.05	&	0.11	&	26.77	&	0.79	&	1	&	0.0665	&	0.0690	\\
10563044-7711393	&	665	&	Cha\_I	&	5.46	&	0.02	&	-21.32	&	0.03	&	2.54	&	0.03	&	16.84	&	0.29	&	1	&	0.9983	&	0.9983	\\
10573004-7620097	&	665	&	Cha\_I	&	1.26	&	0.02	&	-6.73	&	0.04	&	11.44	&	0.03	&	41.03	&	0.14	&	1	&	-1	&	-1	\\
11022491-7733357	&	580	&	Cha\_I	&	5.67	&	0.04	&	-20.87	&	0.07	&	5.46	&	0.06	&	15.67	&	0.33	&	1	&	0.9163	&	0.9164	\\
11100704-7629377	&	580	&	Cha\_I	&	5.57	&	0.28	&	-20.43	&	0.50	&	-3.14	&	0.52	&	14.94	&	0.32	&	0	&	0.9954	&	-1	\\
\hline
\multicolumn{13}{l}{* A Gaia flag of zero indicates targets with potentially unreliable {\it Gaia} data -- $<8$ visiblity periods or renormalised weighted error $<1.4$,}\\
\multicolumn{13}{l}{\citep[see Section~\ref{5.1} and][]{Lindegren2018a}}.
\end{tabular}
\end{table*}

\begin{table*}
\caption{Results of the maximum likelihood analysis of cluster membership . Columns 2 to 7 show the weighted mean central velocity and intrinsic dispersion of the 3D Gaussian distribution fitted to cluster members (see Fig.~\ref{figresults} and Section~\ref{4}). Column~8 shows the proportion of the targets analysed (the Number fitted from Table~\ref{table1}) that are expected cluster members. Columns 9 and 10 show the number of targets with membership probability $p>0.90$ and $p>0.95$ respectively. Note that there are likely additional systematic uncertainties in the velocity dispersions as a result of assumptions about the sample binary properties - see Section~\ref{binarity}.}\label{table4}

\begin{tabular}{lrrrrrrrrr}
\hline
Cluster	&	\multicolumn{3}{c}{Cluster central velocity (km\,s$^{-1}$)}	&	\multicolumn{3}{c}{Intrinsic dispersion of cluster (km\,s$^{-1}$)}	&	Fraction	&	
\multicolumn{2}{c}{Number members}\\
	&	$U_{\rm RA}$	&	$U_{\rm Dec}$	&	$U_{\rm RV}$	&	$D_{\rm RA}$	&	$D_{\rm Dec}$ &	$D_{\rm RV}$	&	members	&	$p>0.9$		&	$p>0.95$	\\
	\hline
Trumpler 14  	&	-85.32$\pm$0.33	&	30.79$\pm$0.31	&	-5.37$\pm$0.80	&	5.62$\pm$0.31	&	5.46$\pm$.27	&	12.92$\pm$0.90	&	0.61$\pm$0.02	&	350	&	279	\\
Chamaeleon I 	&	-20.03$\pm$0.12	&	0.38$\pm$0.15	&	15.75$\pm$0.15	&	0.93$\pm$0.10	&	1.21$\pm$0.14	&	0.97$\pm$0.13	&	0.50$\pm$0.04	&	74	&	72	\\
NGC 6530     	&	8.34$\pm$0.22	&	-12.90$\pm$0.13	&	0.21$\pm$0.17	&	3.57$\pm$0.21	&	2.06$\pm$0.12	&	2.45$\pm$0.17	&	0.36$\pm$0.02	&	327	&	297	\\
NGC 2264     	&	-6.80$\pm$0.09	&	-13.12$\pm$0.05	&	20.33$\pm$0.13	&	1.80$\pm$0.07	&	1.04$\pm$0.05	&	2.44$\pm$0.12	&	0.38$\pm$0.01	&	471	&	439	\\
Rho Ophiuchus	&	-4.43$\pm$0.14	&	-17.26$\pm$0.15	&	-6.33$\pm$0.32	&	0.85$\pm$0.11	&	0.98$\pm$0.12	&	1.43$\pm$0.34	&	0.59$\pm$0.06	&	41	&	41	\\
Lambda Ori   	&	1.93$\pm$0.09	&	-3.51$\pm$0.13	&	26.72$\pm$0.14	&	1.15$\pm$0.08	&	1.50$\pm$0.11	&	1.33$\pm$0.13	&	0.57$\pm$0.03	&	161	&	157	\\
Gamma 2 Vel  	&	-10.81$\pm$0.06	&	15.63$\pm$0.10	&	18.28$\pm$0.16	&	0.83$\pm$0.05	&	1.42$\pm$0.08	&	1.66$\pm$0.15	&	0.45$\pm$0.02	&	206	&	198	\\
NGC 2232     	&	-7.33$\pm$0.05	&	-2.80$\pm$0.06	&	25.40$\pm$0.06	&	0.36$\pm$0.04	&	0.42$\pm$0.06	&	0.10$\pm$0.12	&	0.12$\pm$0.01	&	82	&	80	\\
NGC 2547     	&	-15.98$\pm$0.05	&	7.95$\pm$0.06	&	12.80$\pm$0.09	&	0.60$\pm$0.05	&	0.71$\pm$0.06	&	0.66$\pm$0.08	&	0.62$\pm$0.03	&	159	&	157	\\
IC 4665      	&	-1.58$\pm$0.10	&	-13.93$\pm$0.11	&	-13.75$\pm$0.13	&	0.55$\pm$0.10	&	0.58$\pm$0.12	&	0.37$\pm$0.17	&	0.14$\pm$0.02	&	37	&	33	\\
IC 2602      	&	-12.71$\pm$0.09	&	7.82$\pm$0.10	&	17.58$\pm$0.10	&	0.64$\pm$0.07	&	0.60$\pm$0.09	&	0.27$\pm$0.15	&	0.46$\pm$0.05	&	50	&	49	\\
NGC 2451b    	&	-17.15$\pm$0.11	&	8.34$\pm$0.09	&	15.00$\pm$0.12	&	0.89$\pm$0.08	&	0.61$\pm$0.10	&	0.62$\pm$0.13	&	0.16$\pm$0.02	&	63	&	62	\\
IC 2391      	&	-17.77$\pm$0.09	&	16.68$\pm$0.14	&	14.95$\pm$0.17	&	0.56$\pm$0.07	&	0.82$\pm$0.11	&	0.56$\pm$0.17	&	0.57$\pm$0.06	&	38	&	35	\\
NGC 2451a    	&	-19.36$\pm$0.18	&	13.93$\pm$0.08	&	23.42$\pm$0.08	&	1.08$\pm$0.13	&	0.44$\pm$0.06	&	0.12$\pm$0.12	&	0.12$\pm$0.02	&	37	&	37	\\
NGC 2516     	&	-9.15$\pm$0.04	&	21.93$\pm$0.05	&	23.90$\pm$0.06	&	0.91$\pm$0.04	&	0.92$\pm$0.03	&	0.75$\pm$0.06	&	0.74$\pm$0.02	&	467	&	459	\\
NGC 6067     	&	-20.68$\pm$0.13	&	-27.93$\pm$0.13	&	-38.20$\pm$0.23	&	1.58$\pm$0.11	&	1.69$\pm$0.12	&	1.57$\pm$0.24	&	0.40$\pm$0.02	&	179	&	167	\\
Blanco 1     	&	21.13$\pm$0.05	&	3.05$\pm$0.04	&	6.02$\pm$0.08	&	0.50$\pm$0.04	&	0.40$\pm$0.04	&	0.33$\pm$0.08	&	0.40$\pm$0.03	&	129	&	128	\\
NGC 6259     	&	-11.52$\pm$0.19	&	-32.92$\pm$0.18	&	-33.04$\pm$0.41	&	1.98$\pm$0.17	&	1.83$\pm$0.17	&	3.21$\pm$0.48	&	0.40$\pm$0.03	&	137	&	132	\\
NGC 6705     	&	-17.91$\pm$0.10	&	-47.31$\pm$0.10	&	35.53$\pm$0.16	&	1.88$\pm$0.09	&	1.92$\pm$0.08	&	2.30$\pm$0.17	&	0.59$\pm$0.02	&	540	&	524	\\
NGC 4815     	&	-102.12$\pm$0.46	&	-16.85$\pm$0.38	&	-27.22$\pm$0.72	&	2.58$\pm$0.43	&	2.06$\pm$0.34	&	3.73$\pm$0.74	&	0.51$\pm$0.05	&	50	&	48	\\
NGC 6633     	&	2.18$\pm$0.22	&	-3.02$\pm$0.20	&	-28.18$\pm$0.21	&	1.19$\pm$0.20	&	0.99$\pm$0.18	&	0.77$\pm$0.20	&	0.26$\pm$0.04	&	34	&	31	\\
Trumpler 23  	&	-57.06$\pm$0.39	&	-64.44$\pm$0.27	&	-61.25$\pm$0.27	&	2.01$\pm$0.43	&	1.43$\pm$0.24	&	1.06$\pm$0.32	&	0.51$\pm$0.06	&	39	&	39	\\
NGC 6802     	&	-46.36$\pm$0.28	&	-106.39$\pm$0.40	&	13.29$\pm$0.39	&	1.10$\pm$0.29	&	1.81$\pm$0.56	&	1.54$\pm$0.43	&	0.59$\pm$0.05	&	53	&	50	\\
Berkeley 81  	&	-21.70$\pm$0.45	&	-33.03$\pm$0.48	&	48.11$\pm$0.20	&	1.09$\pm$0.62	&	1.97$\pm$0.67	&	0.59$\pm$0.29	&	0.33$\pm$0.04	&	50	&	49	\\
Ruprecht 134 	&	-20.54$\pm$0.18	&	-30.21$\pm$0.18	&	-40.91$\pm$0.16	&	0.88$\pm$0.22	&	1.09$\pm$0.19	&	0.48$\pm$0.20	&	0.17$\pm$0.02	&	59	&	57	\\
NGC 6005     	&	-57.24$\pm$0.40	&	-53.86$\pm$0.40	&	-24.46$\pm$0.98	&	2.25$\pm$0.49	&	2.09$\pm$0.55	&	4.37$\pm$1.36	&	0.23$\pm$0.03	&	46	&	42	\\
Pismis18     	&	-82.06$\pm$0.31	&	-33.33$\pm$0.37	&	-27.66$\pm$0.50	&	1.22$\pm$0.30	&	1.40$\pm$0.35	&	1.20$\pm$0.71	&	0.30$\pm$0.05	&	24	&	22	\\
Trumpler 20  	&	-139.37$\pm$0.20	&	2.96$\pm$0.19	&	-39.75$\pm$0.17	&	2.36$\pm$0.22	&	2.14$\pm$0.19	&	1.47$\pm$0.24	&	0.41$\pm$0.02	&	157	&	149	\\
NGC 2420     	&	-15.60$\pm$0.10	&	-27.71$\pm$0.09	&	74.63$\pm$0.04	&	1.29$\pm$0.10	&	1.15$\pm$0.08	&	0.61$\pm$0.05	&	0.77$\pm$0.02	&	395	&	391	\\
Berkeley 31  	&	3.79$\pm$1.49	&	-31.74$\pm$0.85	&	56.94$\pm$0.13	&	9.06$\pm$1.75	&	4.44$\pm$1.85	&	0.54$\pm$0.25	&	0.25$\pm$0.03	&	81	&	72	\\
Berkeley 44  	&	-.22.00$\pm$0.33	&	-45.28$\pm$0.32	&	-8.60$\pm$0.21	&	1.03$\pm$0.45	&	0.99$\pm$0.47	&	0.66$\pm$0.25	&	0.54$\pm$0.06	&	43	&	41	\\
NGC 2243     	&	-29.26$\pm$0.13	&	124.35$\pm$0.13	&	59.82$\pm$0.05	&	1.77$\pm$0.13	&	1.74$\pm$0.14	&	0.59$\pm$0.07	&	0.81$\pm$0.02	&	454	&	449	\\\hline
 \end{tabular}
\end{table*}

\section{Probability of cluster membership}
\label{3}
The velocity data in Table~\ref{table3} were used to determine membership probabilities for individual targets using the maximum likelihood method originally proposed by \cite{Pryor1993a} and later updated by \cite{Cottaar2012a} to include the effect of binarity. This technique
assumes that the observed velocities are taken from an intrinsic model broadened by the measurement uncertainties and the effects of unresolved binaries. Given a model specified
by a number of free parameters (see below), the best-fitting model was found by maximising the summed logarithmic likelihood for all stars considered.

\subsection{Observational uncertainty}
\label{3.1}
Tangential velocities from the proper motions are assumed to be unaffected by binarity; in other words, it is assumed that the uncertainties in the tangential velocities are due only to the measurement uncertainties reported in {\it Gaia} DR2, and are described by Gaussian distributions that are scaled with  $\sigma_{\rm VRA}$ and $\sigma_{\rm VDec}$.  It may be that there are additional systematic errors due to unresolved binarity \citep[at separations $<0.1$ arcsec, e.g.][]{Lindegren2018a} affecting the proper motion estimates.  A detailed treatment of this is not possible without knowledge of the sampling and particular scan pattern for each object, but the influence of these possible additional systematic uncertainties is discussed further in Section~\ref{binarity}.

The observational uncertainty in RV can be treated in  a more complex way. Firstly, the distribution of measurement uncertainty in RV is non-Gaussian, being better described by a ($\nu$=6) Student's-t distribution, scaled with the values of $S_{\rm RV}$ shown in Table~\ref{table3} (see Appendix~\ref{appa}). Secondly, for binary stars, the observational uncertainty includes the effects of RV offsets expected from a set of randomly oriented binary systems with a specified distribution of orbital periods, eccentricities, and mass ratios. 
The total likelihood of a target's observed RV is then given by the sum of its likelihood if it were a single star and its likelihood if it were in an unresolved binary. 
\begin{equation}
    \mathscr{L}^{\rm RV} = (1 - f_{\rm B})\mathscr{L}_{\rm S} + f_{\rm B}\mathscr{L}_{\rm B}\, ,
\end{equation}
    where $\mathscr{L}_{\rm S}$ and $\mathscr{L}_{\rm B}$ are the likelihood of single and binary stars and $f_{\rm B}$ is the adopted binary fraction. 

The calculation of the distribution of RV offsets followed the method described by \cite{Cottaar2012a}, adopting $f_{\rm B} = 0.46$, a lognormal period distribution with a mean $\log \rm{period}$ = 5.03 (in days) and dispersion 2.28 dex, and a flat mass ratio distribution for $0.1 < q < 1$ \citep{Raghavan2010a}. The influence of these assumptions is discussed further in Section~\ref{binarity}. For the purposes of calculating the offsets, the primary masses were estimated from the target $\log L/L_{\odot}$ using the Pisa model isochrones with the mass capped at a level equivalent to $\log L/L_{\odot}=1$. A correction for the dilution effect due to the unresolved light from the secondary at the reflex velocity was also made, using the Pisa models to estimate the secondary contribution for a given mass ratio. Note that the amplitude of the binary-induced RV offsets is relatively insensitive to the primary mass, scaling as (mass)$^{1/3}$. 

\subsection{Intrinsic models}
\label{3.2}
The maximum likelihood calculation was made in two stages, first the mean velocity and dispersion of the background population of field stars was characterised for each velocity component using a series of 1D maximum likelihood analyses. These results were then used in a full 3D analysis to determine the likelihood of cluster membership. 

For the majority of clusters we assumed an intrinsic model that is the sum of two Gaussian distributions (cluster plus background)  of unknown central velocity and dispersion. In this case the 1D likelihood was determined as a function of five free parameters, the intrinsic velocity and dispersion of the cluster and background populations and the overall fraction of the observed population that are cluster members.  A more complex model, comprising three Gaussian distributions was  used to fit five of the clusters and gave a significantly higher maximum log likelihood. In three cases this was expected since the clusters are known to lie close in velocity space to a second association or cluster. These three clusters are Gamma Vel \citep{Jeffries2014a}, NGC\,2547 \citep{Sacco2015a} and the cluster pair NGC\,2451a and NGC\,2451b \citep{Hunsch2003a}. We also found that three Gaussian distributions were preferred to fit NGC\,2264 and NGC\,2232. For these five cases the central velocity and broad dispersion of the field population was fixed, so that the 1D likelihood was determined as a function of six free parameters. 

A 3D analysis was then made to determine the intrinsic cluster properties and the membership probabilities of individual targets. In this case the model likelihood was determined as a function of seven free parameters; the cluster velocity and dispersion in each dimension and the fraction of the observed population that are cluster members, $F$, fixing the intrinsic background population velocity and dispersion components at the values found from the 1D analyses (see Appendix~\ref{appb}).
\begin{figure*}
	\begin{minipage}[t]{0.98\textwidth}
	\includegraphics[width = 170mm]{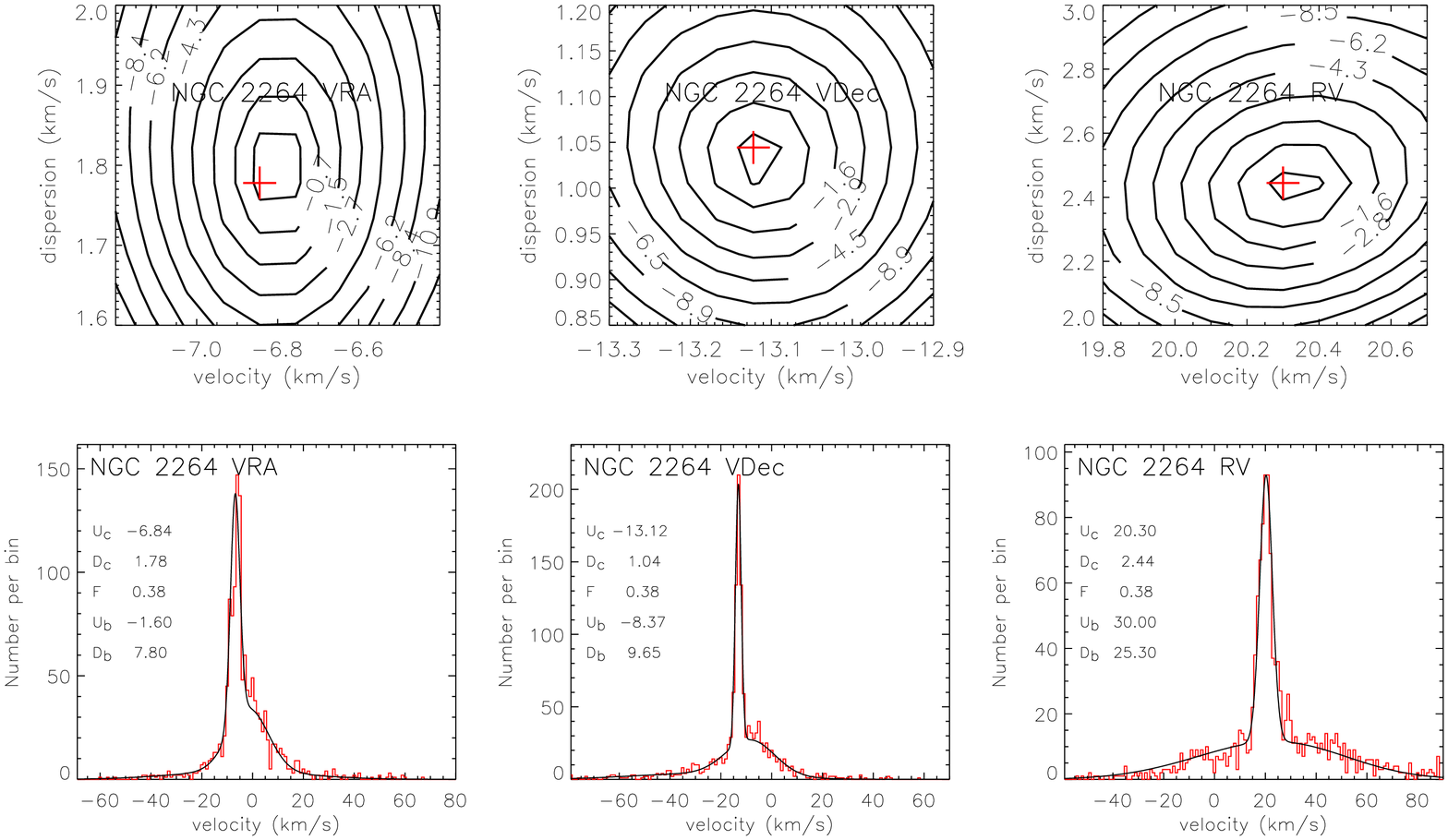}
    \includegraphics[width = 170mm]{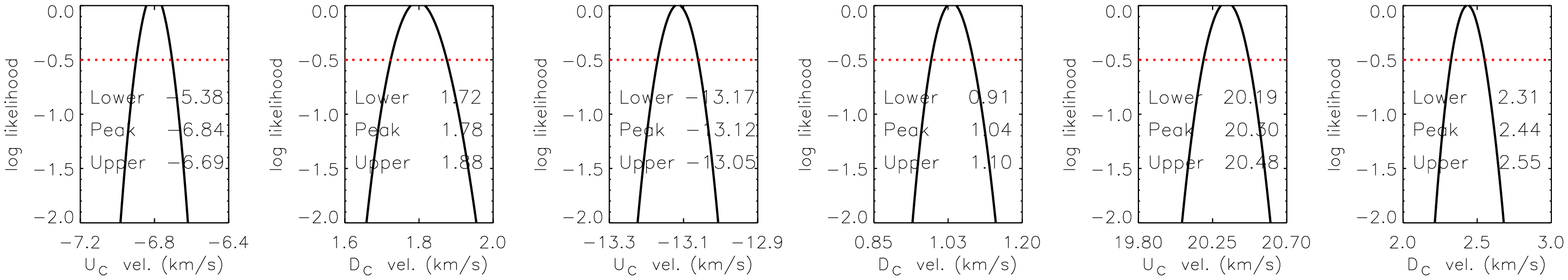}
    \end{minipage}
	\caption{Results of the 3D maximum likelihood analysis of NGC\,2264.  The upper plots shows contours of log likelihood of cluster intrinsic velocity and dispersion relative to the maximum likelihood value located the cross on each plot. The centre plots show histograms of measured velocities together with model probability distributions evaluated at the maximum likelihood values of fitted parameters and a median measurement uncertainty. The lower plots show the variation in log likelihood of each component of cluster mean velocity and intrinsic dispersion.}
	\label{figresults}	
\end{figure*}

\begin{figure*}
	\centering
	\begin{minipage}[t]{\textwidth}
	\centering
	\includegraphics[width = 178mm]{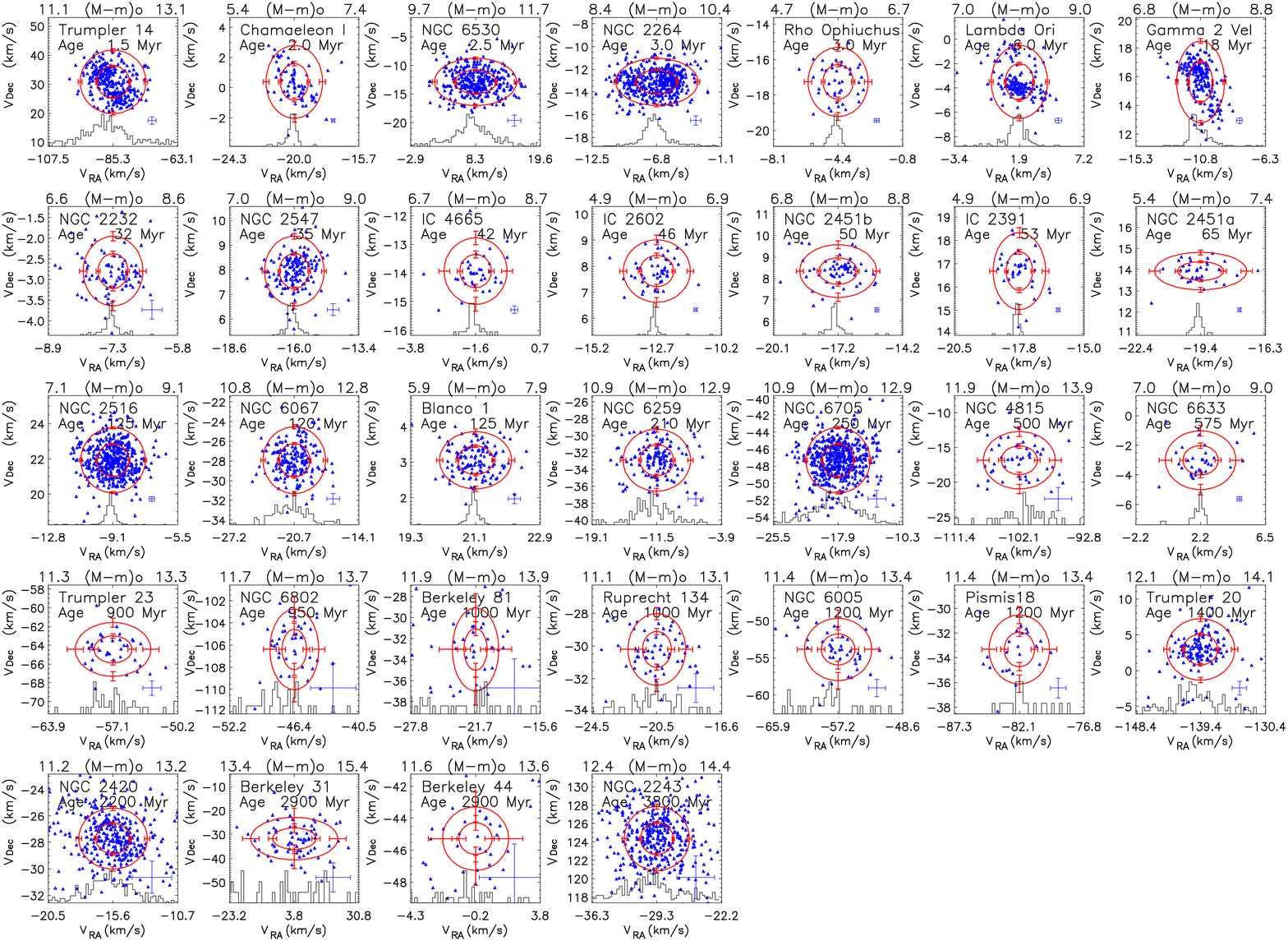}
	\end{minipage}
	\caption{The tangential velocities of cluster members. Blue points show the orthogonal velocity components (in RA and Dec) for targets with a membership probability $p>0.9$ (see Table~\ref{table3}) with the blue cross in the lower right hand of the plot indicating median measurement uncertainties. Red  ellipses show the 1$\sigma$ and 2$\sigma$ intrinsic dispersion of cluster members  (see Table~\ref{table4}) with error bars indicating the rms uncertainty on the calculated dispersion in $V_{\rm RA}$ and $V_{\rm Dec}$.  The black histograms show the distribution of distance modulus for the cluster members over a $\pm 1$\,mag range relative to the cluster center (see scale above each plot).}
	\label{figvtan}
\end{figure*}

\subsection{Likelihood and membership probabilities}
\label{3.3}
We compute the likelihood of a star  being observed with a given velocity vector $\overline{V_i}$ and measurement uncertainty $\overline{\sigma_i}$ as 
\begin{equation}
\mathscr{L}_{i,m,n} = F_{n}\mathscr{L}_{i,m}^{\rm c} + (1-F_{ n})\mathscr{L}_{i,m}^{\rm b}\, ,  \end{equation}
where $\mathscr{L}_{i,m}^{\rm c}$ and $\mathscr{L}_{\rm i,m}^{\rm b}$ are the likelihood of fit of the $i^{th}$ target to the $m^{\rm th}$ cluster and background models respectively and we step through $n$ values for $F$, the value of the fraction of stars that belong to the cluster, giving a total $\log$ likelihood for a fit of all targets to the $m^{\rm th}$ model of
\begin{equation}
\log{\mathscr{L}_{m}}  = \sum_i \sum_{n} \log[F_{n} \,\mathscr{L}_{i,m}^{\rm c} +
(1-F_{n}) \,\mathscr{L}_{i,m}^{\rm b}]\, .
\end{equation}

The uncertainty in RV is independent of the uncertainty in proper motions allowing $\mathscr{L}_{i,m}^{\rm c}$ and $\mathscr{L}_{i,m}^{\rm b}$ to be calculated as the product of the likelihood of fits in RV, $\mathscr{L}_{i,m}^{\rm RV}$, and fits in proper motion space,  $\mathscr{L}_{i,m}^{\rm pm}$. 
Calculation of $\mathscr{L}_{i,m}^{\rm RV}$ takes account of the effects of binarity on measurement uncertainty as described in Section~\ref{3.1}. Calculation of $\mathscr{L}_{i, m}^{\rm pm}$ takes account of the correlated uncertainty  between  $V_{\rm RA}$ and $V_{\rm Dec}$ by using the covariance matrix elements in the {\it Gaia} DR2 dataset \citep{Gaia2016a}. 

Membership probabilities, $p_i$,  are computed as the expectation value  of an individual target being a cluster member summed over the uniform grids of component velocities and dispersions  (a total of $m$ models) and $n$ values of fractional membership $F_n$. The range explored for each parameter is set to be greater than $\pm 5 \sigma$ from the maximum likelihood value of that parameter, but with a minimum value of zero for the velocity dispersions. The probability of the $i^{\rm th}$ target being a cluster member is given by
\begin{equation}
{p}_{i} = \frac{\sum_{n} \sum_{m} F_{n} \mathscr{L}_{i,m}^{\rm c}} {\sum_{n} \sum_{m} F_{n} \mathscr{L}_{i,m}^{\rm c} + \sum_{n} \sum_{m} (1-F_{n}) \mathscr{L}_{i,m}^{\rm b} } \, ,
\label{probmem}
\end{equation}

\subsection{Distance modulus and reddening}
\label{3.4}
A list of probable cluster members (with $p>0.9$) was used to re-evaluate the cluster distance modulus and reddening. The distance was estimated in two steps. First the 3-sigma clipped mean and standard deviation of cluster members was used to estimate an intrinsic dispersion of cluster parallax equal to the standard deviation of cluster members less the RMS parallax uncertainty (subtracted in quadrature). The quadrature sum of the estimated dispersion and the parallax uncertainties were then used as weights to calculate a weighted mean parallax and uncertainty. The uncertainty in this weighted mean was always much less than 10 per cent, so the mean parallax was inverted to yield a cluster distance. Table~\ref{table1} gives the corresponding distance modulus,  $(M-m)_0^{\rm c}$, where two error bars are quoted. The first is the statistical uncertainty and is larger for more distant clusters or those with few members; the second is a systematic uncertainty equivalent to 0.1 mas in parallax, which accounts for possible correlated errors in the parallax zero-point on small spatial scales  \citep{Lindegren2018a}, and which is generally much larger than the statistical uncertainty. Cluster reddening was estimated by comparing the measured $G-K_s$ colours with Pisa model predictions of $(G-K_s)_0$ for main sequence stars (giant members of the older clusters were not included in the calculation) at the target luminosity and the literature age (see Table~\ref{table1}).  $E(B-V)^{\rm c}$ was taken as the median value of the 50 per cent the members with the lowest reddening, since these are more likely to be single stars. The quoted uncertainty in Table~\ref{table1} is the median absolute deviation (MAD) of this subset.

It is important to note that $(M-m)_0^{\rm c}$ and $E(B-V)^{\rm c}$ are scaling constants in the maximum likelihood calculation. Changing them has {\it no direct effect on the membership probabilities}. It is however useful to determine $(M-m)_0^{\rm c}$ with reasonable accuracy in order to compare the distance-dependent mean cluster velocity and dispersion in RA and Dec with those in RV, and an estimate of $E(B-V)^{\rm c}$ is required to compare luminosity and absolute photometric magnitudes with evolutionary models.   Both $(M-m)_0^{\rm c}$ and $E(B-V)^{\rm c}$ have a weak, indirect effect on the membership probabilities, since they affect the estimated mass used to determine the RV offsets of binary stars; but even then, the binary RV offsets scale only as (mass)$^{1/3}$. 

Once a revised cluster distance modulus and reddening were found, the analysis steps described in Sections~\ref{2} and~\ref{3} were iterated. The final values of distance modulus and reddening are reported in Table~\ref{table1}.
 
\section{Results}
\label{4}
Typical results of the maximum likelihood analysis are shown in Fig.~\ref{figresults} for the young cluster NGC\,2264. Similar plots are available for all 32 clusters (Figs. B1--B32) in Appendix~\ref{appb} (online only). The three upper plots in Fig.~\ref{figresults} show contour maps of maximum likelihood as a function of the cluster central velocity and intrinsic dispersion for each velocity component ($V_{\rm RA}$, $V_{\rm Dec}$ and RV) over the range of velocity and dispersion explored. The central cross marks the location of maximum likelihood. Contours mark decreasing levels of log likelihood with respect to this maximum. 

The three central plots show histograms of the number of targets per 1\,km\,s$^{-1}$ bin for each velocity component. Text on the plots shows the maximum likelihood values of cluster velocity ($U_c$) and dispersion ($D_c$), the fraction of targets that are cluster members ($F$), together with the central velocity ($U_b$) and dispersion ($D_b$) of the model distribution of background stars. The black curve shows the  model probability distribution based on the maximum likelihood parameter values, the median of the target measurement uncertainties and the assumed fraction of binary systems  ($f_{\rm B}=0.46$). 

The lower plots show, for each velocity component, the variation in log likelihood as a function of the cluster central velocity and intrinsic velocity dispersion, over the range of parameter values explored in the maximum likelihood calculation. The dotted line marks the 1$\sigma$ level (a log likelihood of -0.5 relative to the maximum value). Text on the plots show the maximum likelihood value of each parameter and values at the upper and lower 1$\sigma$ levels. These probability distributions were used to calculated the weighted mean and rms values of the cluster velocity components and their intrinsic dispersion that are reported in Table~\ref{table4}.

NGC~2264 is one of the  better populated, less distant, clusters with $\sim 40$ per cent of targets being cluster members, producing a clear peak in number density versus all three components of velocity.   This is not always the case. The large relative proper motion uncertainties for more distant clusters, (with $(M-m)^{\rm c}_0>10$) lead to tangential velocity uncertainties that become larger than the intrinsic cluster velocity dispersions. This leads to the cluster members "blending in" to the background and causes larger uncertainties in the tangential velocity dispersions. A well populated example would be NGC~2243 (Fig.~B32), where the peaks in $V_{\rm RA}$ and $V_{\rm Dec}$ are much broader than in Fig.~\ref{figresults}, but still well-defined. In other distant clusters, where fewer members are identified with $p>0.9$ (e.g. Berkeley 31, Fig.~B30), the ability to select members is dominated by a narrow peak in RV, which is not affected by distance. 
For one cluster (NGC\,6005; Fig.~B26) a combination of few high probability members and a probable large intrinsic velocity dispersion made it difficult to separate the cluster from the background and it was not possible to determine a peak in likelihood for $D_{\rm RV}$ versus velocity. In this case the maximum $D_{\rm RV}$ was artificially fixed at 6\,km\,s$^{-1}$ in order to determine cluster membership.

The weighted mean and rms values of cluster central velocity, and the intrinsic cluster dispersion of the models used to determine cluster membership are shown in Table~\ref{table4}. The membership probabilities  of the individual targets in each cluster were calculated from equation~\ref{probmem} and are reported in Table~\ref{table3}. Targets with a membership probability of -1 were not included in the maximum likelihood analysis. Total numbers of targets in each cluster with a membership probability, p~$> 0.9$ and p~$> 0.95$ are also shown in Table~\ref{table4}. There are 5033 targets with $p>0.9$.

The distribution of membership probabilities is a measure of how well the data are able to separate the cluster from the background. In a cluster like NGC~2264 (Fig.~B4) this distribution is quite bi-modal, with most objects either being clear cluster members with $p>0.9$ (the average probability for this subset is $\bar{p}=0.986$) or very unlikely to be cluster members, with $p<0.05$, with relatively few stars in between. This is the case for most of the observed clusters (see Appendix~\ref{appb}), illustrating the power of combining three orthogonal velocity constraints.  There are some exceptions to this very sharply bimodal probability distribution (e.g. Trumpler 14, Fig.~B1; NGC~6530, Fig.~B3; Berkeley 31, Fig.~B30). 

Both Trumpler 14 and NGC~6530 show evidence that a single 3D Gaussian model is a poor representation of the cluster in one of the dimensions (RV in the case of Trumpler 14, although there is a hint of bifurcation in proper motion space too; $V_{\rm RA}$ for NGC~6530). Adding a further Gaussian component for these clusters did not significantly improve the maximum likelihood of the fits, probably because the kinematic substructure is more complex than such a simple model. In these cases, the discrimination between members and non-members may not be optimal and lowers the average probability of cluster members ($p>0.9$) but it does not invalidate the estimated membership probabilities, which simply reflect the fit of the model to the data. Kinematic substructure would be unsurprising in these very young clusters. The GES survey covers an area of Trumpler 14 known to have considerable spatial substructure \citep{Feigelson2011a, Damiani2017a} and kinematic substructure in NGC~6530 has already been noted by \cite{Wright2019a}.

Figure~\ref{figvtan} plots $V_{\rm RA}$ versus $V_{\rm Dec}$ for ($p>0.9$) members of each cluster. Ellipses represent one and two times the cluster velocity dispersions along each axis, with error bars indicating their uncertainties. For the majority of clusters there is a clear grouping of members in proper motion space with $>$90 per cent located within the 2 sigma ellipse. For the most distant clusters (with $(M-m)_0^{\rm c}>12$) the rms uncertainties in tangential velocity are larger than the estimated values of cluster dispersion and so the velocities scatter well beyond the 2 sigma ellipse. These are cases in which case RV becomes the most important parameter  determining the probability of cluster membership, since in principle it is distance-independent. Also shown in Fig.~\ref{figvtan} are histograms indicating the number of members (with $p>0.9$) versus distance modulus over a $\pm 1$ magnitude range relative to $(M-m)_0^{\rm c}$.  Nearby clusters (with $(M-m)_0^{\rm c} <10$)  show a well defined peak, as expected for cluster members. For more distant clusters the uncertainty in the distance modulus of individual targets becomes too large to give a clear indication of cluster membership.  

\section{Discussion}

\subsection{The effect of {\it Gaia} data quality}
\label{5.1}

\begin{figure}
	\centering
	\includegraphics[width = 84mm]{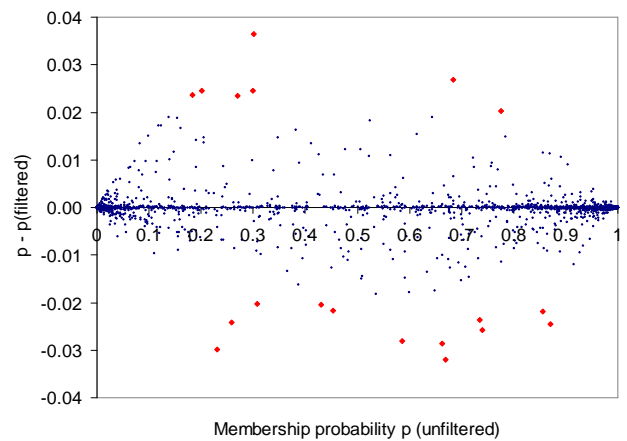}
	\caption{The effect of applying filtering to the {\it Gaia} DR2 on the derived membership probabilities for the 11386 sources in the sample present before and after filtering. Large symbols highlight targets with $p - p_{\rm filtered} > 0.02$.}
	\label{figfilter}
\end{figure}

For the analysis of cluster membership, all reported values of {\it Gaia} DR2 proper motion and parallax (and their uncertainties) were accepted as valid. An alternative approach would be to filter targets with potentially unreliable {\it Gaia} data by requiring all sources to have $\geq 8$ visibility periods \citep{Arenou2018a} and to apply the cut recommended in \cite{Lindegren2018a}, requiring that the "re-normalised unit weighted error" was $< $1.4 \citep[see][for further details]{Wright2019a}.

To assess the effect of additional filtering of the {\it Gaia} astrometry, the membership analysis was repeated after removing all targets which did not pass the tests above. The objects with potentially unreliable {\it Gaia} data (according to these tests) are indicated in Table~\ref{table3}, along with revised membership probabilities calculated using just the targets with "good" {\it Gaia} data.

Applying the filtering reduces the number of valid targets (with all necessary data) by 10 per cent from 12659 to 11386 and the total number of $p>0.9$ members in the 32 clusters  by 7 per cent from 5033 to 4694. A total of 338 targets that were cluster members with $p>0.9$ are rejected as having suspect {\it Gaia} data. Only four targets change from $p<0.9$ to $p>0.9$.  In three cases this is caused by an incremental ($<0.003$) change in $p$. In the fourth case, the target (18043441-2428057) had a GIRAFFE-based RV that was highly discrepant from the cluster centroid which was rejected in favour of the UVES-derived RV of the same star. The UVES measurement was in much better accord with the cluster centroid, yielding $p>0.9$. There is only one case  of movement in the other direction, where a target with $p=0.901$ becomes $p=0.898$ after filtering the {\it Gaia} data.

Figure~\ref{figfilter} shows how $p$ changes before and after filtering the {\it Gaia} data. For the vast majority of sources the changes are very minor. For the subset of 4694 high probability members that originally had $p>0.9$ the maximum change in $p$ is 0.011 and the rms difference is 0.0004. These extremely small changes indicate that the membership probabilities (for targets with $p>0.9$) evaluated using the unfiltered dataset are for practical purposes the same as those calculated if excluding the suspect {\it Gaia} data. 

The reason that the changes in $p$ are so small is that targets that are true cluster members but have potentially unreliable measurements of $V_{\rm RA}$ or $V_{\rm Dec}$ may scatter out of the cluster and appear as background stars. The converse is generally not true. Targets that are true  background stars but have an unreliable measurement of $V_{\rm RA}$ or $V_{\rm Dec}$ are very unlikely to scatter into the right cluster velocity range and appear as cluster members. So the process of membership selection using the full dataset effectively filters out targets with bad $V_{\rm RA}$ and/or $V_{\rm Dec}$ whilst retaining targets that are flagged as having suspect {\it Gaia} data but which actually have $V_{\rm RA}$ and $V_{\rm Dec}$ that agree well with their cluster siblings. By {\it not} filtering the {\it Gaia} data we identify 7 per cent more cluster members with $p>0.9$, with a negligible penalty in terms of additional contamination, as demonstrated in Sections~\ref{5.2} and~\ref{5.3}.

Filtering the {\it Gaia} data does change estimates of the mean cluster velocities and intrinsic dispersions. The average change in mean cluster velocity is 20 per cent of its uncertainty and the average change in intrinsic dispersion is 30 per cent of its measured uncertainty. The exceptions are Lambda Ori and NGC\,2547 which show larger changes in intrinsic dispersion of $\sim 1.5$ times their estimated uncertainty. 

The objective of our analysis is to identify high probability cluster members from their velocities. It is not to characterise the cluster shapes in velocity space. The results in Table~\ref{table4} can only be considered rough estimates of the true intrinsic velocity dispersion, whether or not the {\it Gaia} data are filtered, since the mean velocities and dispersions are the result of fitting the measured velocities in a fixed co-ordinate system to determine the probability of cluster membership. To determine the true cluster shape in velocity space may require a determination of cluster membership that is independent of kinematics and a more general model of the cluster allowing for free rotation of the cluster axes in velocity space, bulk rotation of the cluster,  the finite size of the cluster and "perspective expansion" \citep[see][]{vanLeeuwen2009a, Kuhn2019a}, none of which are explored in this paper and are not crucial to the cluster membership calculations. In addition, the contribution of unresolved binaries will affect the inferred intrinsic velocity dispersion. This does not greatly affect cluster membership calculations, but does result in a further systematic uncertainty in the RV dispersion, where an attempt has been made to account for binarity, and a small overestimate of the tangential velocity dispersion, where it has not (see Section~\ref{binarity}).

\begin{figure}
	\centering
	\includegraphics[width = 84mm]{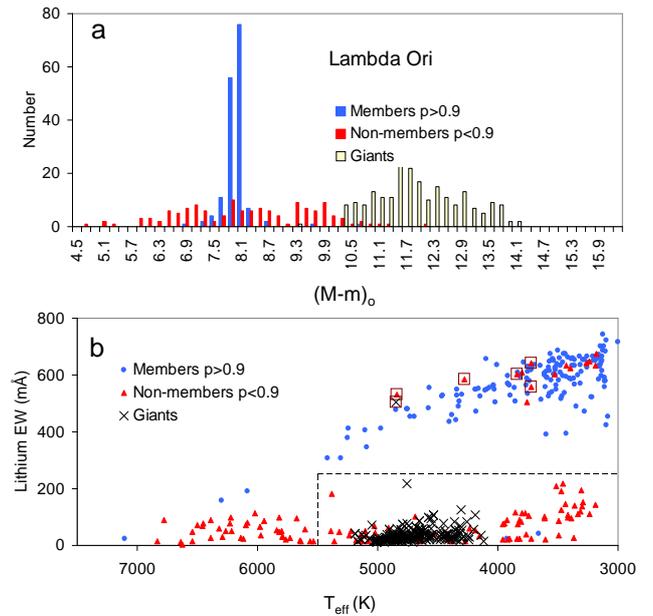}
	\caption{(a) Numbers of probable members ($p>0.9$) and non-members ($p<0.9$) in the Lambda Ori cluster as a function of their distance modulus. Also shown are targets identified as background giants and therefore excluded from the maximum likelihood analysis. (b) The equivalent width of the Li~6708\AA\ feature as a function of $T_{\rm eff}$ for the same targets. The dashed box and the targets marked by squares denote the region discussed in Section~\ref{5.2} and the "false negatives" that remain even after filtering the {\it Gaia} data.}
	\label{figlori}
\end{figure}

\begin{figure}
	\centering
	\includegraphics[width = 84mm]{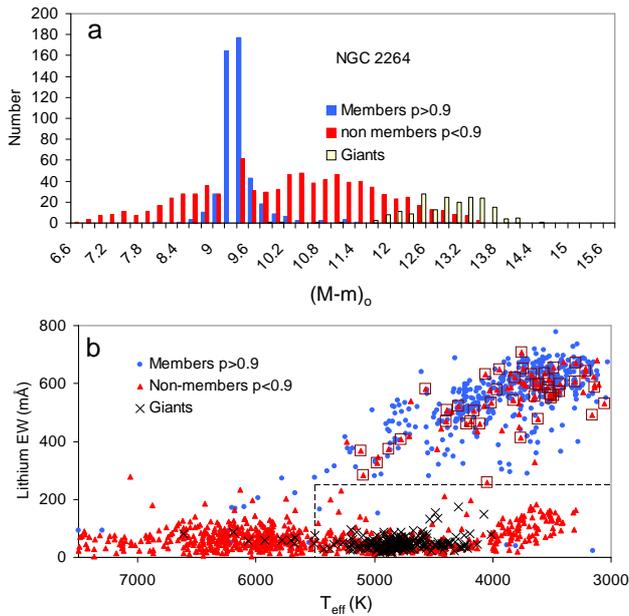}
	\caption{(a) Numbers of probable members ($p>0.9$) and non-members ($p<0.9$) NGC\,2264 as a function of their distance modulus. Also shown are targets identified as background giants and therefore excluded from the maximum likelihood analysis. (b) The equivalent width of the Li~6708\AA\ feature as a function of $T_{\rm eff}$ for the same split of targets. The dashed box and the targets marked by squares denote the region discussed in Section~\ref{5.3} and the "false negatives" that remain even after filtering the {\it Gaia} data.}
	\label{fign2264}
\end{figure}

\subsection{Testing cluster membership: I. The Lambda Ori Cluster}
\label{5.2}
There are a small subgroup of clusters where we can compare the kinemtically determined membership probabilities with a second independent determination of cluster membership. These are nearby young clusters, aged $<10$\,Myr, where members will show a distance modulus close to the cluster mean and genuine low-mass members should show almost undepleted levels of lithium compared with older background field stars \citep[e.g.][]{Jeffries2014a, Jeffries2014b}. For such clusters, distance modulus and Li abundance (or equivalent width of the Li~{\sc i} 6708\AA\ feature, EW(Li)) can be used to identify cluster members and compared with kinematically defined membership.

Figure~\ref{figlori} makes this comparison for the young cluster Lambda Ori. Figure~\ref{figlori}a shows the number of targets as a function of distance modulus for targets with $p>0.9$ and p~$<0.9$ shown separately. Note that only 5 per cent of targets populate the region $0.1<{\rm p}<0.9$ (see Fig.~B6). The distance moduli for individual targets were calculated from the Bayesian distance determinations given by \cite{Bailer2018a} rather than inverting the individual parallaxes. Probable members are tightly grouped in distance modulus, with a dispersion of 0.4\,mag.  Likely non-members are much more dispersed as are targets identified as probable giants on the basis of their $\log g$  which were excluded from the likelihood analysis.  It is notable that, even in this relatively nearby cluster, the use of distance as a membership constraint would not have improved our rejection of non-members significantly. There are only two $>3$ sigma outliers in the top panel of Fig.~\ref{figlori}, both of which have possibly unreliable astrometry (see Section~\ref{5.1}).

Figure~\ref{figlori}b shows GESiDR5 measurements of EW(Li) as a function of $T_{\rm eff}$  for targets with $p>0.9$, $p<0.9$ and for stars rejected as giants. Taking a clear dividing line between Li-rich members and Li-poor non-members with $T_{\rm eff}<5500$\,K and EW(Li)$>250$\,m\AA~ (dashed line in Fig.~\ref{figlori}b), there are 158 members and 91 non-members. There are 2 false positives -- Li-poor targets with $p>0.9$. This compares favourably with an expected number of 1 false positive based on the average probability of $\overline{{p}}=0.995$ for the $p>0.9$ cluster members. Six false negatives are also expected since $\overline{p}=0.065$ for the p~$<0.9$ sample, but we actually see 15 (plus a Li-rich giant).

This comparison of kinematic and Li-based cluster membership  indicates that the kinematic selection process is performing as expected although the higher than expected number of false negatives might suggest either contamination by Li-rich PMS stars from adjacent, dispersed regions of recent star formation at a similar distance, like Orion OB1a \citep{Briceno2019a}, or that the distribution of measurement uncertainties in $V_{\rm RA}$ and $V_{\rm Dec}$ shows a significant non-Gaussian tail. The number of  Li-rich false negatives is reduced to six (highlighted with squares in Fig.~\ref{figlori}b) by filtering for potentially  unreliable {\it Gaia} data (see Section~\ref{5.1}), but it also reduces the number of confirmed Li-rich cluster members from 158 to 140, without changing the number of Li-poor false positives. This confirms the conclusion arrived at in Section~\ref{5.1}; that the membership probabilities for objects classed as members (with high $p$) are reliable and that filtering the {\it Gaia} data merely rejects some false negatives whilst reducing the overall number of cluster members identified.

\subsection{Testing cluster membership: II. NGC 2264}
\label{5.3}
In general we expect narrow distributions of $V_{\rm RA}$, $V_{\rm Dec}$ and RV for cluster members with much broader distributions for the background stars. This is true for the majority of clusters where the best fit dispersion of the apparent\footnote{The conversion from proper motion to tangential velocity assumes all stars are at the distance of the cluster.} velocity of the background population is between 20 and 30\,km\,s$^{-1}$ (see Appendix~\ref{appb}). The obvious exceptions are Gamma Velorum and NGC\,2457, which are in a similar direction to other groups of young stars that are more spatially diffuse but still coherent in velocity \citep{Jeffries2014a, Sacco2015a, Franciosini2018a, CantatGaudin2019a}; and NGC 2541a and NGC 2541b which are a pair of clusters with different distances and kinematics, but with a similar age and observed in the same direction \cite{Hunsch2003a}. Another cluster showing a more complex distribution of background stars is NGC\,2264 where the majority of non cluster members show a narrow dispersion of $\sim 9$\,km\,s$^{-1}$ in $V_{\rm RA}$ and $V_{\rm Dec}$ and a broader 25\,km\,s$^{-1}$ in RV, with a second, less dense background population showing the usual broad velocity distribution in all three components. There are three possible explanations for the observed distribution. 

\begin{itemize}
	\item Stars categorised as non-members are in fact cluster members showing a wide range of velocities in a young, unvirialised, cluster.
  \item NGC\,2264 lies close in velocity to a second cluster or association.
	\item The tangential velocity distribution of background stars in the direction of NGC\,2264 is less dispersed than observed for other clusters.
\end{itemize}
	
Since NGC\,2264 is a young, not too distant, cluster Li abundance and parallax data can be used to find the likely cause. Figure~\ref{fign2264}a show a histogram of target numbers in NGC\,2264 as a function of distance modulus. Stars identified as cluster members ( with $p>0.9$) show a reasonably tight distribution with a dispersion of 0.5\,mag. Stars identified as non-members show a much broader distribution suggesting they are indeed background stars, not members of NGC\.2264 or some second cluster or association at a common distance. Figure~\ref{fign2264}b also shows the EW(Li) of targets in NGC\,2264. Stars identified as cluster members show the high levels of EW(Li) expected for a young cluster aged $\sim 3-5$\,Myr whereas those identified as non-members  or as giants generally show the lower EW(Li) expected of older stars, confirming that the second population are correctly identified as non-members and that the background distribution is much less dispersed in tangential velocity than for most other clusters in our sample. 

For $T_{\rm eff}<5500$\,K, Fig.~\ref{fign2264}b contains 455 members with $p>0.9$ and 414 non-members. There are 7 false-positives -- members with  EW(Li)~$<250$\,m\AA. This is consistent with the average membership probability $\overline{\rm p}=0.986$ for targets with $p>0.9$ which suggests there should be 6 false positives. There are 80 false-negatives identified as non-members but with EW(Li)\,$>250$\,m\AA. Applying the {\it Gaia} quality cuts reduces the number of false negatives to 45, at the expense of reducing the number of members from 455 to 430, but this is still higher than the expected number of 33 false negatives based on $\overline{p}=0.081$ for targets with $p<0.9$, suggesting either that we have not fully accounted for non Gaussian errors in $V_{\rm RA}$, $V_{\rm Dec}$ and/or RV or that the intrinsic model for the cluster as a simple 3D Gaussian in velocity space is inadequate. That the latter is a factor is suggested by the complex spatial and kinematic structure of NGC\,2264 that has already emerged from previous radial velocity studies and early work with {\it Gaia} DR2 \citep{Tobin2015a, Venuti2018a, Buckner2020a}. In any case, like Lambda Ori, the tests above suggest that the membership probabilities of $p>0.9$ members are trustworthy and accurately reflect the amount of contamination in any sample drawn from them.

\begin{figure}
  \centering
	\includegraphics[width = 84mm]{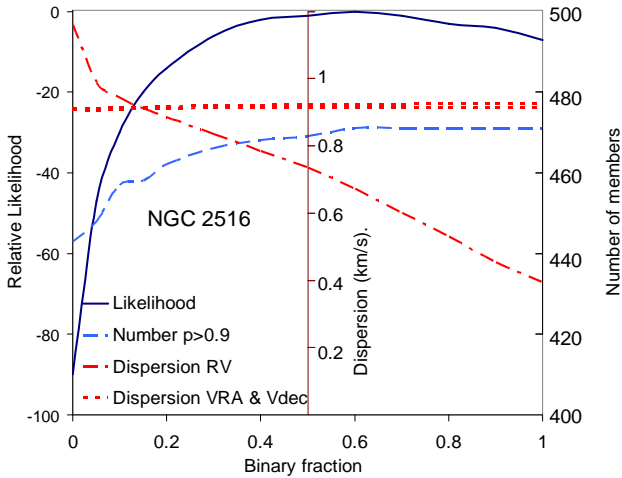}
	\caption{Variation of key parameters with the assumed binary fraction for open cluster NGC\,2516. The black solid line shows the maximum likelihood relative to the peak value, the red dot-dashed line shows the intrinsic cluster RV dispersion, the red dotted line shows that dispersion in $V_{\rm RA}$ and $V_{\rm Dec}$ and the blue dashed line shows the number of targets with a membership probability $p>0.9$.}
	\label{fign2516_1}
\end{figure}

\begin{figure}
  \centering
	\includegraphics[width = 74mm]{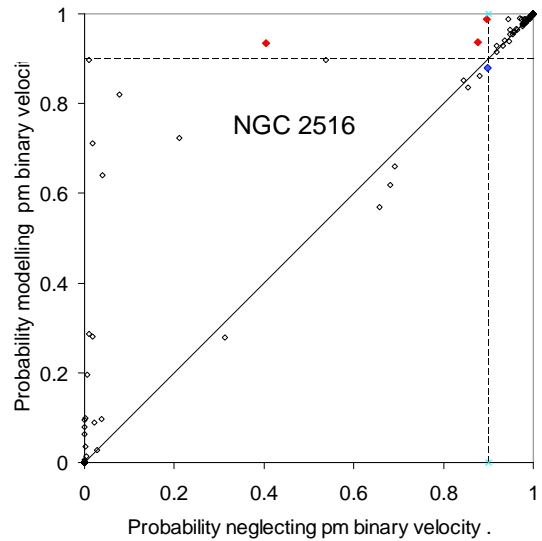}
	\caption{Membership probabilities in NGC\,2516 taking approximate account of the variable offset in proper motion caused by binarity, averaged over the period of {\it Gaia} observations, compared with those (in Table~\ref{table4}) calculated neglecting this effect.}
	\label{fign2516_2}
\end{figure}

\begin{figure*}
	\centering
	\begin{minipage}[ht]{\textwidth}
	\centering
	\includegraphics[width = 178mm]{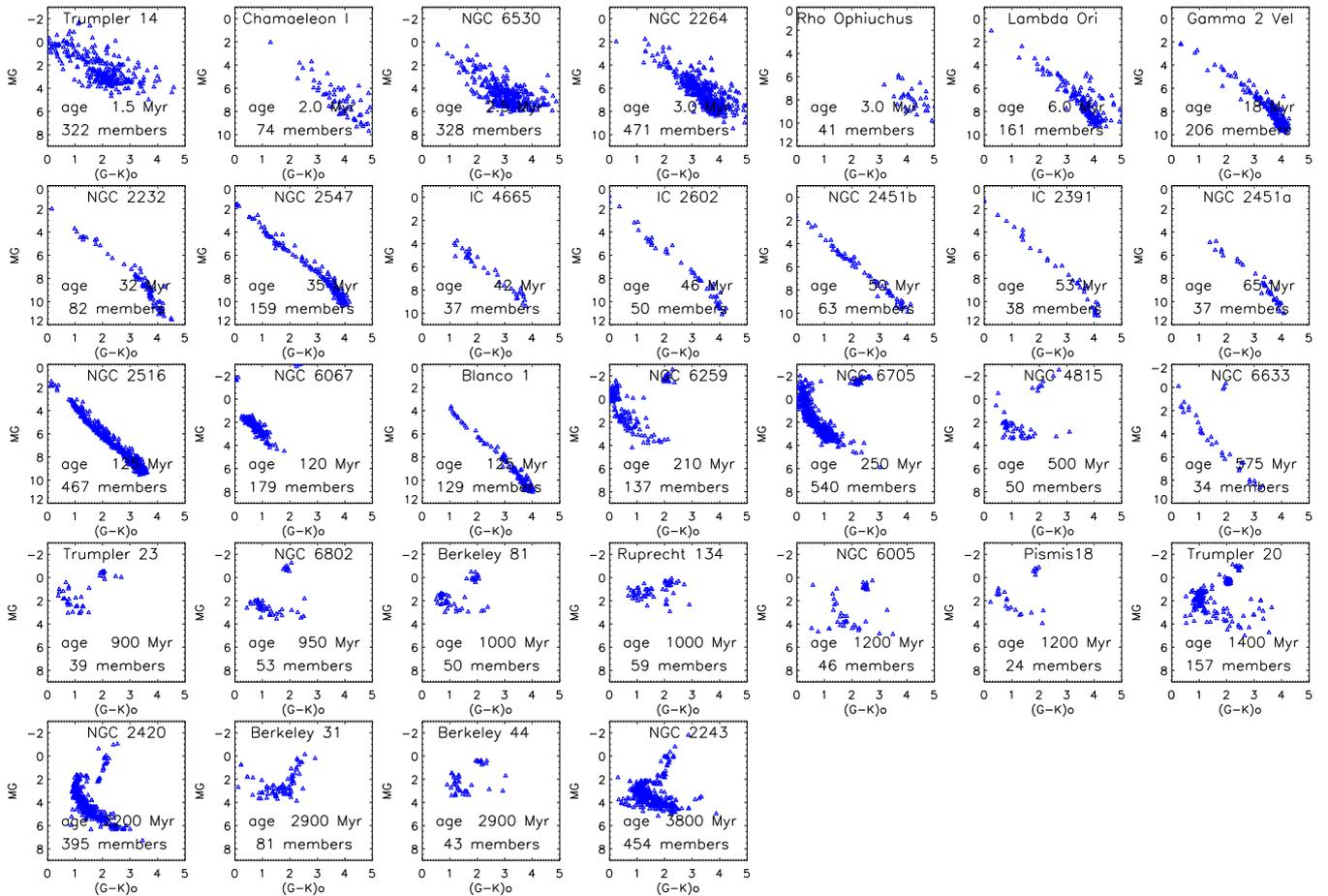}
	\end{minipage}
	\caption{Cluster colour magnitude diagrams showing targets with a membership probability $p>0.9$.}
	\label{figcmd}
\end{figure*}

\begin{figure}
  \centering
	\includegraphics[width = 64mm]{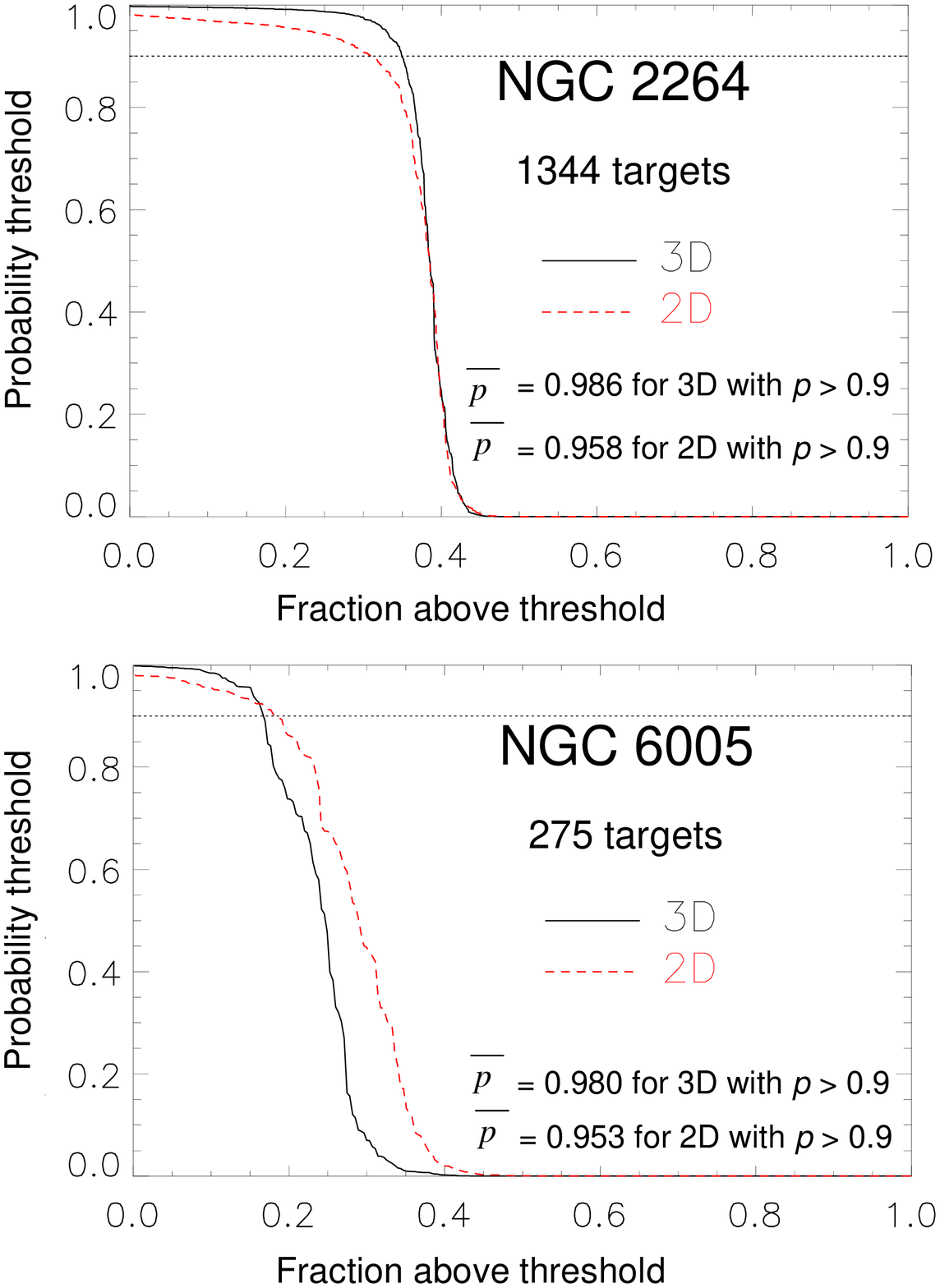}
	\caption{The fraction of targets  with membership probability, $p$ above a threshold level versus the threshold level for NGC\,2264 and NGC\,6530. The solid line shows the results for the standard 3D (proper motions and radial velocity) analysis, whilst the red, dashed lines show the results for a 2D analysis that only uses proper motion. Similar plots are available for all clusters in Appendix~\ref{appb}.}
	\label{fig2d3d}
\end{figure}

\subsection{Binarity}
\label{binarity}
 In this work we have assumed a fixed binary fraction $f_{\rm B}=0.46$ and a separation distribution that is the average for solar-type field stars measured by \cite{Raghavan2010a}. This binary fraction is also similar to that estimated by modelling the photometric deviations from single star isochrones in a number of these clusters \citep[e.g. NGC~2547 and NGC~2516,][]{Jeffries2001a, Jeffries2004a}. However, both the binary fraction and separation distribution (and possibly the mass ratio distribution) are likely to be mass-dependent \citep{Duchene2013a} and may also be different in clusters as a result of mass segregation or dynamical evolution \citep[e.g.][]{Geller2015a, Leiner2015a}.
In principle the binary properties should be treated as a free parameter in the model RV distribution and constrained by the data, since the RV offsets caused by binarity add to the observed RV dispersion of the cluster, and the shape of the RV offset distribution due to binarity is different to the distribution of RV uncertainties \citep{Cottaar2012a}. However in our case there are insufficient data to estimate the binary fraction with any precision (see below) and an accurate estimation would require an exceedingly precise knowledge of the shape, particularly the tails, of the RV uncertainty distribution.

To assess the effect of fixing $f_{\rm B}$, Fig.~\ref{fign2516_1} shows what happens to key parameters from the maximum likelihood analysis if $f_{\rm B}$ is given different values, using NGC\,2516 as an example. This is the best defined of our clusters with a a high number of cluster members and a low background population. The maximum likelihood shows a broad peak centred at $f_{\rm B}\sim 0.5$ but any value between about 0.3 and 0.7 would fit the data equally well. The number of cluster members with $p>0.9$ changes by only $\pm 0.5$ per cent for $0.3 < f_{\rm B} < 0.7$. 
However, changing $f_{\rm B}$ has a stronger effect on estimates of the intrinsic cluster RV dispersion, since the distribution in offsets of RV due to binarity is convolved with $D_{\rm RV}$ in the calculation of maximum likelihood. If broadening due to binarity is increased by increasing $f_{\rm B}$ then less of the observed dispersion needs to be explained by intrinsic dispersion in the cluster. This negative correlation between the intrinsic RV dispersion and the assumed value of $f_{\rm B}$ is clearly shown in Fig.~\ref{fign2516_1}. 

If $0.3 < f_{\rm B} <0.7$ then the best-fit intrinsic RV dispersion changes by $\sim \pm 0.1$\ km\,s$^{-1}$, which is larger than the formal statistical uncertainties on the fit.  For this reason there is a significant additional systematic uncertainty in the values of $D_{\rm RV}$ shown in Table~\ref{table4}. The size of this uncertainty will have a greater or lesser effect on each cluster depending on the size of the intrinsic RV dispersion of the cluster and RV uncertainties compared with the dispersion introduced by binaries. In many clusters (those with velocity dispersions $>1$ km\,s$^{-1}$ or those with large uncertainties in velocity dispersion) it is unimportant. A rule of thumb would be to add or subtract about $0.4$ km\,s$^{-1}$ in quadrature to the quoted value of intrinsic RV dispersion to simulate changing the binary fraction between $0.3< f_{\rm B} <0.7$.

Figure~\ref{fign2516_1} shows the importance of modelling of the offsets in measured RVs of binary stars relative to the barycentre. A similar, but smaller, effect is expected in the tangential velocities due to offsets in proper motion induced by the motion of the photo-centre of binary systems averaged over the observing period of {\it Gaia} \citep[e.g.][]{Lindegren2018a}. To test the significance of this, a simplistic model was developed that includes the additional uncertainty due to offsets in proper motions assuming uniform sampling over the 22 months of {\it Gaia} observations but neglecting any covariance between the astrometric parameters. As in Section~\ref{3.1}, the binary velocity offsets, $V_b$ were calculated for a set of randomly orientated binary systems, with a correction based on the mass ratio of the system to model the influence of the secondary on the motion of the photo-centre. We assume that the effects of the averaging diminish the observed velocity offset roughly as $V_b^{\rm pm} = V_b$\,sinc$(\pi T/P)$,
where $P$ is the binary period and $T$ the total observation time. The net effect on a simulated population is similar to that of binarity on the observed RV distribution, but the effect is smaller, and the tails of the distribution are suppressed, because of the strong averaging effect for shorter period binary systems.

This model was applied to the membership probability calculation for NGC\,2516, using the same distribution of orbital parameters as before and with $f_B=0.46$. Figure~\ref{fign2516_2} shows the comparison of the membership probabilities calculated with and without the effect of binarity on the tangential velocities. Only 6 per cent of stars have a change in membership probability of $>0.01$; these are targets which were outliers in $V_{RA}$ or $V_{Dec}$  in the original calculation. The general trend is to slightly increase membership probabilities; 3 stars (highlighted in red) have their membership probability increased to $>0.9$, while only one star (in blue) moves in the opposite direction.

This simple treatment suggests that the effect of neglecting binary motion on $V_{RA}$ and $V_{Dec}$ may produce a small ($< 1$ per cent) underestimate in the number of cluster members with $p > 0.9$ and is much less important than including the effects of binarity on $RV$ (c.f. the blue dashed line in Fig.~\ref{fign2516_2}). The effect of binarity on the proper motions also has a weaker effect on the inferred tangential velocity dispersions. The inclusion of the effect for NGC\,2516 predicts reduces the estimated intrinsic dispersions, $D_{RA}$ and $D_{Dec}$, from 0.91--0.92~km\,s$^{-1}$ (see Table~\ref{table4}) to  $0.78 \pm 0.04$\,km\,s$^{-1}$ (in better agreement with $D_{\rm RV}$). 

\subsection{The advantages of kinematic membership selection}
\label{5.5}

Whilst there are still caveats in examining the detailed kinematics of the clusters using the analysis presented here (see Sections~\ref{5.1} and ~\ref{binarity}), the fact that the membership probabilities are almost exclusively based on stellar kinematics makes our membership lists a valuable resource for investigating other properties of clusters and the stars within them. The targets in GES clusters were selected mainly\footnote{Many of the bright targets observed by UVES were selected as likely cluster members from previous studies.} on their photometric properties, but with a broad selection in colour-magnitude diagrams that should easily encompass the entire cluster population within the GES magnitude limits. Then, since the membership probabilities here are largely independent of photometry or estimates of stellar chemistry, then the results of this work can serve as inputs to investigate the HR diagram, cluster chemistry, rotation, magnetic activity, light element depletion etc. \citep[e.g.][]{Spina2017a, Randich2018a} but without the concern that results could be biased by using these properties to select members in the first place.

Pursuing these projects is beyond the scope of this paper, but as an illustration Fig.~\ref{figcmd} shows the absolute $G$ versus $(G-K_s)_0$ colour magnitude plots of cluster members (with $p> 0.9$), using the calculated values of distance modulus and reddening shown in Table~\ref{table1}.  A comparison with Fig.~\ref{fighrd} shows the extent to which our membership selection has "cleaned" these diagrams. Many of the clusters now clearly follow a single, age-dependent, isochrone with many less outliers than were seen in HR diagrams of the GES targets in Fig.~\ref{fighrd}. The form of the isochrones are best seen in nearby intermediate age clusters. The scatter is greater on young clusters ($<10$\,Myr) where the photometry is likely affected by differential reddening.  
 
 \subsection{The advantage of 3D over 2D kinematic selection}
 \label{5.6}
 
 The methodology described in Section~\ref{3.3} permits a comparison of how well selection using proper motion alone performs compared with proper motion plus the additional constraints provided by RV. Figure~\ref{fig2d3d} illustrates the distribution of $p$ obtained for two clusters in the sample (similar plots are available for all clusters in Appendix~\ref{appb}) -- NGC~2264 a young cluster ($\sim 5$ Myr) at intermediate distance ($\sim 750$ pc) and NGC~6005 an older
($\sim 1.2$ Gyr) and more distant ($\sim 3$ kpc) cluster. The solid lines indicate the final results of the 3D kinematic selection, whilst the dashed lines indicate 2D selection using tangential velocities alone.  

In almost all the clusters considered here, the addition of RV steepens the transition between objects with high and low $p$. i.e. It reduces the number of objects with intermediate values of $p$ and improves the contrast between members and non-members. Another general feature is that the average value of $p$ for those objects considered to be likely cluster members ($p>0.9$) increases. These increases are small, but highly significant if the aim is to provide secure samples with minimal contamination, since it is this latter statistic that determines the estimated numbers of false positives in the sample. For example in NGC 2264, a sample with $p>0.9$ selected from 2D velocity data would have 20 false positives, whereas adding RV selection reduces this number to 7.

The magnitude of these improvements depends on the size of the background population and the overlap between the cluster kinematics and that of the background. The latter is increased if the peak defined by the cluster in tangential velocity is blurred by the increased uncertainties that accompany greater distance. In contrast, the resolving power of the RV measurements is not directly distance-dependent.

Thus in nearby clusters or where there is relatively little background contamination, the improvements of 3D over 2D selection are very small (e.g. Rho Oph, Fig.~B5; NGC~2516, Fig.~B15; NGC 2243, Fig.~B32). However, when the background is significant and the tangential velocity of the cluster is not distinct from that background, especially in more distant clusters, the improvement in the fidelity of membership selection when adding RV is considerable (e.g. NGC 6530, Fig.~B3; Pismis 18, Fig.~B27; Berkeley 31, Fig.~B30).

\section{Summary}
\label{summary}

In this paper, we have set out a methodology designed to give a secure, rather than complete, set of members for 32 open clusters observed as part of the {\it Gaia}-ESO Survey. After filtering the observed targets to exclude those without the necessary data or which are obvious background stars or giants, membership is assessed solely using the 3D kinematics of the stars. Using a maximum likelihood technique, robust membership probabilities have been computed and in all of the clusters there is a clear separation between the population of high probability members and objects which are most likely to be unrelated to the cluster. The addition of radial velocities improves the ability to separate cluster and background populations over proper motion data alone. This is especially important for the distant clusters where the uncertainties in proper motion are larger than the intrinsic dispersions within the cluster or in clusters where there is significant kinematic overlap between the cluster and background populations.

Tests using independent membership criteria in young clusters suggest that the derived membership probabilities give an accurate indication of the contamination remaining in any sample of high probability cluster members. However, it appears likely that the membership probabilities of some genuine members, that are not classified as such, may be underestimated. The explanation for this may lie in an imperfect understanding of the reliability of some of the kinematic data or of the tails of the radial velocity and proper motion uncertainty distributions. Alternatively it is probable that the simple Gaussian models we have used for the intrinsic velocity distributions are too simplistic to fully reflect the kinematics of the young clusters where we have been able to do these tests.

The results of our investigation are presented in the form of a catalogue of compiled data that includes the membership probability of each star observed towards each GES cluster. There are 5033 high probability ($p>0.9$) members of the 32 clusters, with an average probability of $\bar{p}=0.991$. We also show the RA, Dec and RV components of mean velocity and intrinsic dispersion of the cluster model used to determine  membership probabilities. We caution that these latter results do not fully characterise the shape of clusters in velocity space since any cluster asymmetry does not necessarily align with the chosen axes. 

Since the membership criteria are almost purely kinematic, and independent of stellar photometry and chemistry, then the catalogue will be valuable for investigating other non-kinematic stellar and cluster properties available from the GES data, without having to compromise the investigation by using those properties as membership criteria. Examples include testing stellar evolutionary models using HR and colour-magnitude diagrams or following the evolution of magnetic activity, rotation and light element depletion.

With the final data release of GES due towards the end of 2020 and the improvements expected in {\it Gaia} DR3, it is anticipated that the cluster membership catalogue will be updated in the future to include the full set of clusters observed as part of GES and more reliable and precise astrometric data.

\section{acknowledgements}
RJJ, RDJ and NJW wish to thank the UK Science and Technology Facilities Council
for financial support.  TB was funded by the project grant ’The New Milky Way’ from the Knut and Alice Wallenberg Foundation. 

Based on data products from observations made
with ESO Telescopes at the La Silla Paranal Observatory under programme
ID 188.B-3002. These data products have been processed by the Cambridge
Astronomy Survey Unit (CASU) at the Institute of Astronomy, University
of Cambridge, and by the FLAMES/UVES reduction team at
INAF/Osservatorio Astrofisico di Arcetri. These data have been obtained
from the {\it Gaia}-ESO Survey Data Archive, prepared and hosted by the Wide
Field Astronomy Unit, Institute for Astronomy, University of Edinburgh,
which is funded by the UK Science and Technology Facilities Council.
This work was partly supported by the European Union FP7 programme
through ERC grant number 320360 and by the Leverhulme Trust through
grant RPG-2012-541. We acknowledge the support from INAF and Ministero
dell' Istruzione, dell' Universit\`a' e della Ricerca (MIUR) in the
form of the grant "Premiale VLT 2012". The results presented here
benefit from discussions held during the Gaia-ESO workshops and
conferences supported by the ESF (European Science Foundation) through
the GREAT Research Network Programme.

This work has made use of data from the European Space Agency (ESA) mission
{\it Gaia} https://www.cosmos.esa.int/gaia), processed by the {\it Gaia}
Data Processing and Analysis Consortium (DPAC,
https://www.cosmos.esa.int/web/gaia/dpac/consortium). Funding for the DPAC
has been provided by national institutions, in particular the institutions
participating in the {\it Gaia} Multilateral Agreement.

\section{Data availability statement}
The data underlying this article are available in the GES archive at the Wide
Field Astronomy Unit at Edinburgh University at http//ges/roe.ac.uk/

\bibliographystyle{aa.bst} 
\bibliography{references}

\appendix
\section{Radial velocity measurement precision}
\label{appa}
An empirical estimate of measurement precision is used for GIRAFFE measurements of
RV for the analysis of cluster membership. As described in \cite{Jackson2015a}, 
the empirical measurement precision, $E_{\rm RV}$ is characterised as a Student's t-distribution scaled by an empirical uncertainty $S_{\rm RV}$ which varies with signal-to-noise ratio ($S/N$) 
and projected equatorial velocity ($v\sin i$) 

The scaling constant of uncertainty for short term repeats (e.g. spectra taken consecutively), using the same instrument set up and wavelength calibration, is given by;
\begin{equation}
S_{\rm RV,0} = B\frac{(1+([v\sin i]/C)^2)^{3/4}}{(S/N)}\, ,
\end{equation}
where $B$ is an empirically determined parameter that depends 
on the intrinsic stellar spectrum (largely characterised by the effective temperature) and $C$ is a function of the spectrograph resolving power.

For long-term repeats (e.g, spectra taken on different nights), there is an additional contribution to the measurement uncertainty
due to variations in instrument setup and wavelength calibration, $A$, which adds in 
quadrature to the short term uncertainty, such that the distribution of $E_{\rm RV}$ for 
long-term repeats is characterised by 
\begin{equation}
S_{\rm RV} = \sqrt{A^2+ S_{\rm RV,0}^{2}}\, ,
\end{equation}
  
\cite{Jackson2015a} used data for 9 clusters reported 
in the {\it Gaia}-ESO Survey (GES) data release iDR2/3 to determine empirical values 
for $A$, $B$ and $C$. In this case both $A$ and $C$ were treated as constants over the whole analysis.

The analysis of \cite{Jackson2015a} has been repeated here using data for the 32 clusters 
from GESiDR5 to determine appropriate expressions 
for $A$, $B$ and $C$. This  has required two modifications to the analysis: 
first, the use of a reduced value of $v\sin i$ to account for changes in instrument resolving power over time; and second, the scaling of constant $A$ with $S/N$ 
as $A=A_0 + A_1/(S/N)$ in order to fit data from more distant clusters which show lower average levels of $S/N$.

\subsection{Calculation of the reduced projected equatorial velocity}
\label{appa1}

The GES pipeline used to estimate $v\sin i$ for GESiDR5 data assumes a fixed spectral resolving power, $R=17000$ for filter HR15n. In practice, the effective resolution of spectra observed using the HR15n filter,  measured from the line width of arc-lamp spectra has varied with time over the period of the GES observations, falling from $R \sim 15000$ in January 2012 to $R \sim$13000 in February 2015 after which a new focusing procedure for the instrument produced a consistent level of $R \sim 17000$. As a result the pipeline values of projected equatorial velocity ($V_{\rm ROT}$) are higher than the true value of $v\sin i$ for observations made before February 2015. The effect is most pronounced for the slowest rotating stars where a $V_{\rm ROT}$ of $\sim 12$\,km\,s$^{-1}$ is reported. To correct for the reduction in $R$ below the expected level a reduced value of $v\sin i$ is used to determine the effect of rotational velocity on measurement precision of $RV$;  
 \begin{equation}
	v\sin i = \sqrt{V_{\rm ROT}^2 - V_{\rm cor}^2} \ \,\,{\rm for} \,\, V_{\rm ROT} > V_{\rm cor}\, ,
\end{equation}
where 
\begin{equation}
V_{\rm cor} = 0.895c\,\sqrt{\frac{1}{R^2} - \frac{1}{17000^2}}\, ,
\end{equation}
$c$ is the speed of 
light and $R$ is the resolving power over the period when the cluster was observed (values $R$ are shown in Table~\ref{table1}).

\subsection{Fitted parameters}
\label{appa2}
Data for 30,000 short-term repeats and 4,400 long-term repeats were analysed to determine empirical values of $A$, $B$ and $C$ in equations A1 and A2 giving (in units of km\,s$^{-1}$):
\begin{eqnarray}
A & = & 0.09 + 10.0/(S/N)\, ; \\\nonumber
B & = & 4.12 + 242.6\log({T_{\rm eff}/3600\,{\rm K})}\, ; \\\nonumber
C & = & 0.895c/R\, .\nonumber
\end{eqnarray}

\section{Results for individual clusters.}
\label{appb}
\begin{figure*}
\begin{minipage}[t]{0.98\textwidth}
\centering
\includegraphics[width = 145mm]{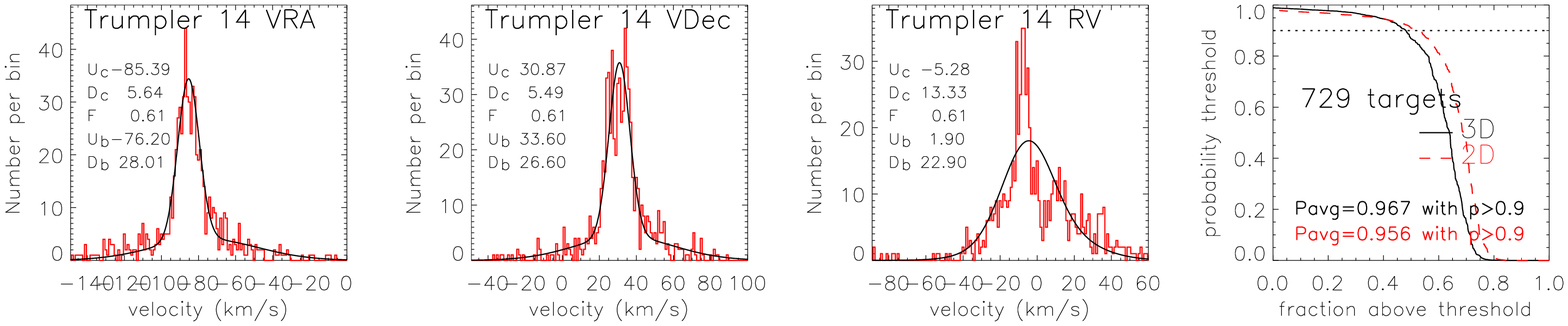}\\
\includegraphics[width = 145mm]{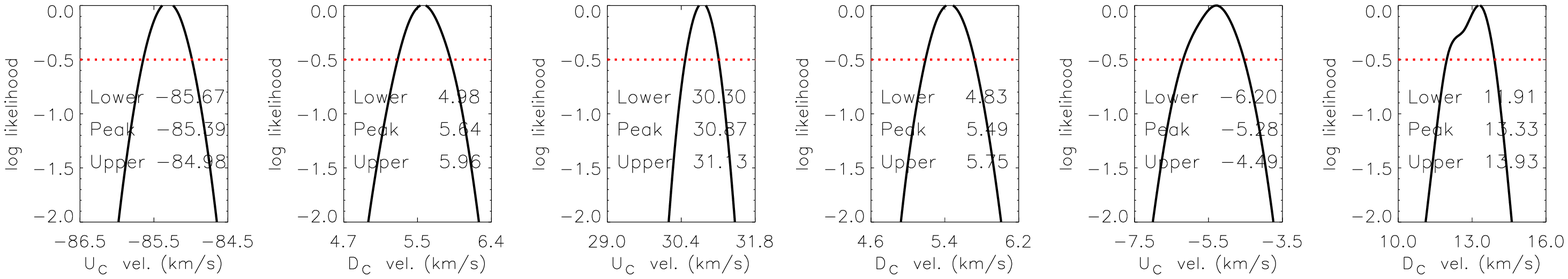}
\end{minipage}
\caption{Open cluster Trumpler 14: Results of the maximum likelihood analysis. The upper row of plots show histograms in each of the three velocity components ($V_{\rm RA}$, $V_{\rm Dec}$ and RV), together with model probability distributions evaluated at the maximum likelihood values of the fitted parameters, using a median measurement uncertainty. The text on the plots reports the best fitting (i.e. at the peak of the likelihood distribution) values for the cluster velocity ($U_{\rm c}$, in km\,s$^{-1}$), the intrinsic dispersion of the cluster velocity ($D_{\rm c}$), the fraction of objects assigned to the cluster population ($F$) and the velocity and dispersion of the background population ($U_{\rm b}$, $D_{\rm b}$).
The fourth plot in the upper row shows fraction of targets with membership probability, $p$ above a threshold level versus the threshold level and reports the average probability for members with $p>0.9$. Results are shown for our standard 3D analysis and for a 2D analysis that only takes account of proper motion velocities. The lower row of plots show the variation in log likelihood distribution for each component of the cluster velocity and its intrinsic dispersion (see Section 3.3 for further detail).}

\label{figB:1}
\end{figure*}

\begin{figure*}
\begin{minipage}{0.98\textwidth}
\centering
\includegraphics[width = 145mm]{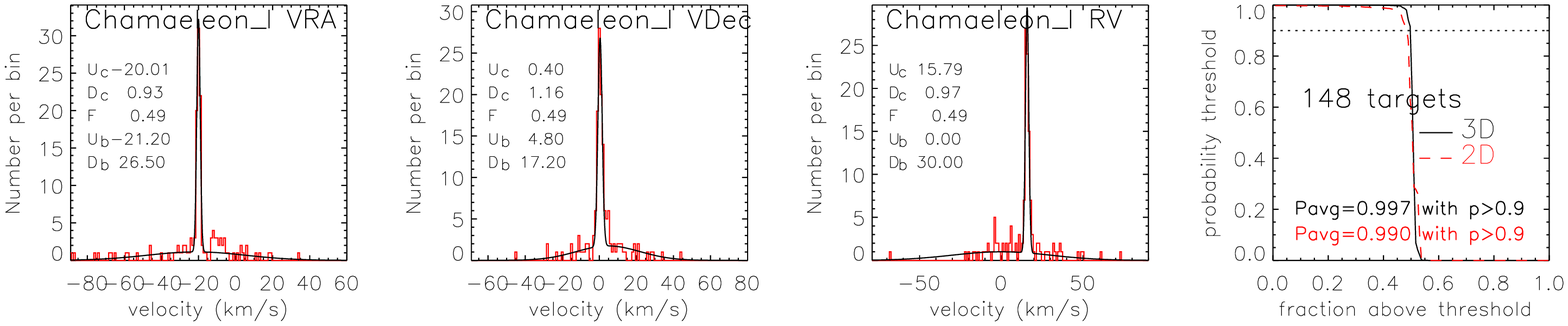}\\
\includegraphics[width = 145mm]{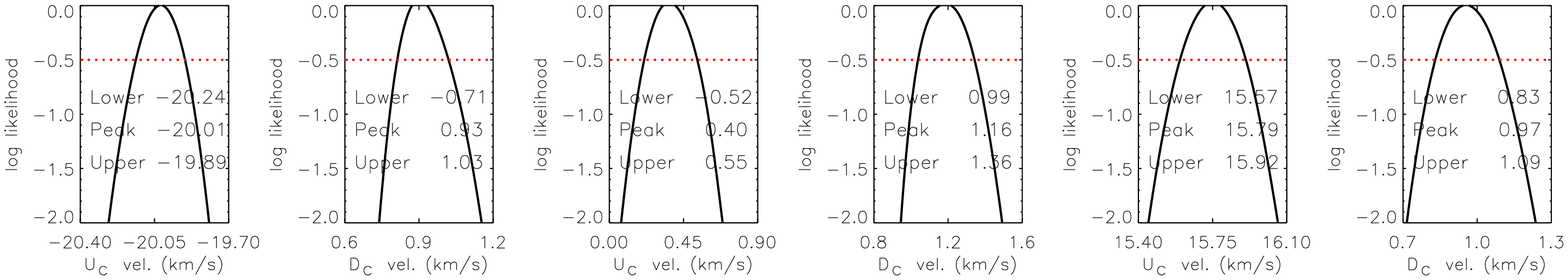}
\end{minipage}
\caption{Open cluster Chamaeleon I: Results of the maximum likelihood analysis. See Fig.~\ref{figB:1} for a detailed description of the individual plots.}
\label{figB:2}
\end{figure*}

\begin{figure*}
\begin{minipage}[t]{0.98\textwidth}
\centering
\includegraphics[width = 145mm]{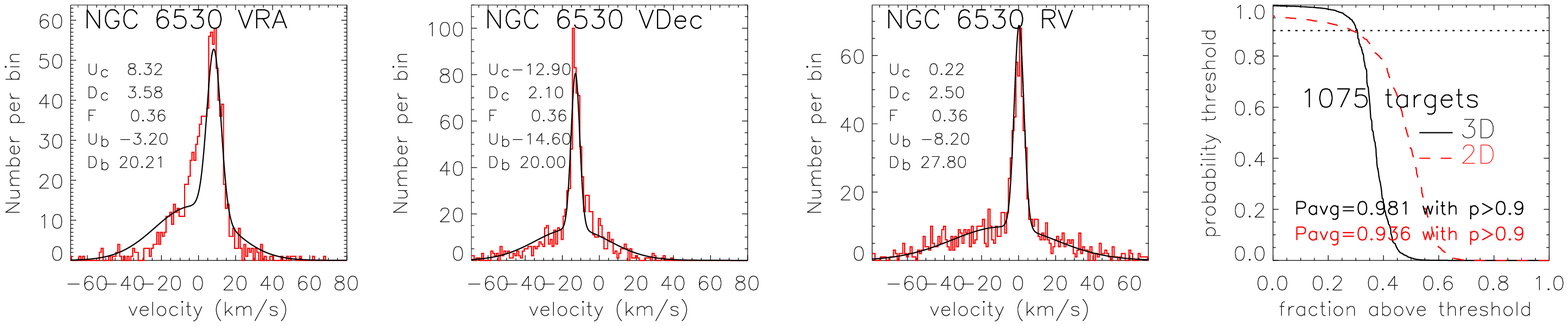}\\
\includegraphics[width = 145mm]{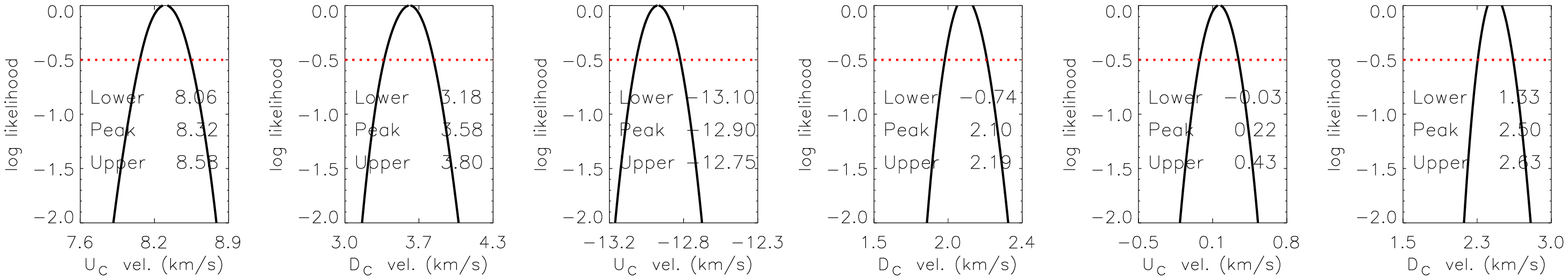}
\end{minipage}
\caption{Open cluster NGC 6530: Results of the maximum likelihood analysis. See Fig.~\ref{figB:1} for description of individual plots.}
\label{figB:3}
\end{figure*}

\begin{figure*}
\begin{minipage}{0.98\textwidth}
\centering
\includegraphics[width = 145mm]{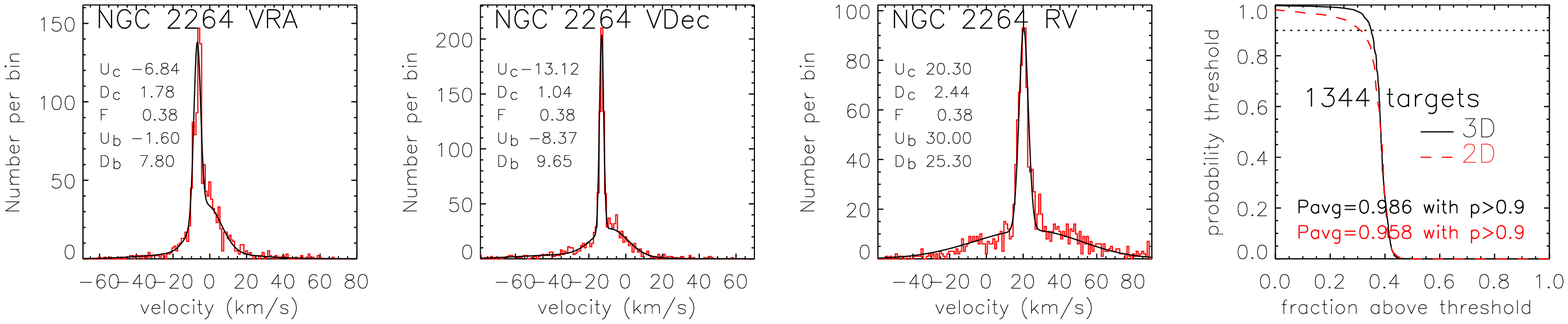}\\
\includegraphics[width = 145mm]{NGC2264_bestfit_3d.eps}
\end{minipage}
\caption{Open cluster NGC 2264: Results of the maximum likelihood analysis. See Fig.~\ref{figB:1} for description of individual plots.}
\label{figB:4}
\end{figure*}

\begin{figure*}
\begin{minipage}[t]{0.98\textwidth}
\centering
\includegraphics[width = 145mm]{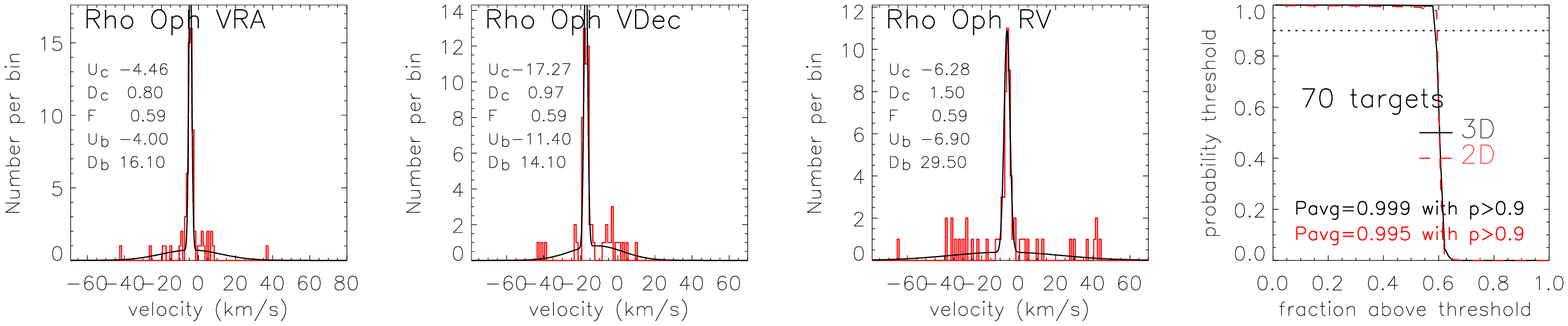}\\
\includegraphics[width = 145mm]{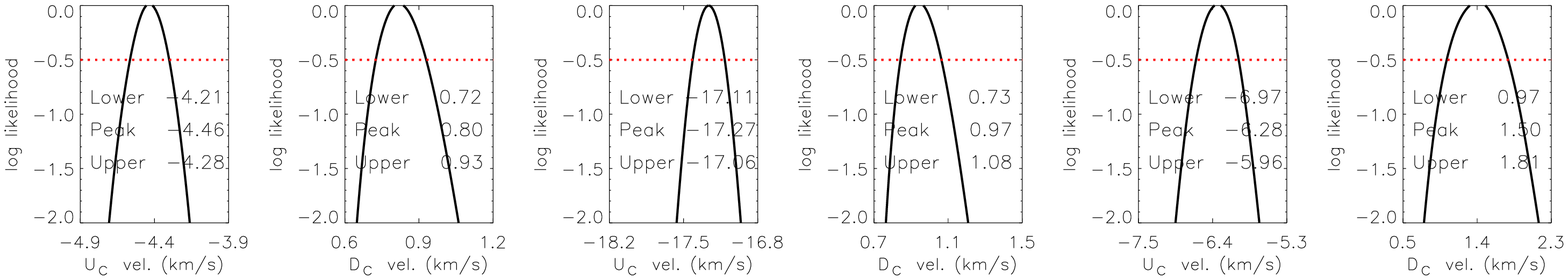}
\end{minipage}
\caption{Open cluster Rho Ophiuchus: Results of the maximum likelihood analysis. See Fig.~\ref{figB:1} for description of individual plots.}
\label{figB:5}
\end{figure*}

\begin{figure*}
\begin{minipage}{0.98\textwidth}
\centering
\includegraphics[width = 145mm]{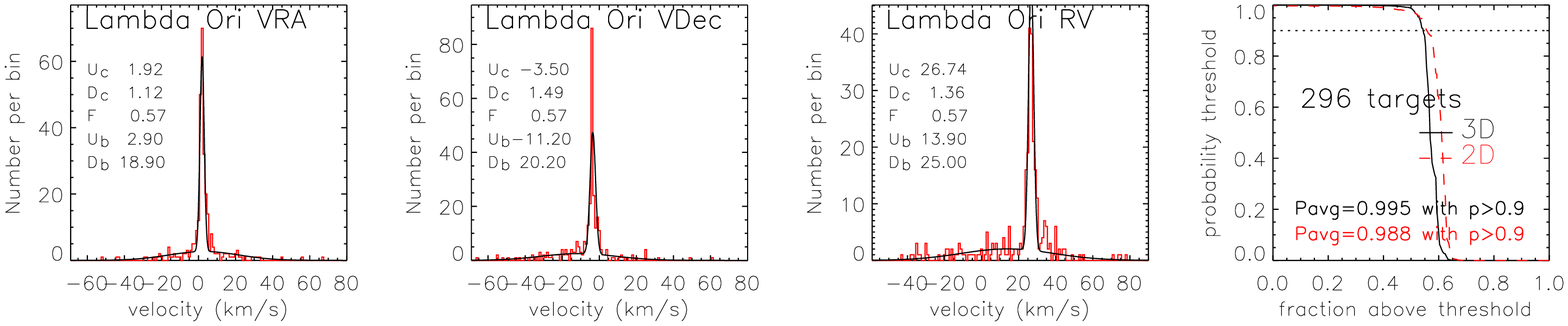}\\
\includegraphics[width = 145mm]{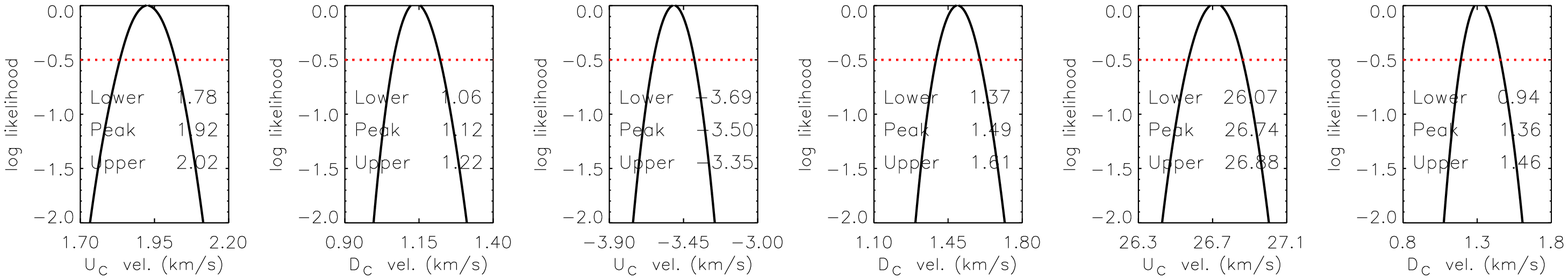}
\end{minipage}
\caption{Open cluster Lambda Ori: Results of the maximum likelihood analysis. See Fig.~\ref{figB:1} for description of individual plots.}
\label{figB:6}
\end{figure*}

\begin{figure*}
\begin{minipage}[t]{0.98\textwidth}
\centering
\includegraphics[width = 145mm]{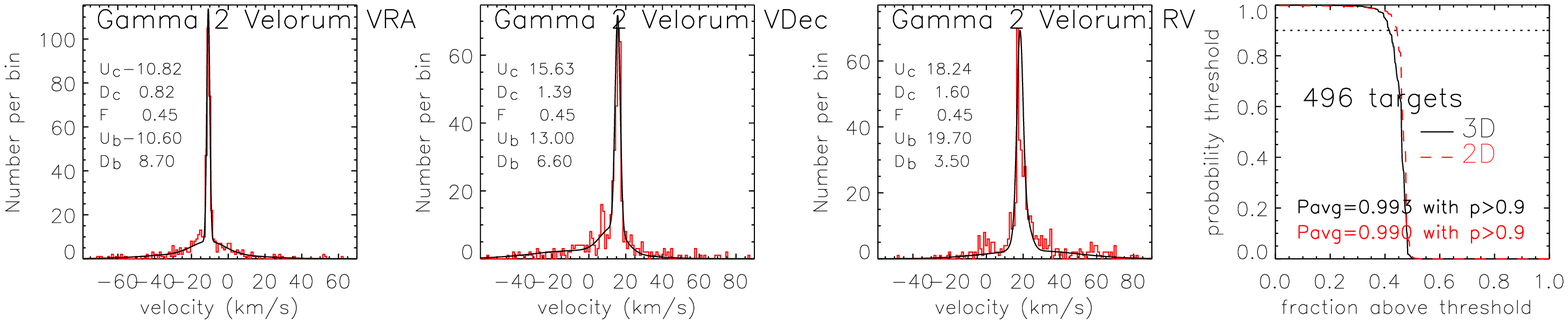}\\
\includegraphics[width = 145mm]{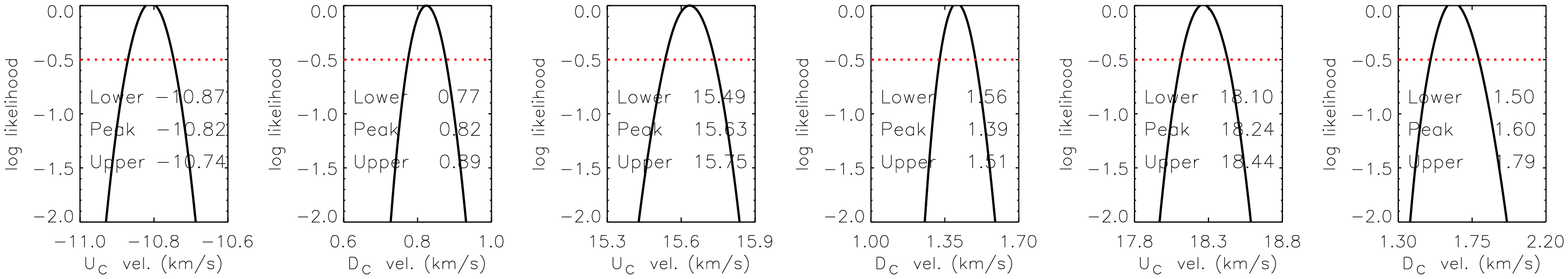}
\end{minipage}
\caption{Open cluster Gamma 2 Vel: Results of the maximum likelihood analysis. See Fig.~\ref{figB:1} for description of individual plots.}
\label{figB:7}
\end{figure*}

\begin{figure*}
\begin{minipage}{0.98\textwidth}
\centering
\includegraphics[width = 145mm]{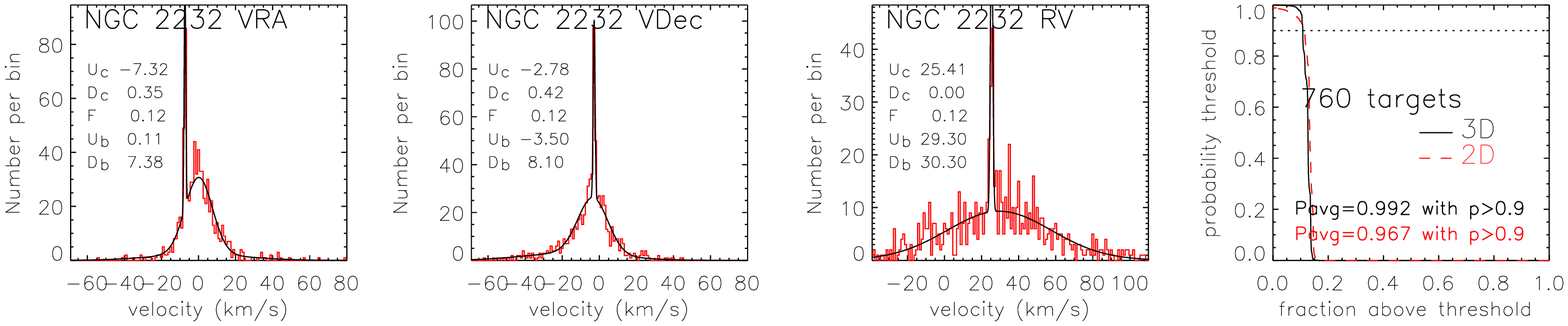}\\
\includegraphics[width = 145mm]{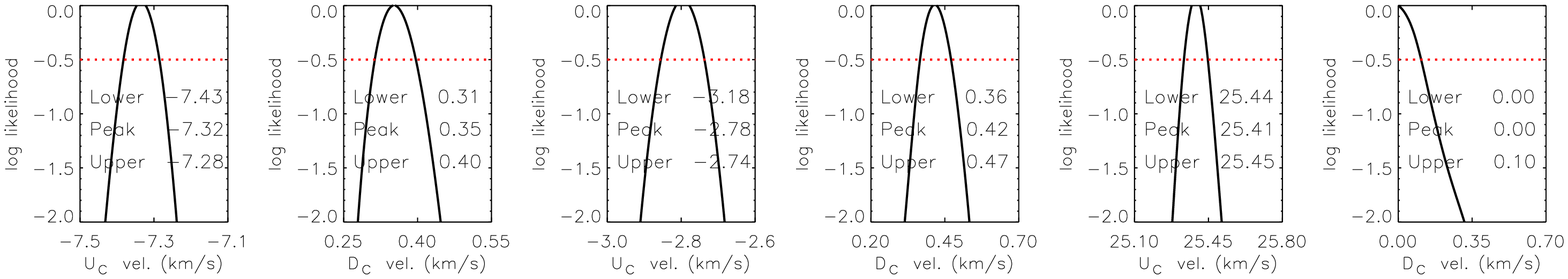}
\end{minipage}
\caption{Open cluster NGC 2232: Results of the maximum likelihood analysis. See Fig.~\ref{figB:1} for description of individual plots.}
\label{figB:8}
\end{figure*}

\begin{figure*}
\begin{minipage}[t]{0.98\textwidth}
\centering
\includegraphics[width = 145mm]{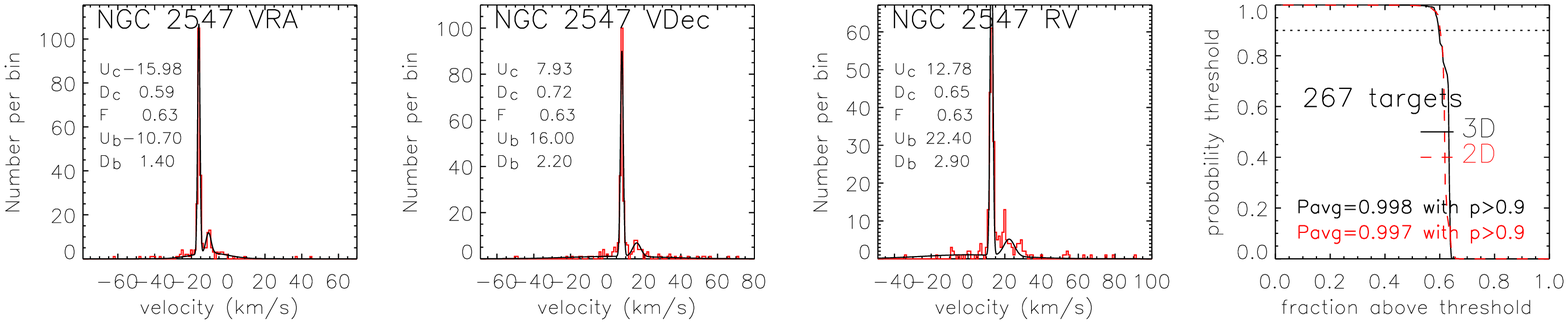}\\
\includegraphics[width = 145mm]{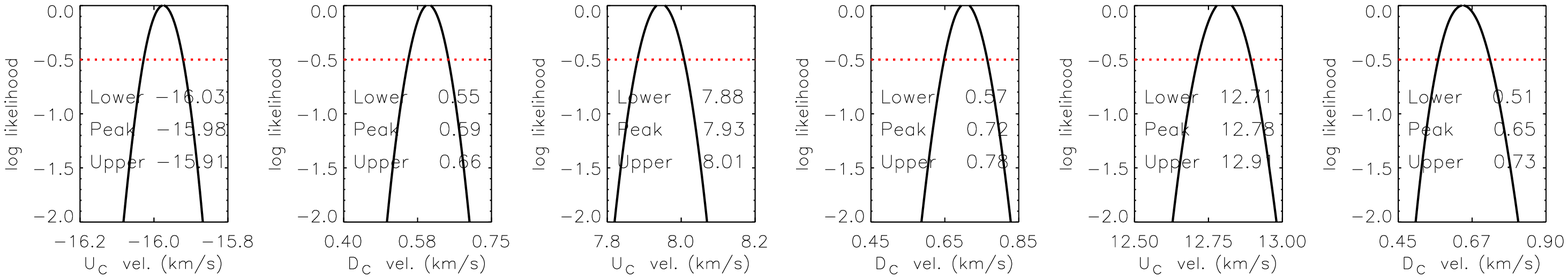}
\end{minipage}
\caption{Open cluster NGC 2547: Results of the maximum likelihood analysis. See Fig.~\ref{figB:1} for description of individual plots.}
\label{figB:9}
\end{figure*}

\begin{figure*}
\begin{minipage}{0.98\textwidth}
\centering
\includegraphics[width = 145mm]{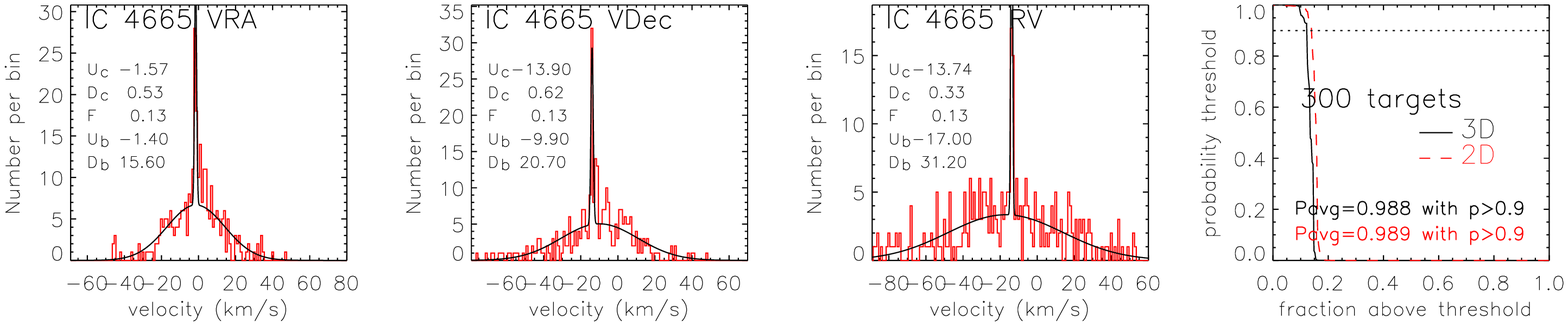}\\
\includegraphics[width = 145mm]{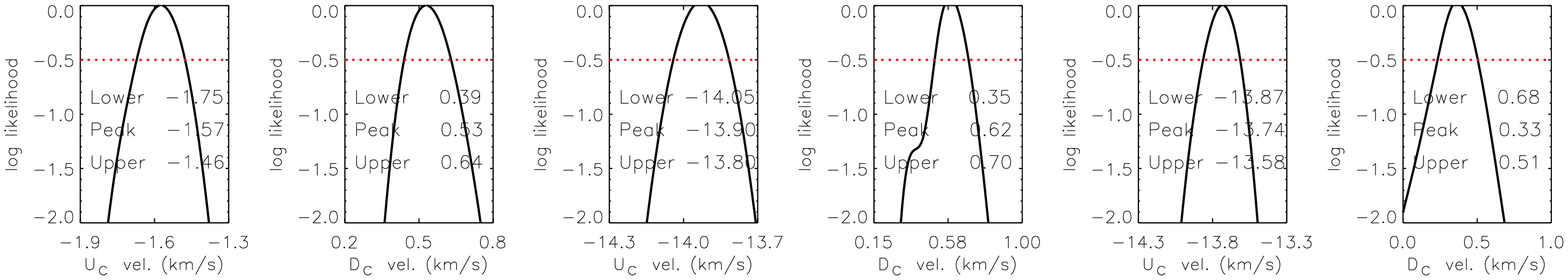}
\end{minipage}
\caption{Open cluster IC 4665: Results of the maximum likelihood analysis. See Fig.~\ref{figB:1} for description of individual plots.}
\label{figB:10}
\end{figure*}

\begin{figure*}
\begin{minipage}[t]{0.98\textwidth}
\centering
\includegraphics[width = 145mm]{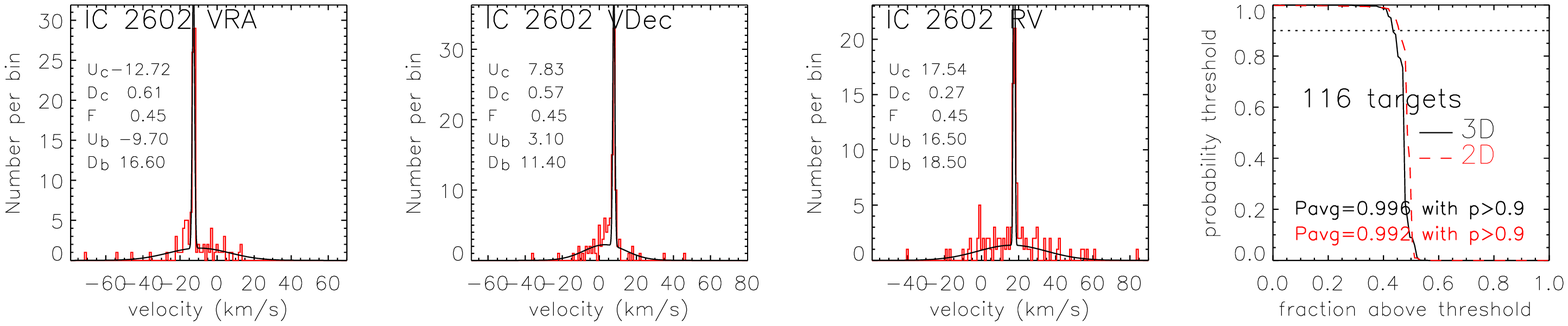}\\
\includegraphics[width = 145mm]{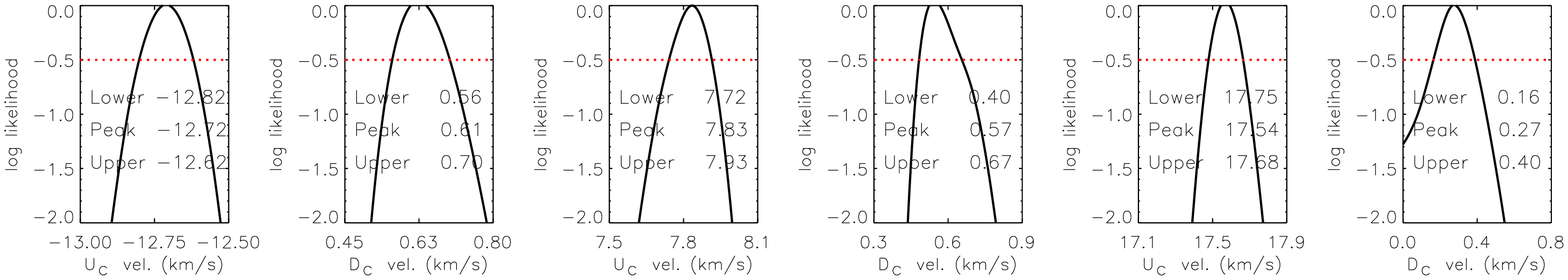}
\end{minipage}
\caption{Open cluster IC 2602: Results of the maximum likelihood analysis. See Fig.~\ref{figB:1} for description of individual plots.}
\label{figB:11}
\end{figure*}

\begin{figure*}
\begin{minipage}{0.98\textwidth}
\centering
\includegraphics[width = 145mm]{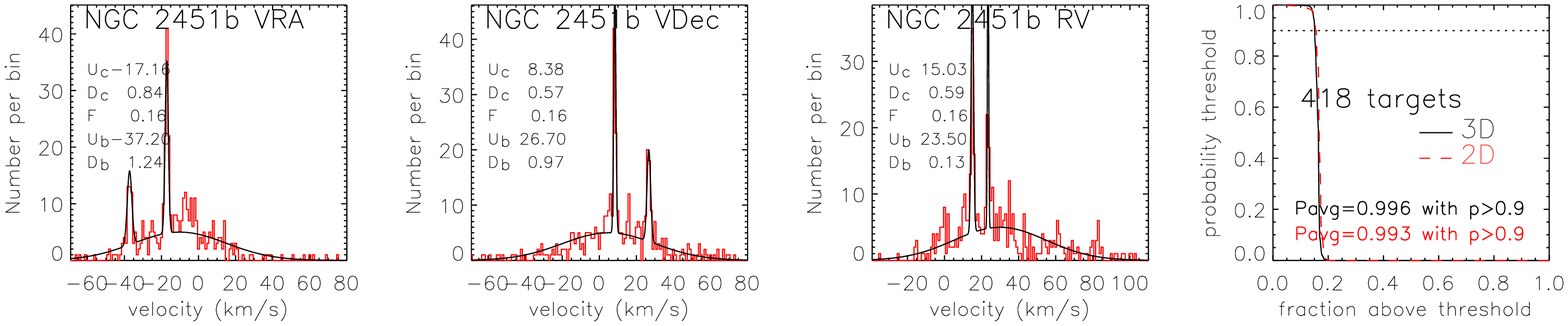}\\
\includegraphics[width = 145mm]{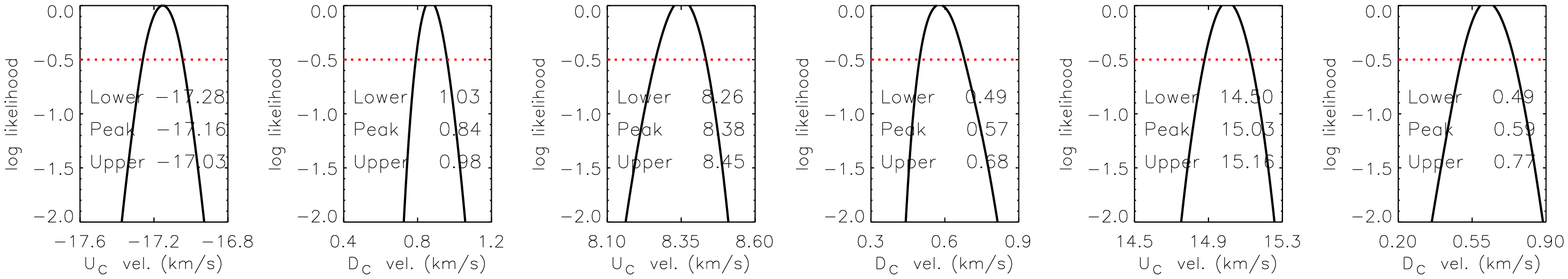}
\end{minipage}
\caption{Open cluster NGC 2451b: Results of the maximum likelihood analysis. See Fig.~\ref{figB:1} for description of individual plots.}
\label{figB:12}
\end{figure*}

\begin{figure*}
\begin{minipage}[t]{0.98\textwidth}
\centering
\includegraphics[width = 145mm]{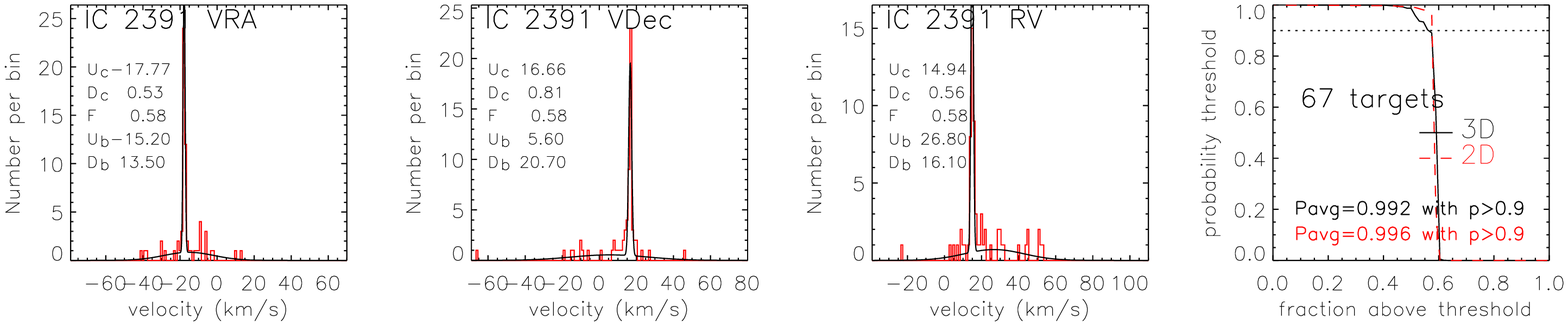}\\
\includegraphics[width = 145mm]{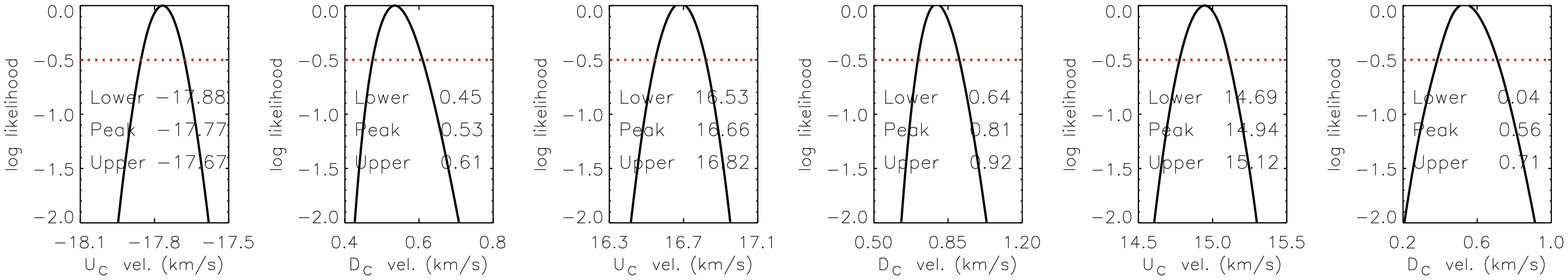}
\end{minipage}
\caption{Open cluster IC 2391: Results of the maximum likelihood analysis. See Fig.~\ref{figB:1} for description of individual plots.}
\label{figB:13}
\end{figure*}

\begin{figure*}
\begin{minipage}{0.98\textwidth}
\centering
\includegraphics[width = 145mm]{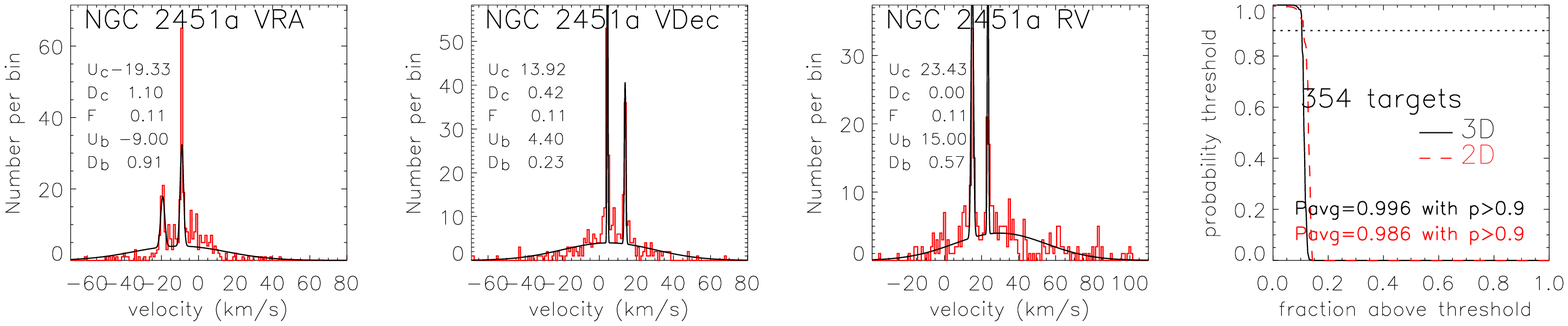}\\
\includegraphics[width = 145mm]{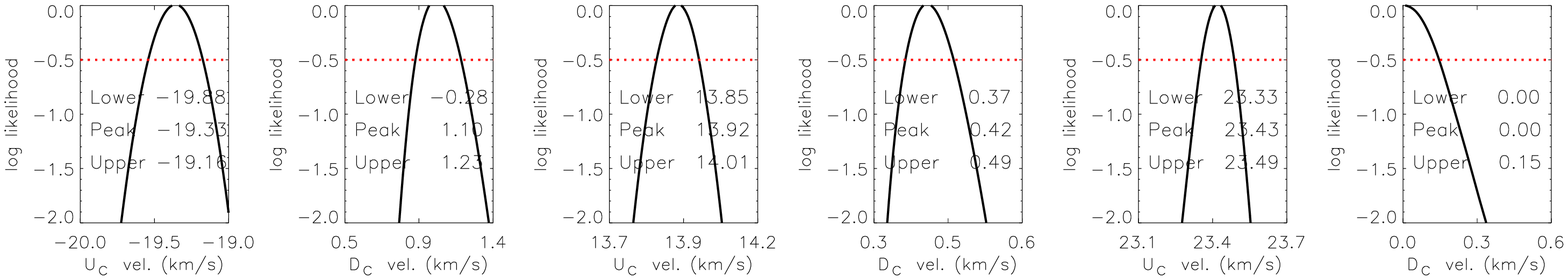}
\end{minipage}
\caption{Open cluster NGC 2451a: Results of the maximum likelihood analysis. See Fig.~\ref{figB:1} for description of individual plots.}
\label{figB:14}
\end{figure*}

\begin{figure*}
\begin{minipage}[t]{0.98\textwidth}
\centering
\includegraphics[width = 145mm]{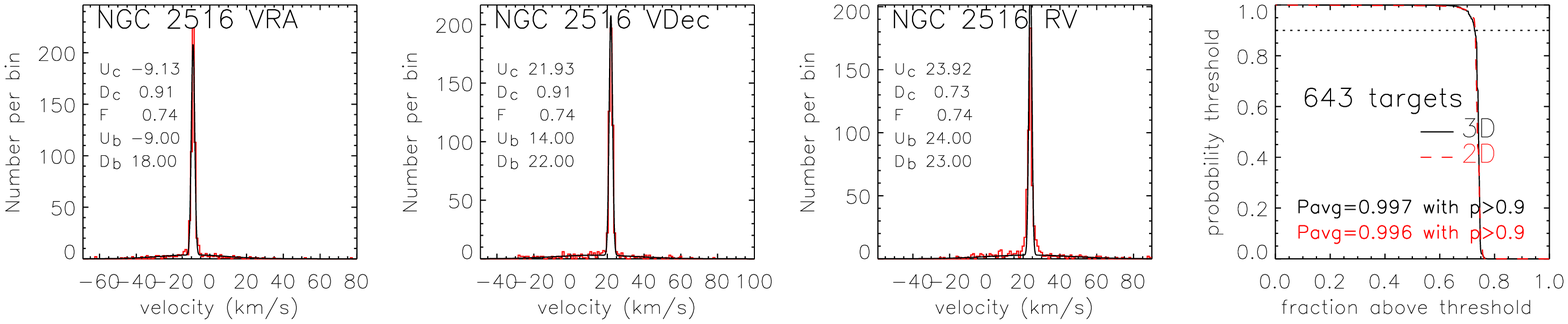}\\
\includegraphics[width = 145mm]{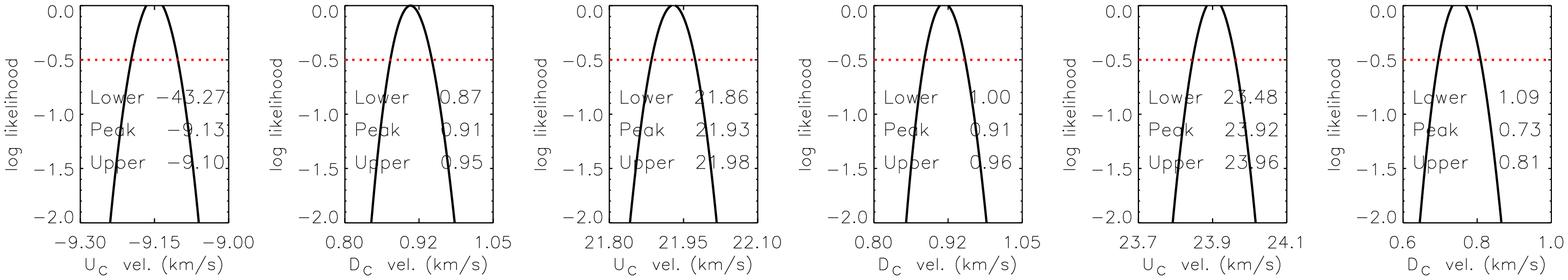}
\end{minipage}
\caption{Open cluster NGC 2516: Results of the maximum likelihood analysis. See Fig.~\ref{figB:1} for description of individual plots.}
\label{figB:15}
\end{figure*}

\begin{figure*}
\begin{minipage}{0.98\textwidth}
\centering
\includegraphics[width = 145mm]{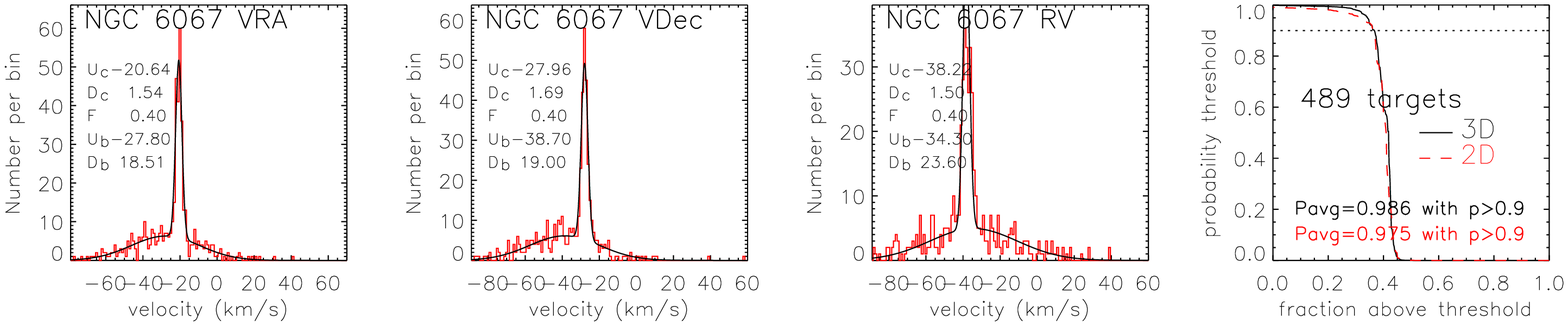}\\
\includegraphics[width = 145mm]{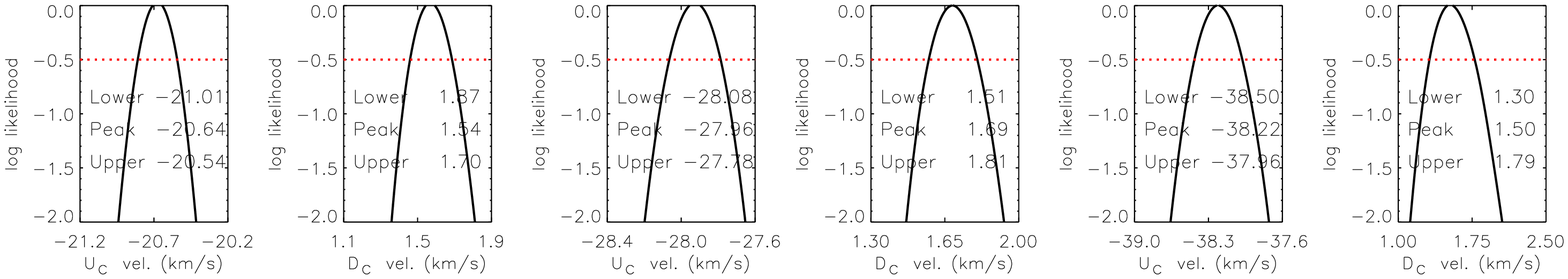}
\end{minipage}
\caption{Open cluster NGC 6067: Results of the maximum likelihood analysis. See Fig.~\ref{figB:1} for description of individual plots.}
\label{figB:16}
\end{figure*}

\begin{figure*}
\begin{minipage}[t]{0.98\textwidth}
\centering
\includegraphics[width = 145mm]{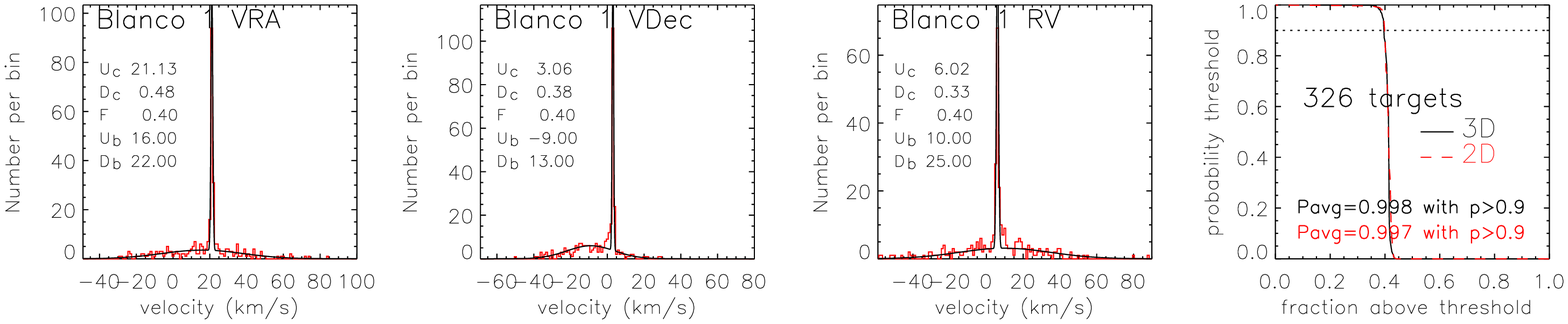}\\
\includegraphics[width = 145mm]{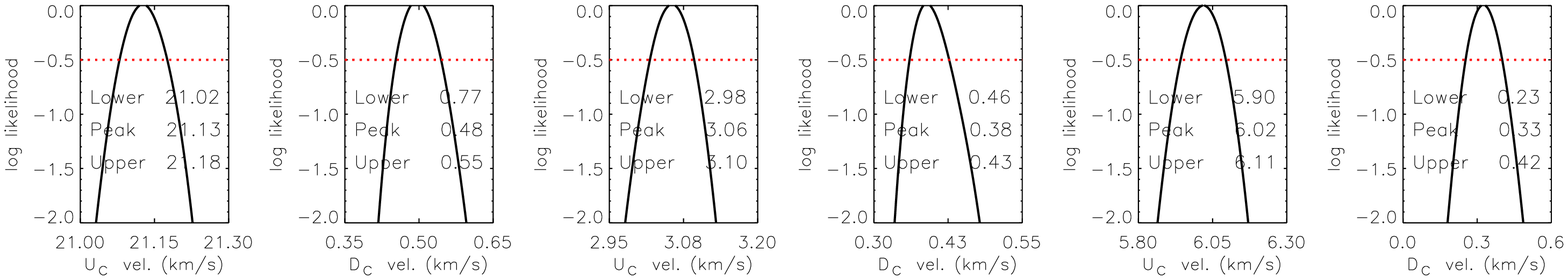}
\end{minipage}
\caption{Open cluster Blanco 1: Results of the maximum likelihood analysis. See Fig.~\ref{figB:1} for description of individual plots.}
\label{figB:17}
\end{figure*}

\begin{figure*}
\begin{minipage}{0.98\textwidth}
\centering
\includegraphics[width = 145mm]{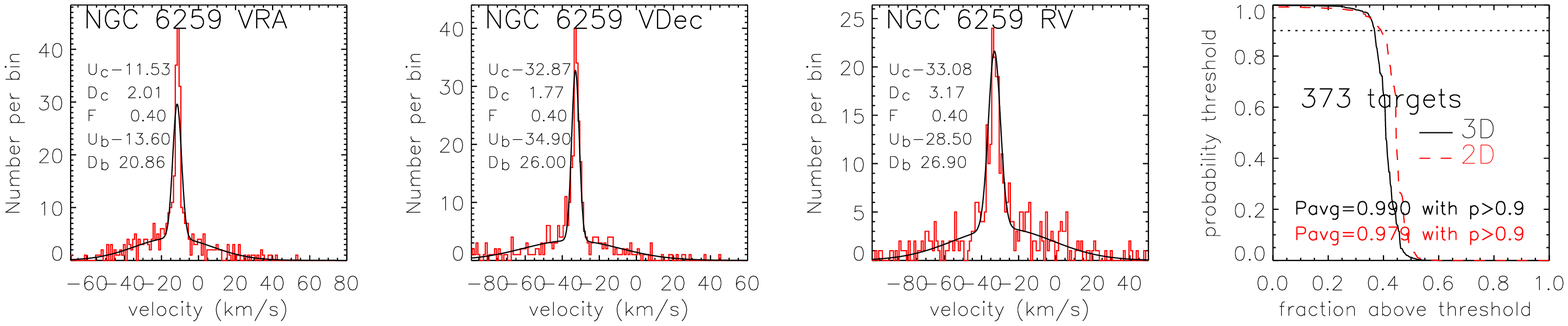}\\
\includegraphics[width = 145mm]{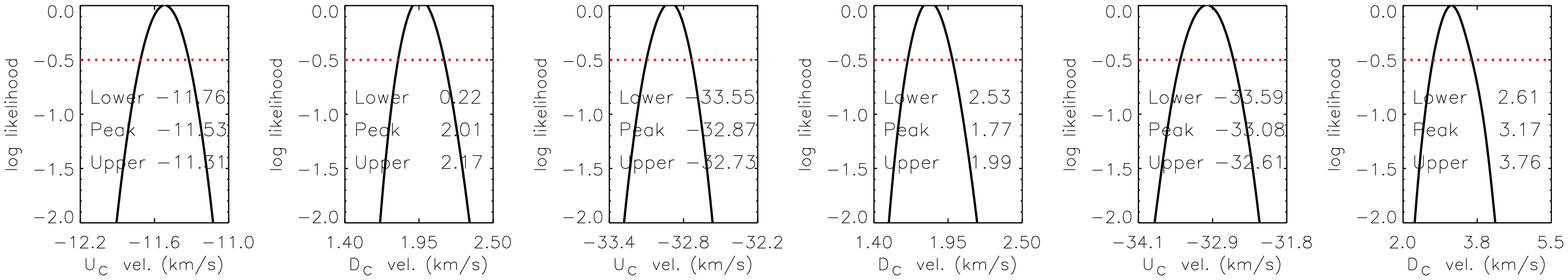}
\end{minipage}
\caption{Open cluster NGC 6259: Results of the maximum likelihood analysis. See Fig.~\ref{figB:1} for description of individual plots.}
\label{figB:18}
\end{figure*}

\begin{figure*}
\begin{minipage}[t]{0.98\textwidth}
\centering
\includegraphics[width = 145mm]{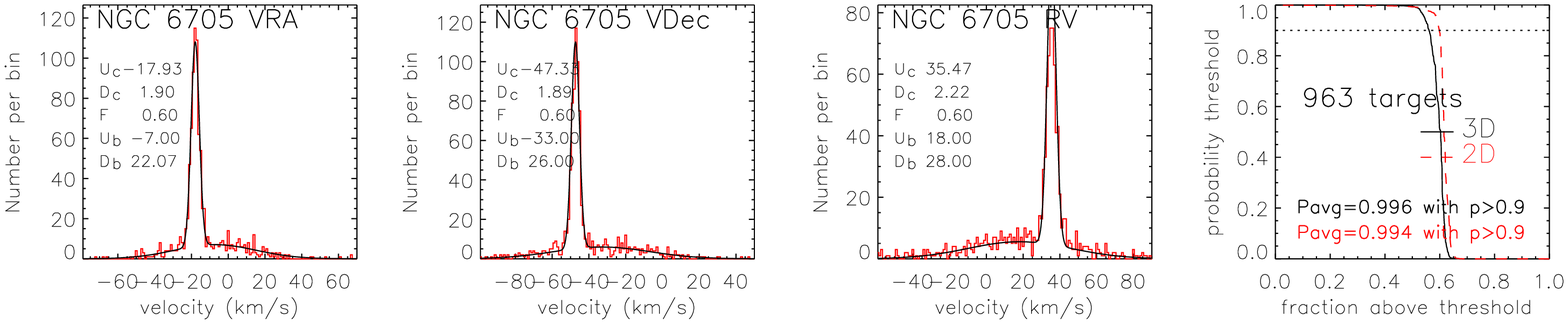}\\
\includegraphics[width = 145mm]{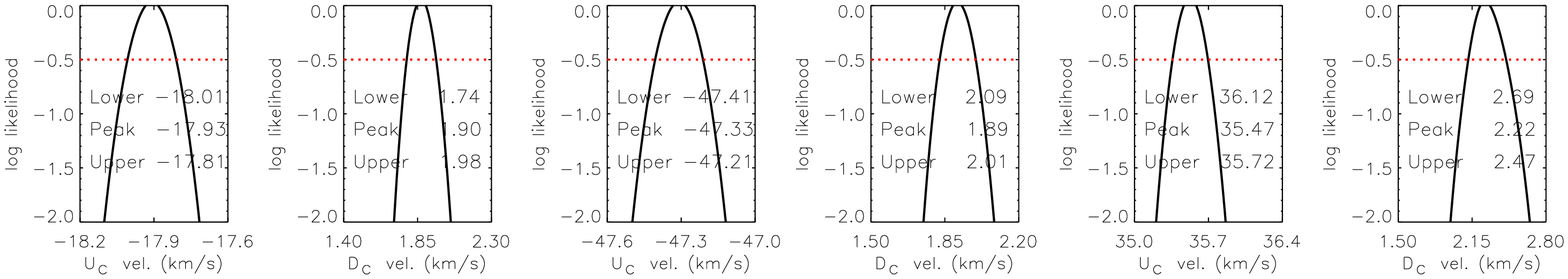}
\end{minipage}
\caption{Open cluster NGC 6705: Results of the maximum likelihood analysis. See Fig.~\ref{figB:1} for description of individual plots.}
\label{figB:19}
\end{figure*}

\begin{figure*}
\begin{minipage}{0.98\textwidth}
\centering
\includegraphics[width = 145mm]{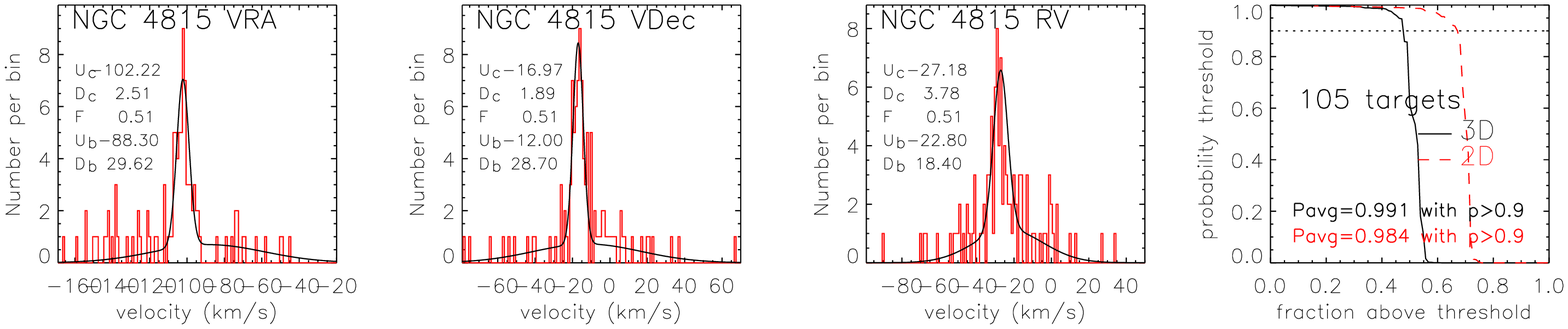}\\
\includegraphics[width = 145mm]{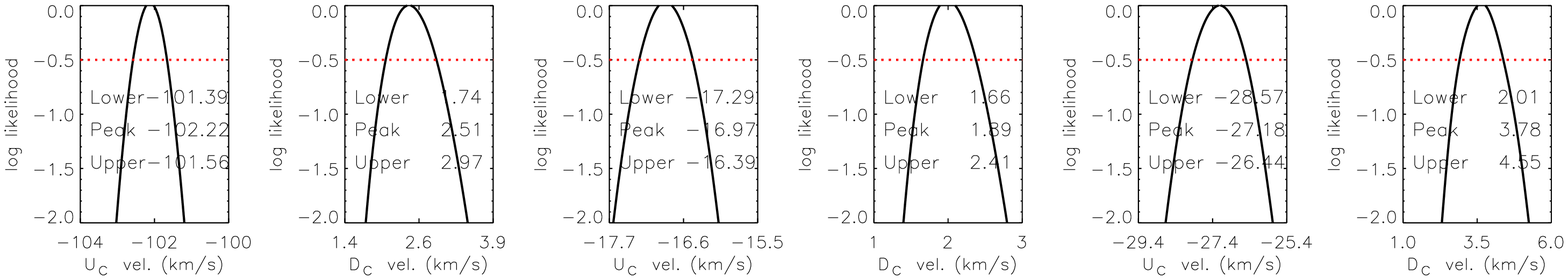}
\end{minipage}
\caption{Open cluster NGC 4815: Results of the maximum likelihood analysis. See Fig.~\ref{figB:1} for description of individual plots.}
\label{figB:20}
\end{figure*}

\begin{figure*}
\begin{minipage}[t]{0.98\textwidth}
\centering
\includegraphics[width = 145mm]{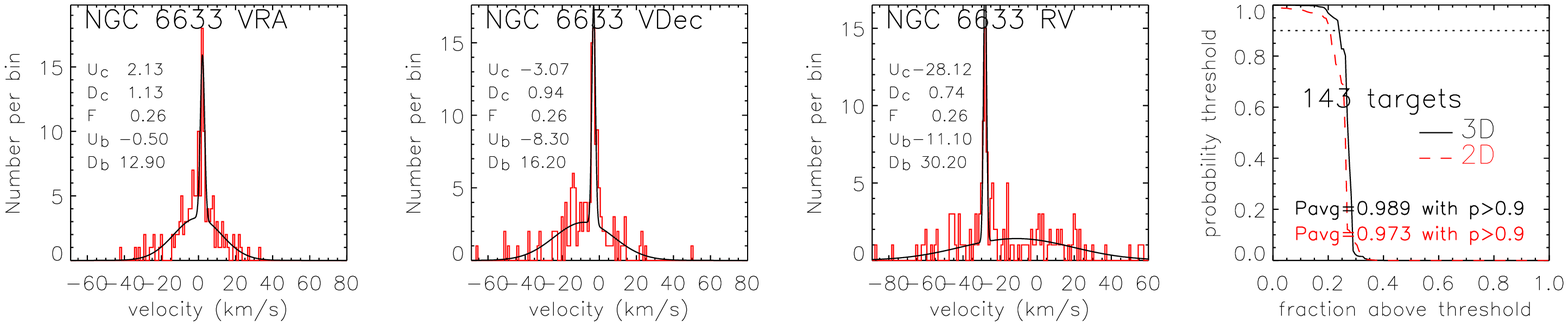}\\
\includegraphics[width = 145mm]{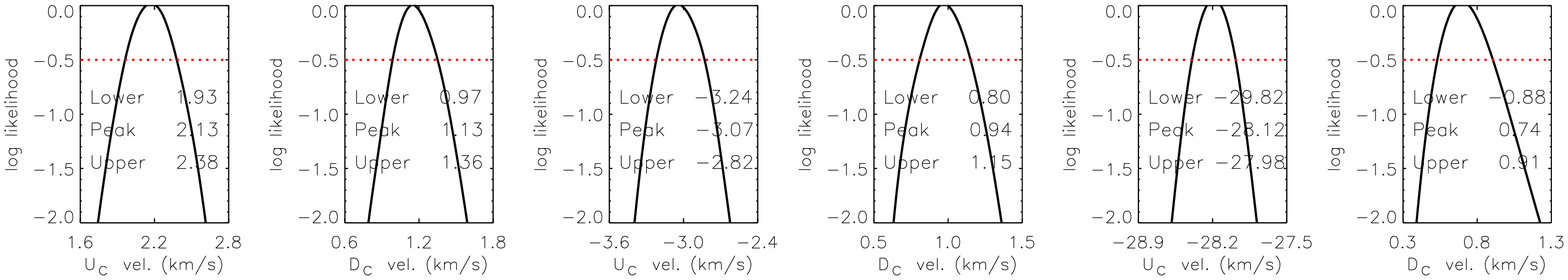}
\end{minipage}
\caption{Open cluster NGC 6633: Results of the maximum likelihood analysis. See Fig.~\ref{figB:1} for description of individual plots.}
\label{figB:21}
\end{figure*}

\begin{figure*}
\begin{minipage}{0.98\textwidth}
\centering
\includegraphics[width = 145mm]{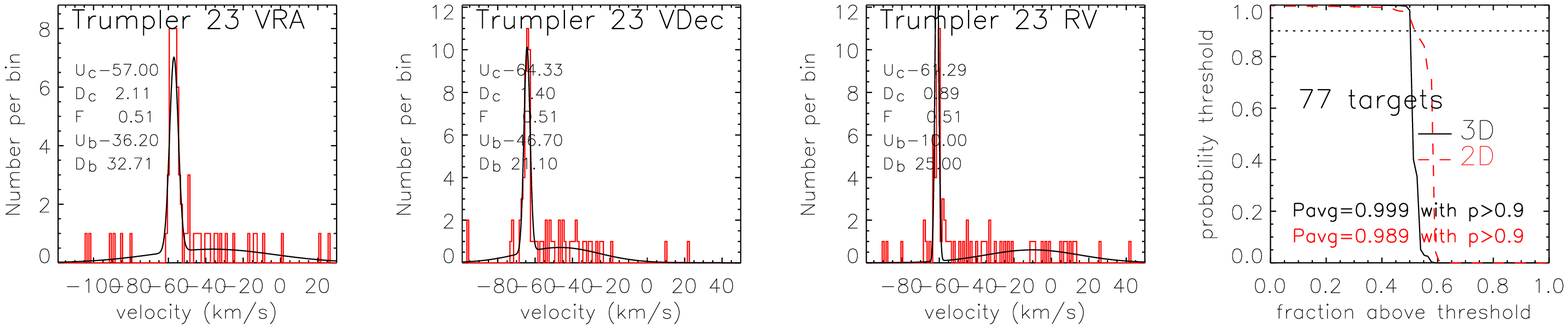}\\
\includegraphics[width = 145mm]{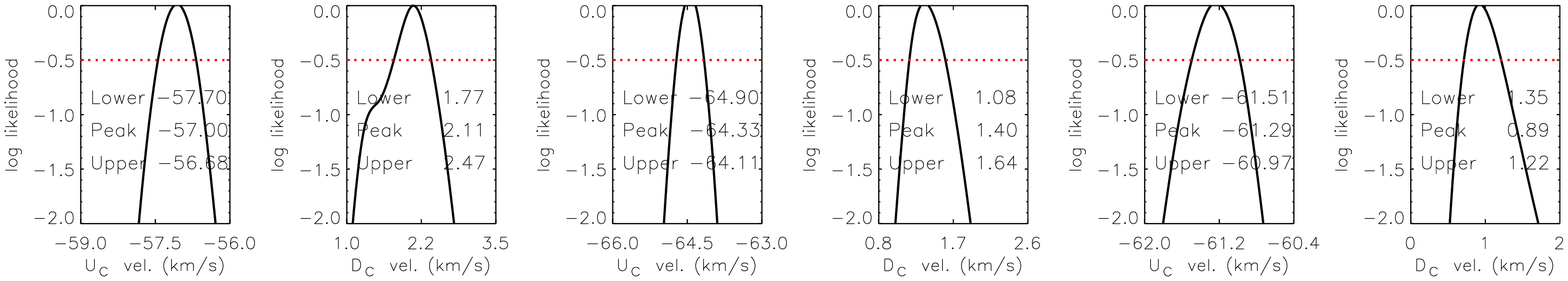}
\end{minipage}
\caption{Open cluster Trumpler 23: Results of the maximum likelihood analysis. See Fig.~\ref{figB:1} for description of individual plots.}
\label{figB:22}
\end{figure*}

\begin{figure*}
\begin{minipage}[t]{0.98\textwidth}
\centering
\includegraphics[width = 145mm]{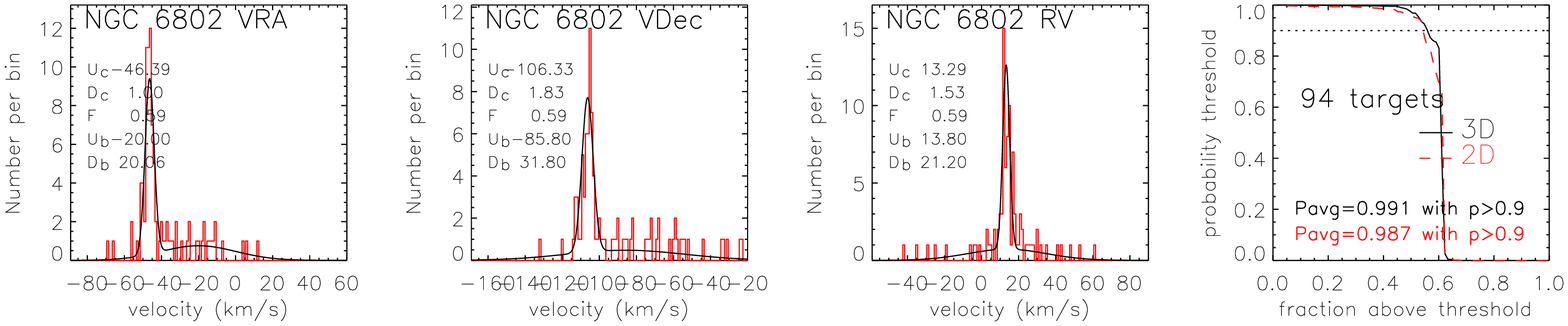}\\
\includegraphics[width = 145mm]{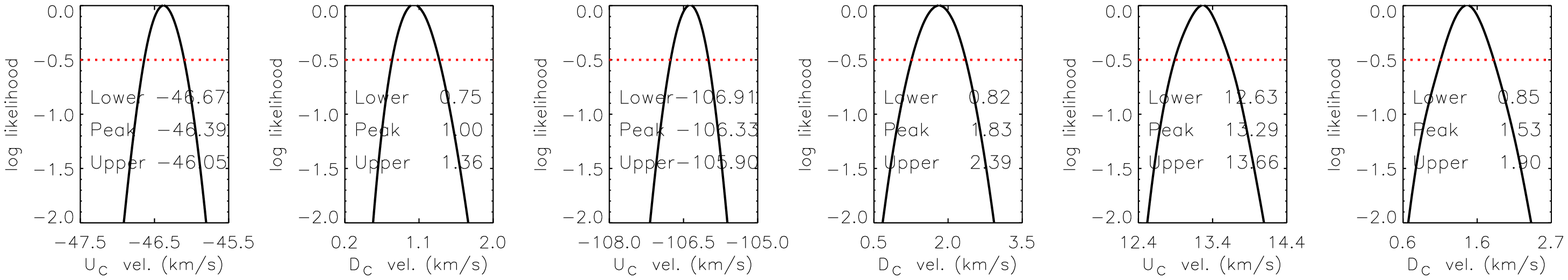}
\end{minipage}
\caption{Open cluster NGC 6802: Results of the maximum likelihood analysis. See Fig.~\ref{figB:1} for description of individual plots.}
\label{figB:23}
\end{figure*}

\begin{figure*}
\begin{minipage}{0.98\textwidth}
\centering
\includegraphics[width = 145mm]{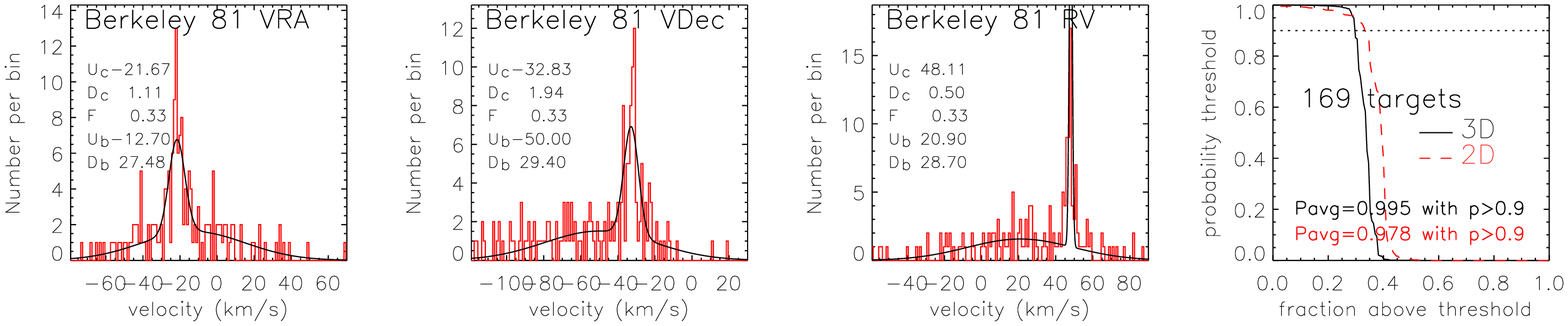}\\
\includegraphics[width = 145mm]{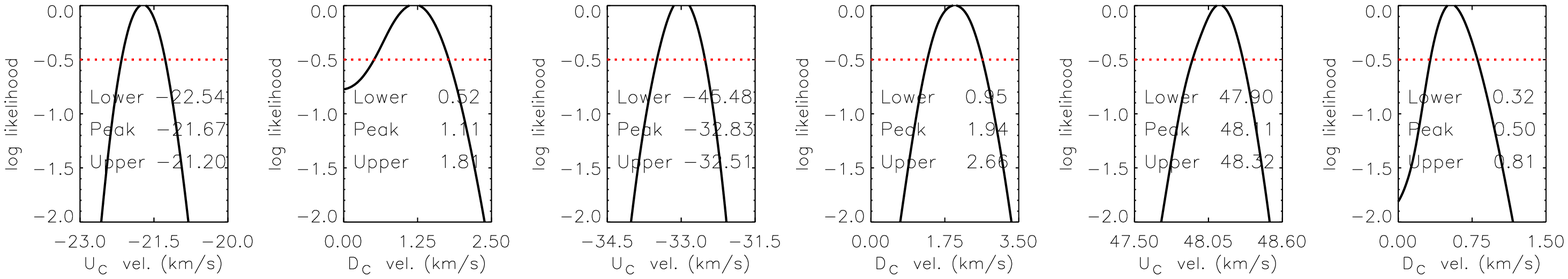}
\end{minipage}
\caption{Open cluster Berkeley 81: Results of the maximum likelihood analysis. See Fig.~\ref{figB:1} for description of individual plots.}
\label{figB:24}
\end{figure*}

\begin{figure*}
\begin{minipage}[t]{0.98\textwidth}
\centering
\includegraphics[width = 145mm]{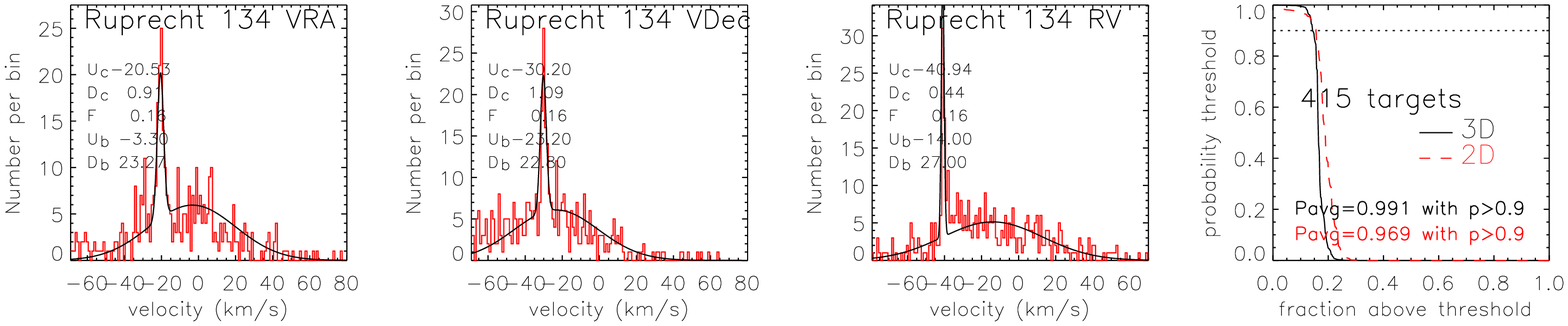}\\
\includegraphics[width = 145mm]{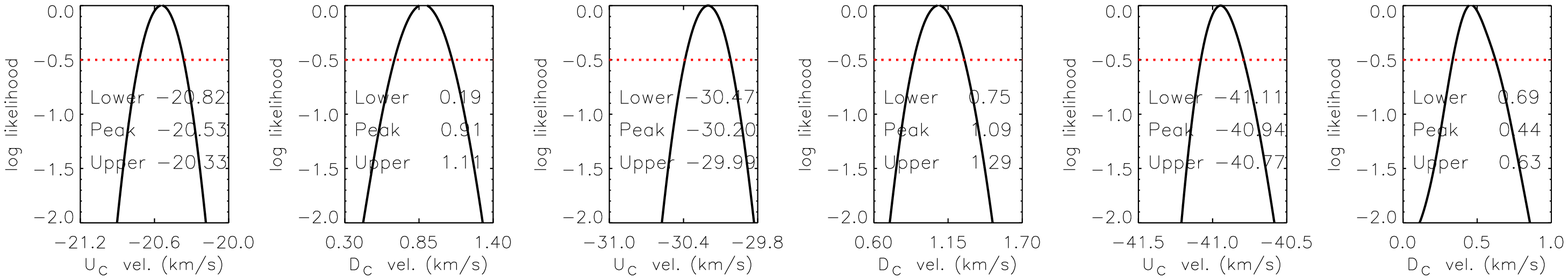}
\end{minipage}
\caption{Open cluster Ruprecht 134: Results of the maximum likelihood analysis. See Fig.~\ref{figB:1} for description of individual plots.}
\label{figB:25}
\end{figure*}

\begin{figure*}
\begin{minipage}{0.98\textwidth}
\centering
\includegraphics[width = 145mm]{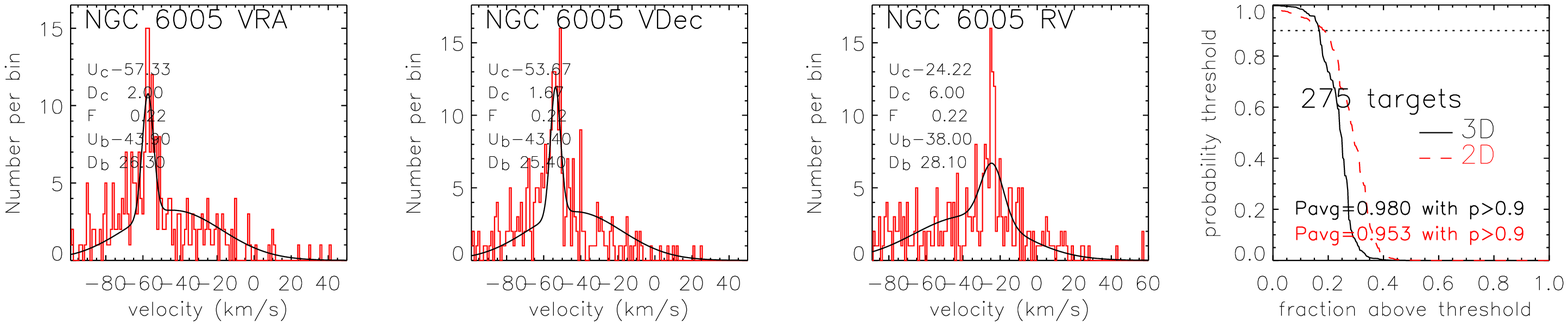}\\
\includegraphics[width = 145mm]{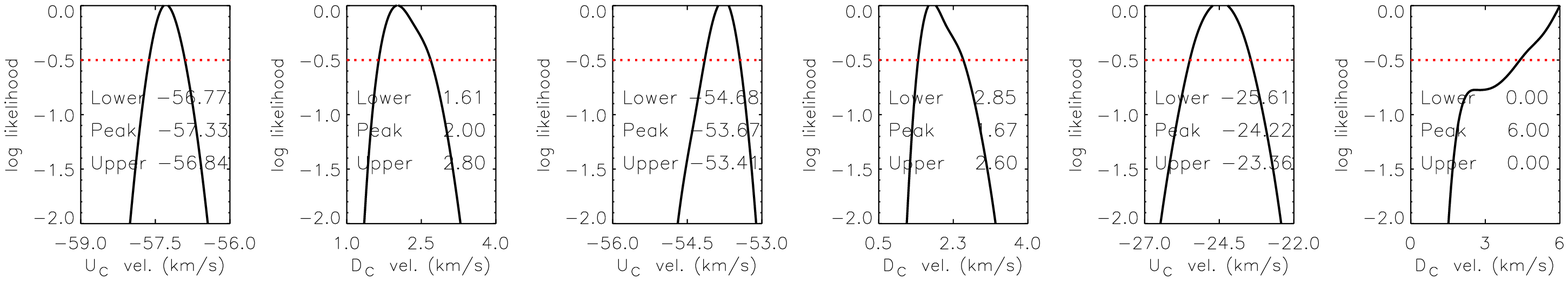}
\end{minipage}
\caption{Open cluster NGC 6005: Results of the maximum likelihood analysis. See Fig.~\ref{figB:1} for description of individual plots.}
\label{figB:26}
\end{figure*}

\begin{figure*}
\begin{minipage}[t]{0.98\textwidth}
\centering
\includegraphics[width = 145mm]{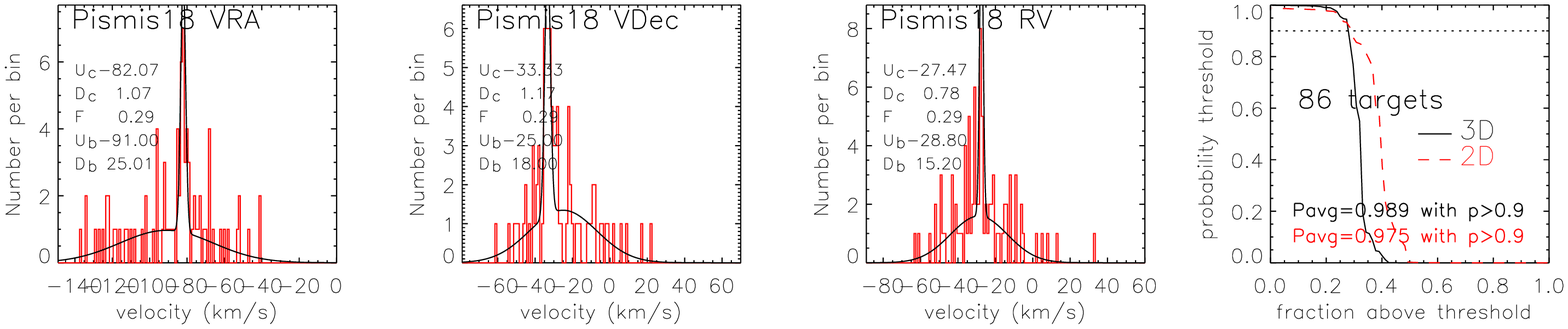}\\
\includegraphics[width = 145mm]{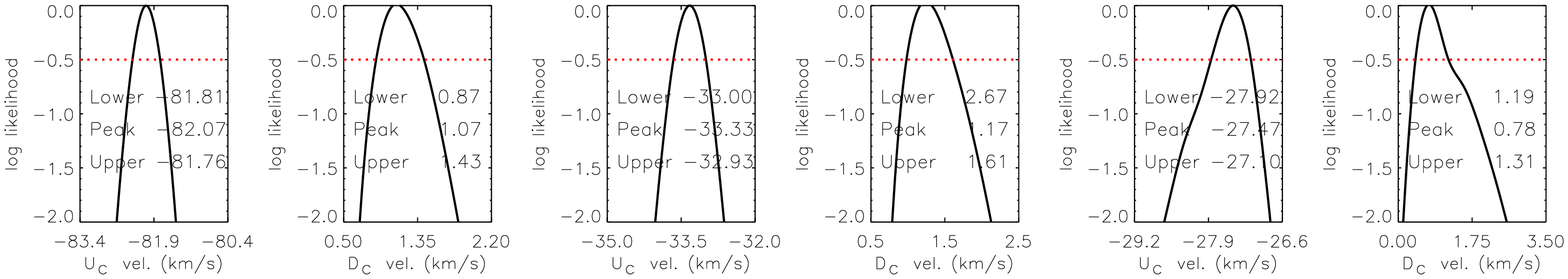}
\end{minipage}
\caption{Open cluster Pismis18: Results of the maximum likelihood analysis. See Fig.~\ref{figB:1} for description of individual plots.}
\label{figB:27}
\end{figure*}

\begin{figure*}
\begin{minipage}{0.98\textwidth}
\centering
\includegraphics[width = 145mm]{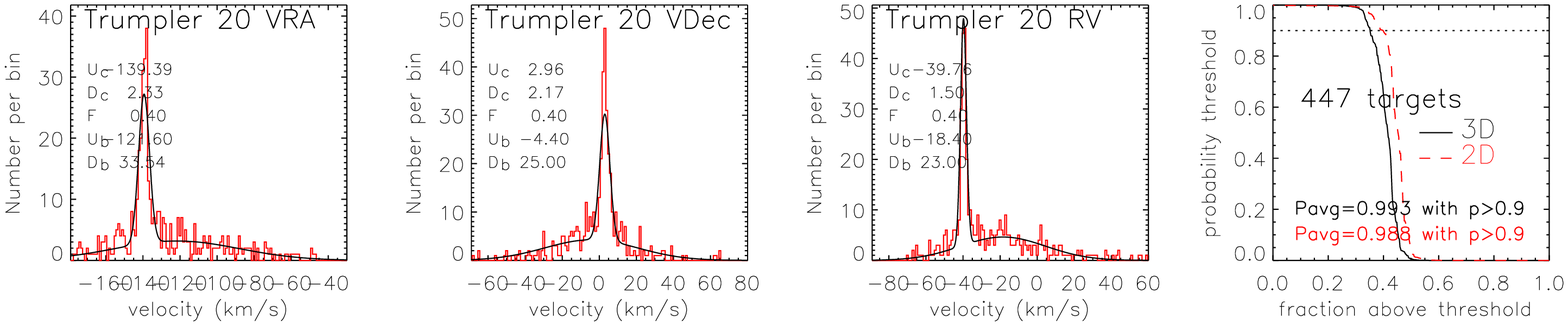}\\
\includegraphics[width = 145mm]{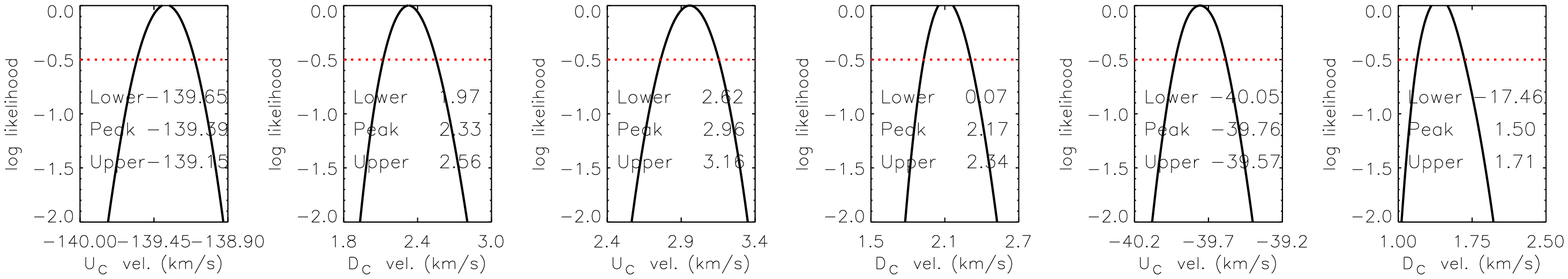}
\end{minipage}
\caption{Open cluster Trumpler 20: Results of the maximum likelihood analysis. See Fig.~\ref{figB:1} for description of individual plots.}
\label{figB:28}
\end{figure*}

\begin{figure*}
\begin{minipage}[t]{0.98\textwidth}
\centering
\includegraphics[width = 145mm]{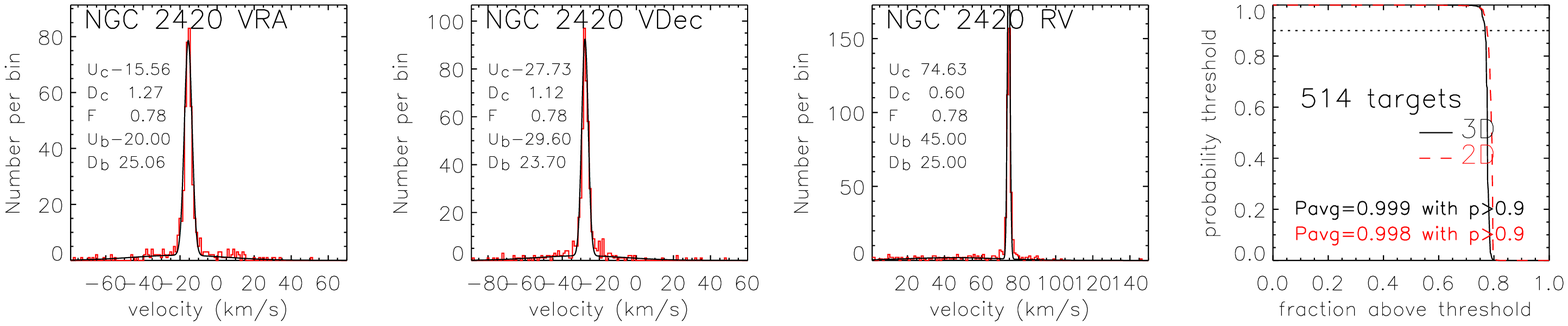}\\
\includegraphics[width = 145mm]{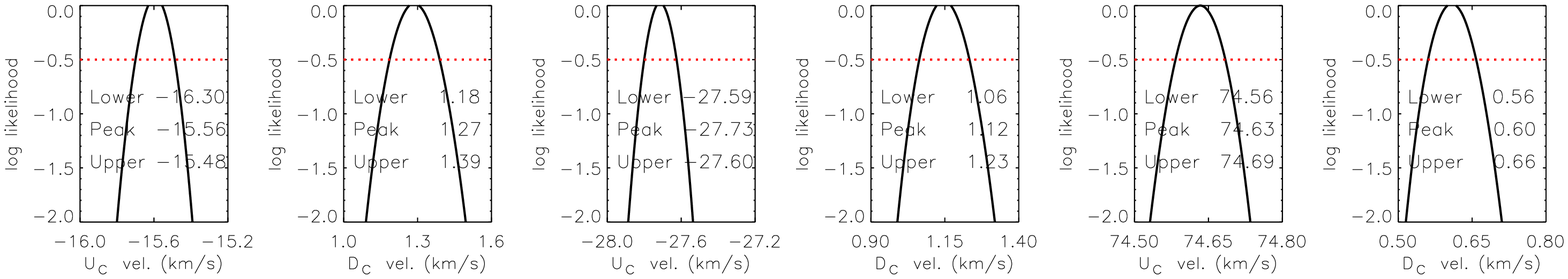}
\end{minipage}
\caption{Open cluster NGC 2420: Results of the maximum likelihood analysis. See Fig.~\ref{figB:1} for description of individual plots.}
\label{figB:29}
\end{figure*}

\begin{figure*}
\begin{minipage}{0.98\textwidth}
\centering
\includegraphics[width = 145mm]{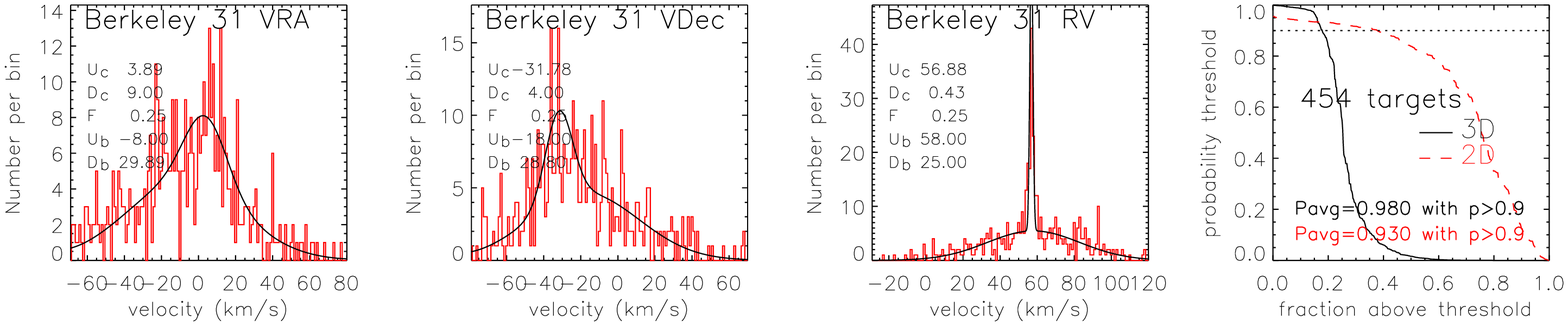}\\
\includegraphics[width = 145mm]{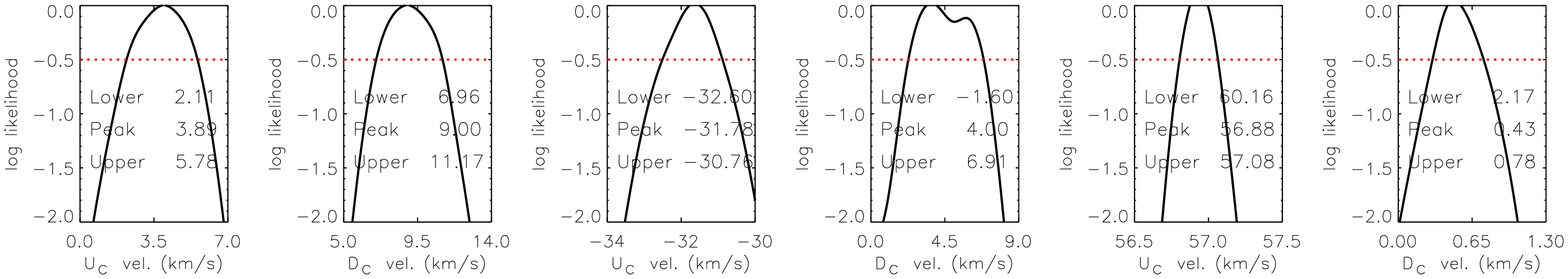}
\end{minipage}
\caption{Open cluster Berkeley 31: Results of the maximum likelihood analysis. See Fig.~\ref{figB:1} for description of individual plots.}
\label{figB:30}
\end{figure*}

\begin{figure*}
\begin{minipage}[t]{0.98\textwidth}
\centering
\includegraphics[width = 145mm]{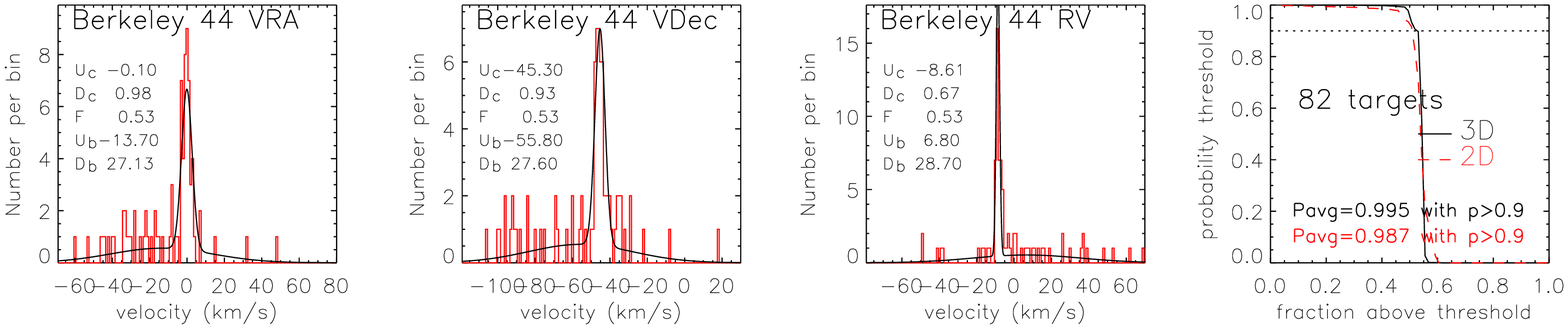}\\
\includegraphics[width = 145mm]{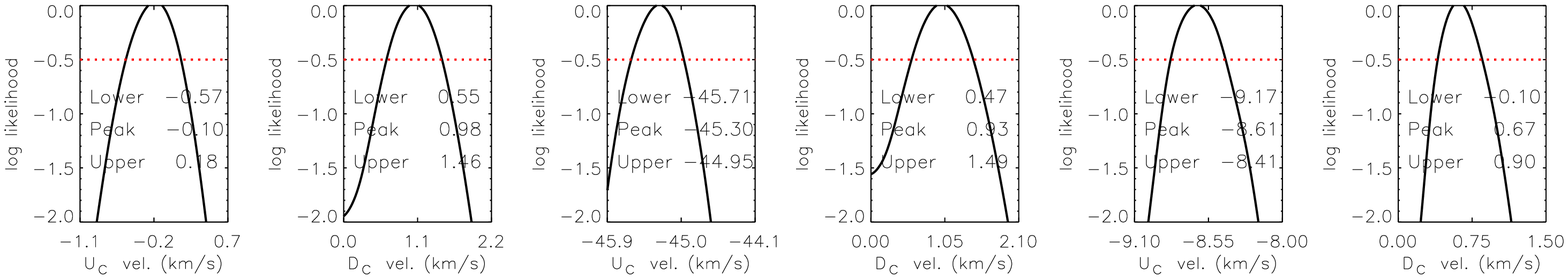}
\end{minipage}
\caption{Open cluster Berkeley 44: Results of the maximum likelihood analysis. See Fig.~\ref{figB:1} for description of individual plots.}
\label{figB:31}
\end{figure*}

\begin{figure*}
\begin{minipage}{0.98\textwidth}
\centering
\includegraphics[width = 145mm]{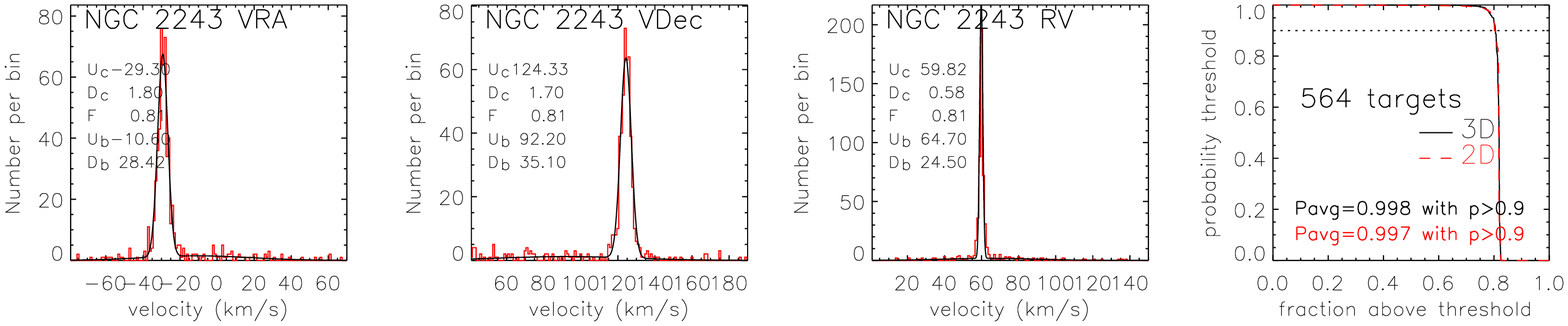}\\
\includegraphics[width = 145mm]{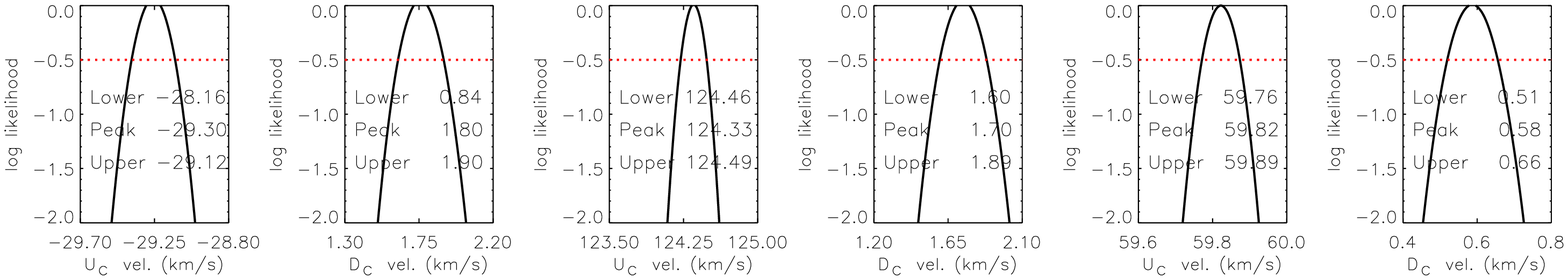}
\end{minipage}
\caption{Open cluster NGC 2243: Results of the maximum likelihood analysis. See Fig.~\ref{figB:1} for description of individual plots.}
\label{figB:32}
\end{figure*}

Figures~\ref{figB:1}--\ref{figB:32} (available online only) graphically show the results of the maximum likelihood analysis procedure, along with the distribution of membership probabilities (described in Sections~\ref{3.2} and~\ref{3.3}), for each of the 32 GES clusters considered in this paper.

\label{lastpage}
\clearpage

\newpage

\newpage

\newpage

\newpage

\clearpage
\newpage

\newpage

\newpage

\newpage

\newpage

\newpage

\end{document}

%% file: latex_macros.tex
\newcommand{\be}{\begin{equation}}
\newcommand{\ee}{\end{equation}}
\newcommand{\bd}{\begin{displaymath}}
\newcommand{\ed}{\end{displaymath}}